\documentclass[journal]{IEEEtran}
 
\usepackage{cite}
\usepackage[dvips]{graphicx}
\usepackage[cmex10]{amsmath}
\usepackage{color}
\usepackage{bm}
\usepackage{theorem} 
\usepackage{amscd}  
\usepackage{amssymb}
\usepackage {subfigmat}  
\usepackage{multirow}

{\theorembodyfont{\normalfont\it}
\theoremheaderfont{\normalfont\bf}
\newtheorem{thm}{ Theorem}
\newtheorem{dfn}[thm]{ Definition}
\newtheorem{lmm}[thm]{ Lemma}

\newtheorem{crl}[thm]{ Corollary}
\newtheorem{asm}[thm]{ Assumption}
\newtheorem{prp}[thm]{ Proposition}
\newtheorem{cjt}[thm]{ Conjecture}}

{\theorembodyfont{\normalfont}
\theoremheaderfont{\normalfont\it}
\newtheorem{prf}{ Proof:}}

{\theorembodyfont{\normalfont}
\theoremheaderfont{\normalfont\it}
}

{\theorembodyfont{\normalfont}
\theoremheaderfont{\normalfont\it}
}

{\theorembodyfont{\normalfont}
\theoremheaderfont{\normalfont\it}
}

{\theorembodyfont{\normalfont}
\theoremheaderfont{\normalfont\it}
}

{\theorembodyfont{\normalfont}
\theoremheaderfont{\normalfont\it}
}

{\theorembodyfont{\normalfont}
\theoremheaderfont{\normalfont\it}
\newtheorem{rmk}{ Remark.}}

\newcommand{\bra}[1]{\mbox{$\langle#1|$}}

\newcommand{\ket}[1]{\mbox{$|#1\rangle$}}

\newcommand{\inpro}[2]{\mbox{$\left\langle#1|#2\right\rangle$}}
\newcommand{\outpro}[2]{\mbox{$\ket{#1}\!\bra{#2}$}}

\newcommand{\proj}[1]{\mbox{$\ket{#1}\!\bra{#1}$}}

\newcommand{\alg}[1]{\begin{align}#1\end{align}}

\newcommand{\nn}{\nonumber}
\newcommand{\ca}[1]{{\mathcal #1}}
\newcommand{\mbb}[1]{{\mathbb #1}}
\newcommand{\mfk}[1]{{\mathfrak #1}}

\newcommand{\bthm}[1]{\begin{thm}\label{thm:#1}}
\newcommand{\ethm}{\end{thm}}
\newcommand{\rthm}[1]{\ref{thm:#1}}
\newcommand{\rThm}[1]{Theorem \ref{thm:#1}}
\newcommand{\blmm}[1]{\begin{lmm}\label{lmm:#1}}
\newcommand{\elmm}{\end{lmm}}
\newcommand{\rLmm}[1]{Lemma \ref{lmm:#1}}

\newcommand{\bdfn}[1]{\begin{dfn}\label{dfn:#1}}
\newcommand{\edfn}{\end{dfn}}

\newcommand{\rDfn}[1]{Definition \ref{dfn:#1}}
\newcommand{\basm}[1]{\begin{asm}\label{asm:#1}}
\newcommand{\easm}{\end{asm}}
\newcommand{\bprp}[1]{\begin{prp}\label{prp:#1}}
\newcommand{\eprp}{\end{prp}}

\newcommand{\rPrp}[1]{Proposition \ref{prp:#1}}
\newcommand{\bcrl}[1]{\begin{crl}\label{crl:#1}}
\newcommand{\ecrl}{\end{crl}}

\newcommand{\rCrl}[1]{Corollary \ref{crl:#1}}
\newcommand{\bcjt}[1]{\begin{cjt}\label{cjt:#1}}
\newcommand{\ecjt}{\end{cjt}}

\newcommand{\bprf}{\begin{prf}}
\newcommand{\eprf}{\end{prf}}
\newcommand{\brmk}{\begin{rmk}}
\newcommand{\ermk}{\end{rmk}}
\newcommand{\laeq}[1]{\label{eq:#1}}
\newcommand{\req}[1]{(\ref{eq:#1})}

\newcommand{\QED}{\hfill$\blacksquare$}
\newcommand{\lsec}[1]{\label{sec:#1}}
\newcommand{\rsec}[1]{\ref{sec:#1}}
\newcommand{\rSec}[1]{Section \ref{sec:#1}}
\newcommand{\lapp}[1]{\label{app:#1}}
\newcommand{\rapp}[1]{\ref{app:#1}}
\newcommand{\rApp}[1]{Appendix \ref{app:#1}}

\newcommand{\bitem}{\begin{itemize}}
\newcommand{\entem}{\end{itemize}}
\newcommand{\benum}{\begin{enumerate}}
\newcommand{\ennum}{\end{enumerate}}

\newcommand{\otm}{\otimes}

\newcommand{\argmax}{\mathop{\rm arg~max}\limits}

\newcommand{\beq}{\begin{eqnarray}}
\newcommand{\eeq}{\end{eqnarray}}

\newcommand{\tr}{\mathrm{Tr}}

\begin{document}

\title{One-Shot Triple-Resource Trade-Off in Quantum Channel Coding}

\author{Eyuri Wakakuwa, and Yoshifumi Nakata \\ (Report number: YITP-20-61)

}

\maketitle

\begin{abstract}
We analyze a task in which classical and quantum messages are simultaneously communicated via a noisy quantum channel, assisted with a limited amount of shared entanglement. 
We derive direct and converse bounds for the one-shot capacity region, represented by the smooth conditional entropies and the error tolerance. 
The proof is based on the randomized partial decoupling theorem, which is a generalization of the decoupling theorem. 
The two bounds match in the asymptotic limit of infinitely many uses of a memoryless channel and coincide with the previous result obtained by Hsieh and Wilde. 
Direct and converse bounds for various communication tasks are obtained as corollaries, both for the one-shot and asymptotic scenarios. 
\end{abstract}

\begin{IEEEkeywords}
Quantum Channel Capacity, One-Shot, Decoupling
\end{IEEEkeywords}

\section{Introduction}

One of the major goals of quantum communication theory is to investigate the ultimate capacities of a noisy quantum channel for transmitting classical and quantum information (see, e.g., \cite{wildetext,nielsentext}).
In an asymptotic limit of infinitely many uses of a memoryless channel, 
the classical capacity of a noisy quantum channel was obtained by Holevo \cite{holevo98}, Schumacher and Westmoreland \cite{schumacher97};
the quantum capacity by Lloyd \cite{lloyd1997capacity}, Shor \cite{shor2002quantum} and Devetak \cite{devetak2005private};
the entanglement-assisted classical capacity by Bennet {\it et al.} \cite{bennett1999entanglement,bennett2002entanglement} and Shor \cite{shor2004classical};
the entanglement-assisted quantum capacity by Devetak {\it et al.} \cite{devetak2004family,deve08};
and the capacity of a quantum channel for simultaneously transmitting classical and quantum messages was obtained by Devetak and Shor \cite{devetak2005capacity}.
One of the most general results in this direction was shown by Hsieh and Wilde \cite{hsieh2010entanglement}, which provides a complete characterization of the triple trade-off among the resources of classical communication, quantum communication, and entanglement in quantum channel coding.
As proved in \cite{hsieh2010entanglement}, the capacity theorems in the special cases mentioned above are obtained from their result as corollaries.

A modern approach addresses the same communication tasks under the assumption that the channel is used only once, which is referred to as the {\it one-shot} regime.
Along this line, the classical capacity of a noisy quantum channel was first investigated by Mosonyi {\it et al.} \cite{mosonyi2009generalized} and was developed by Renes {\it et al.} \cite{renes2011noisy} and Wang {\it et al.} \cite{wang2012one};
the quantum capacity was obtained by Buscemi {\it et al.}\cite{buscemi2010quantum};
the entanglement-assisted classical and quantum capacities by Datta {\it et al.} \cite{datta2012one};
and the capacity region for simultaneously transmitting classical and quantum information was obtained by Salek {\it et al.} \cite{salek2019one}.
In the asymptotic limit of infinitely many uses of a memoryless channel, all these results (except \cite{mosonyi2009generalized}) are known to coincide with the preexisting results presented above.
Numerous studies have been done to develop these approaches further (see e.g.~\cite{dupuis2014decoupling,datta2011apex,datta2013smooth,matthews2014finite,datta2016second,tomamichel2016quantum,qi2018applications,anshu2018building,anshu2019near}).
However, the one-shot capacity theorem that reduces in the asymptotic limit to the result by Hsieh and Wilde \cite{hsieh2010entanglement} has not been obtained so far.

In this paper, we derive a one-shot generalization of the three-dimensional rate region obtained in \cite{hsieh2010entanglement}. 
Namely, we consider a task in which classical and quantum messages are transmitted via a noisy quantum channel with the assistance of a limited amount of shared entanglement, all in the one-shot scenario. 
The main result is that we derive inner and outer bonds for the one-shot rate region, which are represented in terms of the smooth conditional entropy of the channel.
To our knowledge, this is the first time that trade-off relations that apply simultaneously to all the three resources of classical communication, quantum communication, and shared entanglement has been obtained in the one-shot channel coding scenario.
In the asymptotic limit of infinitely many uses of the channel, the direct and converse bounds match and coincide with the achievable rate region obtained in \cite{hsieh2010entanglement}.
As proved in \cite{hsieh2010entanglement}, the aforementioned capacity theorems in the special cases, such as the Holevo-Schumacher-Westmoreland theorem \cite{holevo98,schumacher97} and the Lloyd-Shor-Devetak theorem \cite{lloyd1997capacity,shor2002quantum,devetak2005private}, are obtained from their result as corollaries.
Hence, our one-shot result also recovers the capacity theorems in the asymptotic limit in those special cases.
Besides, the direct and converse bounds for various communication tasks in the one-shot scenario readily follow from our result.
It should be noted, however, that in special cases, our result is not necessarily as strong as those in the previous literature.
To compare our results quantitatively with those in the previous literature is left as a future work.

The proof is based on the notion of {\it randomized partial decoupling} \cite{wakakuwa2021one}, which is a generalization of decoupling \cite{berry05,horo07,ADHW2009,patr08,DBWR2010,majenz2017catalytic}.
Here, we consider a scenario in which a bipartite quantum state on system $AR$ is subject to a unitary operation on $A$, followed by the action of a linear completely positive (CP) map. Unlike the usual setting of decoupling, we assume that
the subsystem is decomposed into a direct-sum form, and 
the unitary is chosen at random from the set of unitaries that are block-diagonal under the decomposition. 
In \cite{wakakuwa2021one}, we proved the {\it randomized partial decoupling theorem}, which shows that the distance between the final state and the averaged one is bounded in terms of smooth conditional entropies of the initial state and the channel.
The previous results of the one-shot decoupling theorem \cite{DBWR2010} and the dequantization theorem \cite{dupuis2014decoupling}
are obtained from this result as corollaries, up to changes in the smoothing and error parameters (see Section III D in \cite{wakakuwa2021one} for the details).
The current paper applies the randomized partial decoupling theorem to the channel coding scenario for proving the one-shot capacity theorems.

This paper is organized as follows. 
In \rSec{prelimi}, we introduce notations and definitions that will be used throughout this paper. 
In \rSec{mainresults}, we present formulations of the problem and state the main results. 
In \rSec{RPD}, we summarize the statement of the randomized partial decoupling theorem.
The proofs of the main theorems are provided in Section \rsec{direct} and \rsec{converse}.
In \rSec{iid}, we apply the one-shot results to the asymptotic scenario of infinitely many uses of the channel and obtain a three-dimensional achievable rate region that is equivalent to the one obtained by Hsieh and Wilde.
Proof of the asymptotic result is provided in \rSec{PRFasymptotic}.
In \rSec{comparison}, we apply our results to special cases where one or two of the three resources are assumed to be zero and compare the results to those in the previous literature.
Conclusions are presented in \rSec{cncl}.

\section{Preliminaries}
\lsec{prelimi}

We summarize notations and definitions that will be used throughout this paper. 

\subsection{Notations}
We denote the set of linear operators and that of Hermitian operators on a Hilbert space $\ca{H}$ by $\ca{L}(\ca{H})$ and ${\rm Her}(\ca{H})$, respectively.
For positive semidefinite operators, density operators and sub-normalized density operators, we use the following notations, respectively:
\begin{align}
&
\ca{P}(\ca{H}) = \{\rho \in {\rm Her}(\ca{H}) : \rho \geq 0 \},
\\
&
\ca{S}_=(\ca{H}) = \{\rho \in \ca{P}(\ca{H}) : \tr [\rho]=1 \},
\\
&
\ca{S}_{\leq}(\ca{H}) = \{\rho \in \ca{P}(\ca{H}) : \tr [\rho] \leq 1 \}.
\end{align}
A Hilbert space associated with a quantum system $A$ is denoted by ${\mathcal H}^A$, and its dimension is denoted by $d_A$. A system composed of two subsystems $A$ and $B$ is denoted by $AB$. When $M$  and $N$ are linear operators on ${\mathcal H}^A$ and ${\mathcal H}^B$, respectively, we denote $M\otimes N$ as $M^A\otimes N^B$ for clarity. In the case of pure states, we often abbreviate $|\psi\rangle^A\otimes|\phi\rangle^B$ as $|\psi\rangle^A|\phi\rangle^B$. 
For $X^{AB} \in \ca{L}(\ca{H}^{AB})$, $X^{A}$ represents ${\rm Tr}_B[X^{AB}]$.  
We denote $|\psi\rangle\!\langle\psi|$ simply by $\psi$.
The maximally entangled state between $A$ and $A'$, where $\ca{H}^{A} \cong \ca{H}^{A'}$, is defined by
\alg{
\ket{\Phi}^{AA'}:=\frac{1}{\sqrt{d_A}}\sum_{\alpha=1}^{d_A}\ket{\alpha}^A\ket{\alpha}^{A'}
}
with respect to a fixed orthonormal basis $\{\ket{\alpha}\}_{\alpha=1}^{d_A}$.

The identity operator is denoted by $I$. 
We denote $(M^A\otimes I^B)\ket{\psi}^{AB}$ as $M^A\ket{\psi}^{AB}$ and $(M^A\otimes I^B)\rho^{AB}(M^A\otimes I^B)^{\dagger}$ as $M^A\rho^{AB}M^{A\dagger}$. 
When ${\mathcal T}$ is a supermap from $\ca{L}(\ca{H}^{A})$ to $\ca{L}(\ca{H}^{B})$, we denote it by $\ca{T}^{A \rightarrow B}$. When $A = B$, we use $\ca{T}^{A}$ for short.
We also denote $({\mathcal T}^{A \rightarrow B} \otimes{\rm id}^C)(\rho^{AC})$ by ${\mathcal T}^{A \rightarrow B} (\rho^{AC})$.  
When a supermap is given by a conjugation of a unitary $U^A$ or a linear operator $W^{A \rightarrow B}$, we denote it by its calligraphic font as 
$
\ca{U}^{A}(X^A):= (U^{A }) X^A (U^{A })^{\dagger}
$
and
$
\ca{W}^{A \rightarrow B}(X^A):= (W^{A \rightarrow B}) X^A (W^{A \rightarrow B})^{\dagger}
$.
In that case, the adjoint map of $\ca{W}^{A \rightarrow B}$ is defined by $\ca{W}^{\dagger B \rightarrow A}(\cdot):=(W^{A \rightarrow B})^\dagger(\cdot)(W^{A \rightarrow B})$.

For any linear CP map $\ca{T}^{A\rightarrow B}$, there exist a finite dimensional quantum system $E$ and a linear operator $\Gamma_{\ca{T}}^{A\rightarrow BE}$ such that $\ca{T}^{A\rightarrow B}(\cdot)={\rm Tr}_E[\Gamma_{\ca{T}}(\cdot)\Gamma_{\ca{T}}^\dagger]$.
The operator $\Gamma_{\ca{T}}$ is called the Stinespring dilation of $\ca{T}^{A\rightarrow B}$ \cite{stinespring1955positive},
and the linear CP map defined by ${\rm Tr}_B[\Gamma_{\ca{T}}(\cdot)\Gamma_{\ca{T}}^\dagger]$ is called the {\it complementary map} of $\ca{T}^{A\rightarrow B}$.
With a slight abuse of notation, we denote the complementary map by $\ca{T}^{A\rightarrow E}$.

\subsection{Norms and Distances}

For a linear operator $X$, the trace norm is defined as $|\! | X |\! |_1 = \tr[ \sqrt{X^{\dagger}X}]$. 
The trace distance between two unnormalized states $\rho,\rho'\in\ca{P}(\ca{H})$ is defined by $\frac{1}{2}\|\rho-\rho'\|_1$. 
For subnormalized states $\rho,\rho'\in\ca{S}_\leq(\ca{H})$, the generalized fidelity and the purified distance are defined by
\alg{
&
\bar{F}(\rho,\rho')
:=
\|\sqrt{\rho}\sqrt{\rho'}\|_1
+
\sqrt{(1-{\rm Tr}[\rho])(1-{\rm Tr}[\rho'])},
\\
&
P(\rho,\rho')
:=
\sqrt{1-\bar{F}(\rho,\rho')^2},
\laeq{dfnPD}
}
respectively \cite{tomamichel2010duality}.
The trace distance and the purified distance are related as
 \alg{
\frac{1}{2}\|\rho-\rho'\|_1
\leq
P(\rho,\rho')
\leq
\sqrt{2\|\rho-\rho'\|_1}
\laeq{relTDPD}
}
for any $\rho,\rho'\in\ca{S}_\leq(\ca{H})$.
The epsilon ball of a subnormalized state $\rho\in\ca{S}_\leq(\ca{H})$ is defined by
\begin{align}
\ca{B}^\epsilon(\rho):=\{\rho'\in\ca{S}_\leq(\ca{H})|\:P(\rho,\rho')\leq\epsilon\}.
\label{eq:epsilon}
\end{align}

\subsection{One-shot entropies}

For any subnormalized state $\rho\in\ca{S}_\leq(\ca{H}^{AB})$ and normalized state $\varsigma\in\ca{S}_=(\ca{H}^{B})$, define
\alg{
&
H_{\rm min}(A|B)_{\rho|\varsigma} := \sup \{ \lambda \in \mathbb{R}| 2^{-\lambda} I^A \otimes \varsigma^B \geq \rho^{AB} \}, \label{eq:condmins} \\
&
H_{\rm max}(A|B)_{\rho|\varsigma} := \log{\|\sqrt{\rho^{AB}}\sqrt{I^A\otm\varsigma^B}\|_1^2}.
\label{eq:condmaxs}
}
The conditional min- and max- entropies (see e.g.~\cite{T16}) are defined by
\begin{align}
H_{\rm min}(A|B)_{\rho}& := \sup_{\varsigma^B \in \ca{S}_=(\ca{H}^B)}H_{\rm min}(A|B)_{\rho|\varsigma},\laeq{condmin} \\
H_{\rm max}(A|B)_{\rho}& := \sup_{\varsigma^B \in \ca{S}_=(\ca{H}^B)}H_{\rm max}(A|B)_{\rho|\varsigma},\laeq{condmax} 
\end{align}
respectively. 
The smooth conditional min- and max-entropies are defined by
\begin{align}
H_{\rm min}^\epsilon(A|B)_{\rho}& := \sup_{\hat{\rho}^{AB} \in \ca{B}^\epsilon(\rho)}H_{\rm min}(A|B)_{\hat\rho},\laeq{condminsm} \\
H_{\rm max}^\epsilon(A|B)_{\rho}& := \inf_{\hat{\rho}^{AB} \in \ca{B}^\epsilon(\rho)}H_{\rm max}(A|B)_{\hat\rho}\laeq{condmaxsm} 
\end{align}
for a smoothing parameter $\epsilon\geq0$.  
The properties of the smooth conditional entropies used in this paper are summarized in \rApp{PSE}.

\subsection{Choi-Jamiolkowski representation}

Let $\ca{T}^{A \rightarrow B}$ be a linear supermap from  $\ca{L}(\ca{H}^A)$ to $\ca{L}(\ca{H}^B)$ and let $\Phi^{AA'}$ be the maximally entangled state between $A$ and $A'$. A linear operator $\mfk{J}(\ca{T}^{A \rightarrow B})\in\ca{L}(\ca{H}^{AB})$ defined by $\mfk{J}(\ca{T}^{A \rightarrow B}) := \ca{T}^{A' \rightarrow B}(\Phi^{AA'})$ is called the {\it Choi-Jamio\l kowski representation of $\ca{T}$} \cite{J1972,C1975}. The representation is an isomorphism. For an operator $X^{AB} \in \ca{L}(\ca{H}^{AB})$, the inverse map is given by
\alg{
\mfk{J}_{A}^{-1}(X^{AB}) (\varsigma^A) = d_A \tr_A \bigl[ (\varsigma^{A^T} \otimes I^B) X^{AB} \bigr],
\laeq{CJinverse}
}
where $A^T$ denotes the transposition of $A$ with respect to the Schmidt basis of $\Phi^{AA'}$.
When $\ca{T}$ is completely positive, $\mfk{J}(\ca{T}^{A \rightarrow B})$ is an unnormalized state on $AB$ and is called the {\it Choi-Jamio\l kowski state of $\ca{T}$}.

\subsection{Haar measure}

For a unitary group of finite degree, there exists the unique left- and right- unitarily invariant probability measure, known as the Haar measure. We denote it by ${\sf H}$. The Haar measure satisfies the property that, for any unitary $U$ and a set of unitaries $\mathcal{V}$,
\begin{equation}
{\sf H}(U \mathcal{V}) = {\sf H}(\mathcal{V}U) = {\sf H}(\mathcal{V}).
\end{equation}
When a unitary $U$ is chosen uniformly at random with respect to the Haar measure, we denote it by $U \sim {\sf H}$.

\begin{figure*}[t]
\begin{center}
\includegraphics[bb={0 0 1141 404}, scale=0.37]{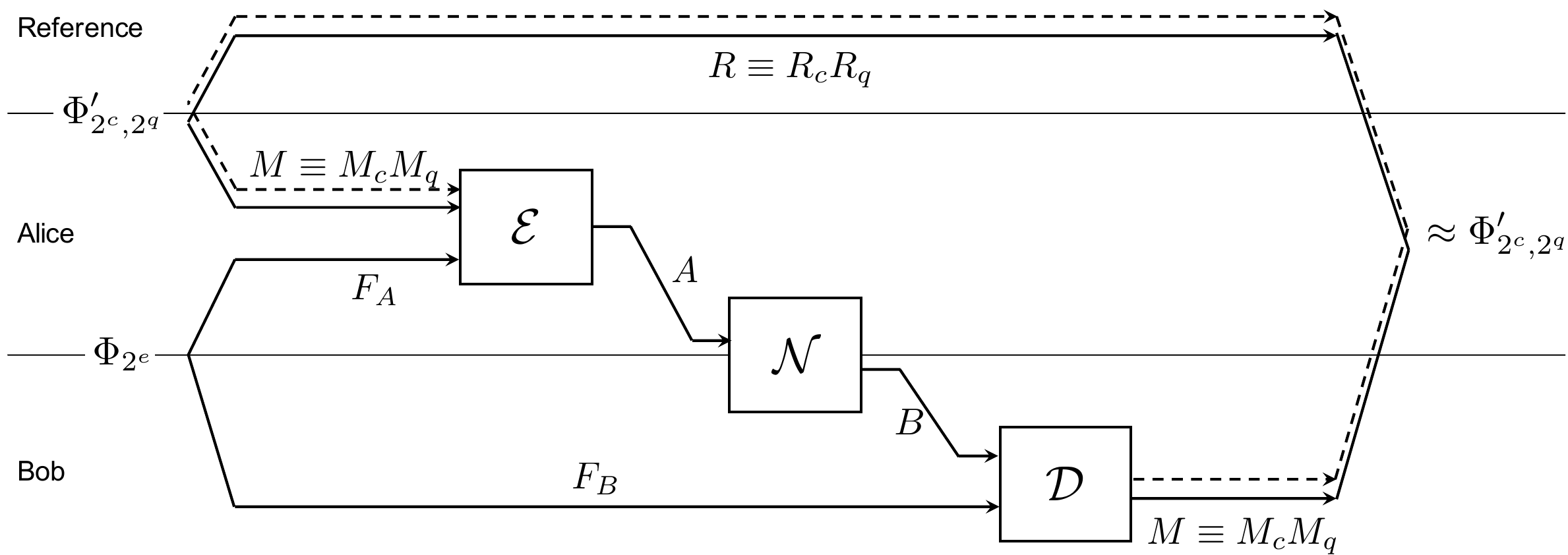}
\end{center}
\caption{
One-shot channel coding is depicted.
The normal arrows represent quantum systems, and the dotted arrows represent the classical part of the source state.
}
\label{fig:oneshotchannelcoding}
\end{figure*}

\section{One-Shot Capacity Theorems}
\lsec{mainresults}

Consider a noisy quantum channel $\ca{N}^{A \rightarrow B}$.
Suppose that the sender, Alice, transmits $c$-bits of classical and $q$-qubits of quantum messages simultaneously to the receiver, Bob, through a noisy quantum channel assisted by $e$-ebits of shared entanglement (see Figure \ref{fig:oneshotchannelcoding}). 
We assume that Bob initially has no side information about the messages. 
Our goal is to obtain the conditions for this task to be achievable within error tolerance $\delta$, in terms of $c$, $q$ and $e$.

To be more precise, let $M_q$ be a $2^q$-dimensional quantum system that represents the quantum message, and $R_q$ be a reference system with the same dimension that are inaccessible to Alice and Bob.
The quantum message is described by the maximally entangled state $\Phi_{2^q}^{M_qR_q}$ with Schmidt rank $2^q$.
With $F_A$ and $F_B$ being $2^e$-dimensional quantum registers, the entanglement resource is given by the maximally entangled state $\Phi_{2^e}^{F_AF_B}$ with Schmidt rank $2^e$.
A communication protocol is represented by a set of encoding CPTP maps $\{\ca{E}_j^{M_qF_A\rightarrow A}\}_{j=1}^{2^c}$ and a decoding instrument $\{\ca{D}_j^{BF_B\rightarrow M_q}\}_{j=1}^{2^c}$.
Let $m$ and $\hat{m}$ be the original classical message and the decoded one, respectively.
The probability that the decoded message is $\hat{m}=k$ when the original message is $m=j$ is
\alg{
p(\hat{m}= k|m=j)
=
{\rm Tr}[\ca{D}_k \!\circ\! \ca{N} \!\circ\! \ca{E}_j (\Phi_{2^q}^{M_qR_q} \!\otm\! \Phi_{2^e}^{F_AF_B})].
} 
When the decoded message is $\hat{m}=k$ and the original message is $m=j$, the state of the quantum part after the decoding is given by
\alg{
\hat{\Phi}_{j,k}^{M_qR_q}
:=
\frac{1}{p(\hat{m}= k|m=j)}\ca{D}_k \!\circ\! \ca{N} \!\circ\! \ca{E}_j (\Phi_{2^q}^{M_qR_q} \!\otm\! \Phi_{2^e}^{F_AF_B}).
}
The average probability of error in decoding the classical message is calculated to be
\alg{
\delta_C
:=
\frac{1}{2^c}
\sum_{j=1}^{2^c}
p(\hat{m}\neq j|m=j),
\laeq{oss}
}
and that for the quantum message is
\alg{
\delta_Q
&
:=\frac{1}{2^c}
\sum_{j,k=1}^{2^c}
p(\hat{m}= k|m=j)
\left\|
\hat{\Phi}_{j,k}^{M_qR_q} \! -\! \Phi_{2^q}^{M_qR_q}
\right\|_1.
\laeq{oss2}
}
Both the classical and quantum errors shall be small, i.e., we want $\delta_C$ and $\delta_Q$ to be below a certain threshold $\delta>0$.

A rigorous definition of a code is given as follows.
Here, we treat the classical and quantum parts of the message as one large quantum system, because it is more convenient in our analysis.

\bdfn{oneshotcode}
Consider the setting described above.
Let $M_c$ and $R_c$ be $2^c$-dimensional quantum system with a fixed orthonormal basis $\{\ket{j}\}_{j=1}^{2^c}$.
We denote $M_cM_q$ by $M$ and $R_cR_q$ by $R$ for brevity.
Let $\Phi_{2^c,2^q}'^{MR}$ be a source state defined by
\alg{
\Phi_{2^c,2^q}'^{MR}=\frac{1}{2^c}\sum_{j=1}^{2^c}\proj{j}^{M_c}\otm\proj{\Phi_{2^q}}^{M_qR_q}\otm\proj{j}^{R_c}.
\laeq{sourcestate}
}
A pair of an encoding CPTP map $\ca{E}^{MF_A\rightarrow A}$ and a decoding CPTP map $\ca{D}^{BF_B\rightarrow M}$ is called a $(c,q,e,\delta)$ code for the channel $\ca{N}$ if it holds that
\alg{
\left\|
 \ca{D} \circ \ca{N} \circ \ca{E} (\Phi_{2^c,2^q}'^{MR}\otm\Phi_{2^e}^{F_AF_B})  - \Phi_{2^c,2^q}'^{MR}
\right\|_1
\leq 
\delta \laeq{tsuzuiteku}
}
and
\alg{
&
\ca{E}^{MF_A\rightarrow A}=\ca{E}^{MF_A\rightarrow A}\circ\ca{C}^{M_c},
\laeq{encdeccls}\\
&\ca{D}^{BF_B\rightarrow M}=\ca{C}^{M_c}\circ\ca{D}^{BF_B\rightarrow M},
\laeq{encdecclst}
}
where $\ca{C}$ is the completely dephasing operation on $M_c$ with respect to the basis $\{\ket{j}\}_{j=1}^{2^c}$.
\edfn

\noindent
Note that the condition \req{tsuzuiteku} is equivalent to the condition that $\delta_C,\delta_Q\leq\delta$, up to a constant multiplication.
The correspondence between the encoding and decoding operations is given by
\alg{
&
\ca{E}_j^{M_qF_A\rightarrow A}(\cdot)
=
\ca{E}^{MF_A\rightarrow A}( \proj{j}^{M_c}\otm(\cdot)^{M_qF_A}),
\\
&
 \ca{D}_j^{BF_B\rightarrow M_q}(\cdot)
 =
 \bra{j}^{M_c}\ca{D}^{BF_B\rightarrow M}(\cdot) \ket{j}^{M_c}
}
and
\alg{
&
\ca{E}^{MF_A\rightarrow A}(\cdot)
=
\sum_{j=1}^{2^c}\ca{E}_j(\bra{j}^{M_c}(\cdot)\ket{j}^{M_c}),
\\
&
 \ca{D}^{BF_B\rightarrow M}(\cdot)
 =
 \sum_{j=1}^{2^c}
 \proj{j}^{M_c}\otm\ca{D}_j^{BF_B\rightarrow M_q}(\cdot).
}

It should be noted that the capacity theorems obtained in terms of the average probability of error, as in \req{oss}, \req{oss2} and \req{tsuzuiteku}, are translated into those based on the worst-case error, up to halving of the message length. See, for example, Corollary 1 in \cite{buscemi2010quantum} and Theorem 11 in \cite{datta2012one} for the quantum part and Lemma 1 in \cite{renes2011noisy} for the classical part. The latter is known as the expurgation trick (see e.g.~\cite{wildetext}).

\begin{figure*}[t!]
\begin{center}
\includegraphics[bb={0 40 1224 286}, scale=0.3]{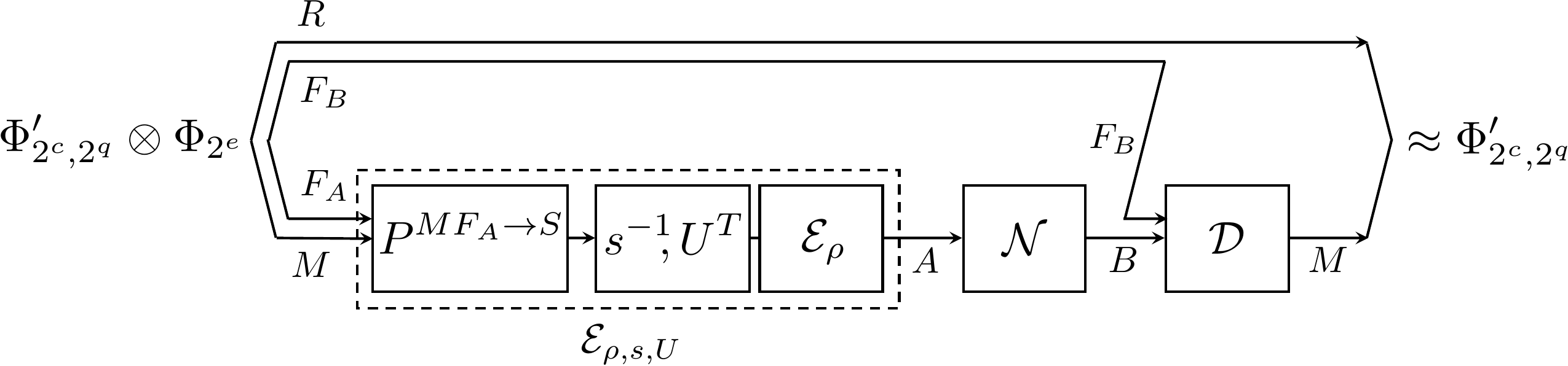}
\end{center}
\caption{
The protocol for communication over the channel $\ca{N}$ constructed in terms of randomized partial decoupling is depicted.
The encoding operation $\ca{E}_{\rho,s,U}$ is composed of (i) a linear isometry $P^{MF_A\rightarrow S}$ that embeds the message system $M$ and Alice's share of the entanglement resource $F_A$ to a larger system $S$, (ii) the permutation $s$ and the unitary $U$ that appear in randomized partial decoupling, and (iii) a linear CPTP map $\ca{E}_\rho$ that is obtained from $\rho^{SA}$ by the Choi-Jamiolkowski correspondence. 
The explicit form of the decoder $\ca{D}$ is left open, because we only prove the {\it existence} of a proper decoder $\ca{D}$ in the proof of the direct part.
}
\label{fig:directprotocol}
\end{figure*}

\subsection{Channel Capacity with Limited Entanglement}

First, we consider the situation in which the amount of the resource of shared entanglement is limited.

\bdfn{achievableratelim}
A triplet $(c,q,e)$ is said to be achievable within the error $\delta$ for the channel $\ca{N}^{A\rightarrow B}$ if there exists a $(c,q,e,\delta)$ code for $\ca{N}^{A\rightarrow B}$.
\edfn

\noindent
The direct part is represented by the following theorem.  
The proof is based on the direct part of randomized partial decoupling (\rThm{SmoothExMarkov}), and will be provided in \rSec{direct}.
A protocol that achieves the direct bound is depicted in Figure \ref{fig:directprotocol}.

\bthm{OSDcomp}
Let $S_r$ be a finite dimensional quantum system and let $S_c$ be a quantum system with a fixed orthonormal basis $\{\ket{j}\}_{j=1}^{d_{S_c}}$ such that $d_{S_c}\geq2$.
We denote $S_cS_r$ by $S$.
Consider a state in the form of
\alg{
\rho^{SA}=\frac{1}{d_{S_c}}\sum_{j=1}^{d_{S_c}}\proj{j}^{S_c}\otm\rho_j^{S_rA}, 
\laeq{yapzet23}
 }
where$\{\rho_j\}_{j=1}^{d_{S_c}}$ is a set of normalized states on $S_rA$
such that $\rho^{S}$ is the full-rank maximally mixed state on $S$.
For any such $S_c$, $S_r$, $\rho^{SA}$, any $\delta_1,\delta_2>0$ and $\epsilon\geq0$, a triplet $(c,q,e)$ is achievable within the error 
\alg{
\delta=
2\sqrt{\sqrt{\delta_1}+\sqrt{\delta_2}+4\epsilon}
\laeq{thesixthings}
}
 for the channel $\ca{N}^{A\rightarrow B}$
if $d_{S_c}\geq2^c$ and the following three inequalities hold: 
\alg{
q+e
&
\leq
\log{d_{S_r}},
\laeq{righton23}\\
c+q-e
&
\leq
-H_{\rm max}^{\epsilon}(S|B)_{\ca{N}(\rho)}
\nn\\
&
\quad\quad\quad\quad
+\log{(d_{S_c}-1)}
+\log{\delta_1},
\laeq{rightonf23}
\\
q-e
&
\leq
-H_{\rm max}^{\epsilon}(S_r|BS_c)_{\ca{N}(\rho)}
+\log{\delta_2}.
\laeq{rightoff23}
}
The same statement holds in the cases of $(c=0,d_{S_c}=1)$ and $(q=e=0,d_{S_r}=1)$.
In the former case, the condition \req{rightonf23} is removed and $\delta_1$ in \req{thesixthings} is assumed to be zero.
In the latter, the condition \req{rightoff23} is removed and $\delta_2$ in \req{thesixthings} is considered to be zero.
\ethm

\noindent
The converse part is stated by the following theorem, which will be proved in \rSec{converse} based on the converse part for randomized partial decoupling (\rThm{converse}).

\bthm{converseonehyb}
Suppose that a triplet $(c,q,e)$ is achievable within the error $\delta$ for the channel $\ca{N}^{A\rightarrow B}$.
Then, there exist a quantum system $S$ satisfying $d_S\leq2^{c+q+e}$ and a state $\rho^{SA}$ such that the following conditions hold.
First, $S$ is composed of finite dimensional quantum systems $S_c$ and $S_r$,
where $S_c$ is equipped with an orthonormal basis $\{\ket{j}\}_{j=1}^{d_{S_c}}$. 
Second, the state $\rho^{SA}$ is in the form of
\alg{
\rho^{SA}=\frac{1}{d_{S_c}}\sum_{j=1}^{d_{S_c}}\proj{j}^{S_c}\otm\rho_j^{S_rA},
\laeq{yapururu}
 } 
where $\{\rho_j\}_{j=1}^{d_{S_c}}$ is a set of normalized states on $S_rA$
and $\rho^{S}$ is the full-rank maximally mixed state on $S$.
Third, for any $\iota\in(0,1]$, it holds that
\alg{
q+e
&
\leq
\log{d_{S_r}},
\laeq{convineq1}\\
c+q-e
&
\leq
-H_{\rm max}^{\lambda}(S|B)_{\ca{N}(\rho)}
+\log{d_{S_c}}-
\log{\iota},
\laeq{convineq2}\\
q-e
&
\leq
-H_{\rm max}^{\lambda'}(S_r|BS_c)_{\ca{N}(\rho)}-\log{\iota}.
\laeq{convineq3}
}
The smoothing parameters $\lambda$ and $\lambda'$ are given by
\alg{
\lambda:=
&
2\sqrt{\iota+2x^2}
+x
+2x^2,
\laeq{smlam}\\
\lambda':=
&
\sqrt{4\sqrt{\iota+2x}+2\sqrt{x}+(4\sqrt{\iota+8}+24) x}
\laeq{smlamp}
}
and $x:=2\sqrt[8]{\delta}$.
\ethm

\subsection{Channel Capacity with Free Entanglement}

Second, we consider the situation in which the resource of shared entanglement is freely available.
Following \cite{datta2012one}, we assume that the entanglement resource is given in the form of the maximally entangled state.

\bdfn{achievablerateunlim}
A pair $(c,q)$ is said to be achievable within the error $\delta$ for the channel $\ca{N}^{A\rightarrow B}$ with the assistance of entanglement if there exists $e\geq0$ such that a triplet $(c,q,e)$ is achievable within the error $\delta$ for $\ca{N}^{A\rightarrow B}$.
\edfn

\noindent
The direct and converse bounds for the scenario of free entanglement immediately follow from the direct bound and the converse bound for the case of limited entanglement, i.e., from \rThm{OSDcomp} and \rThm{converseonehyb}.

\bcrl{OSDcompunlim}
For any $\epsilon\in[0,1/2)$ and $\delta'\in(0,1-2\epsilon]$, a pair $(c,q)$ is achievable within the error
\alg{
\delta=
2\sqrt{\sqrt{2\delta'}+\sqrt{\delta'}+4\epsilon}
\laeq{thesixthingsunlim}
}
for the channel $\ca{N}^{A\rightarrow B}$  with the assistance of entanglement, 
if there exist a quantum system $S$ and a state $\rho^{SA}$ 
such that $\rho^{S}$ is the full-rank maximally mixed state on $S$ and the following inequality holds: 
\alg{
c+2q
\leq
\log{d_{S}}-H_{\rm max}^{\epsilon}(S|B)_{\ca{N}(\rho)}
+\log{\delta'}.
\laeq{rightonf23unl}
}
\end{crl}

\bcrl{converseonehybunlim}

Suppose that a pair $(c,q)$ is achievable within the error $\delta$ for the channel $\ca{N}^{A\rightarrow B}$ with the assistance of entanglement.
Then, there exist a quantum system $S$ and a state $\rho^{SA}$ such that $\rho^{S}$ is the full-rank maximally mixed state on $S$
and for any $\iota\in(0,1]$, it holds that
\alg{
c+2q
\leq
\log{d_{S}}-H_{\rm max}^{\lambda}(S|B)_{\ca{N}(\rho)}
-
\log{\iota}.
}
The smoothing parameter $\lambda$ is given by \req{smlam}.
\end{crl}

\noindent
{\bf Proof of Corollaries:}
\rCrl{converseonehybunlim} immediately follows from Inequalities \req{convineq1} and \req{convineq2} in \rThm{converseonehyb}.
To prove \rCrl{OSDcompunlim} from \rThm{OSDcomp},
suppose that there exists a state $\rho^{SA}$ that satisfies the conditions in \rCrl{OSDcompunlim}.
Let $S_c'$ be a system such that $d_{S_c'}\geq2^c$.
Define $S_r':=S$, $S':=S_c'S_r'$ and
consider a state
\alg{
\rho'^{S'A}
:=
\frac{1}{d_{S_c'}}
\sum_{j=1}^{d_{S_c'}}\proj{j}^{S_c'}\otm\rho^{S_r'A}.
}
Due to the property of the smooth max entropy for product states (\rLmm{Hmaxprod}), we have
\alg{
H_{\rm max}^{\epsilon}(S_r'|BS_c')_{\ca{N}(\rho')}
&
=
H_{\rm max}^{\epsilon}(S|B)_{\ca{N}(\rho)}
\laeq{wantit2}\\
&
\geq
H_{\rm max}^{\epsilon}(S'|B)_{\ca{N}(\rho')}
-
\log{d_{S_c'}}.
\laeq{wantit}
}
It follows from \req{wantit2} and \req{rightonf23unl} that
\alg{
q-\log{d_{S_r'}}
\leq
-H_{\rm max}^{\epsilon}(S_r'|BS_c')_{\ca{N}(\rho')}
+\log{\delta'}-c-q.
}
Thus, there exists $e\in\mbb{R}$ such that
\alg{
\begin{cases}
q+e
&\leq 
\log{d_{S_r'}},
\\
c+q-e
&\leq
-H_{\rm max}^{\epsilon}(S_r'|BS_c')_{\ca{N}(\rho')}
+\log{\delta'}.
\laeq{mainichi}
\end{cases}
}
We may assume that $e\geq0$, since $q\leq \log{d_{S_r'}}$.
This is because the dimension bound for the smooth max entropy (see \rLmm{tomerare} in \rApp{PSE}) imply
\alg{
2q
\leq
\log{d_{S}}-H_{\rm max}^{\epsilon}(S|B)_{\ca{N}(\rho)}
\leq2\log{d_{S}}
+
\log{\left(\frac{\delta'}{1-2\epsilon}\right)}.
\nn
}
The second inequality in \req{mainichi} leads to
\alg{
c+q-e
&\leq
\!-\!H_{\rm max}^{\epsilon}(S'|B)_{\ca{N}(\rho')}
\!+\!\log{(d_{S_c'}\!-\!1)}\!+\!\log{2\delta'}
\laeq{mainichi2}
}
due to Inequality \req{wantit} and the relation $d_{S_c'}/(d_{S_c'}-1)\leq2$,
and to \alg{
q-e
\leq
-H_{\rm max}^{\epsilon}(S_r'|BS_c')_{\ca{N}(\rho')}
+\log{\delta'}
\laeq{mainichi3}
}
since $c\geq0$.
Combining \req{mainichi2}, \req{mainichi3} and the first inequality in \req{mainichi} with \rThm{OSDcomp}, we complete the proof.
\QED

\begin{figure*}[t]
\begin{center}
\includegraphics[bb={0 40 1369 582}, scale=0.3]{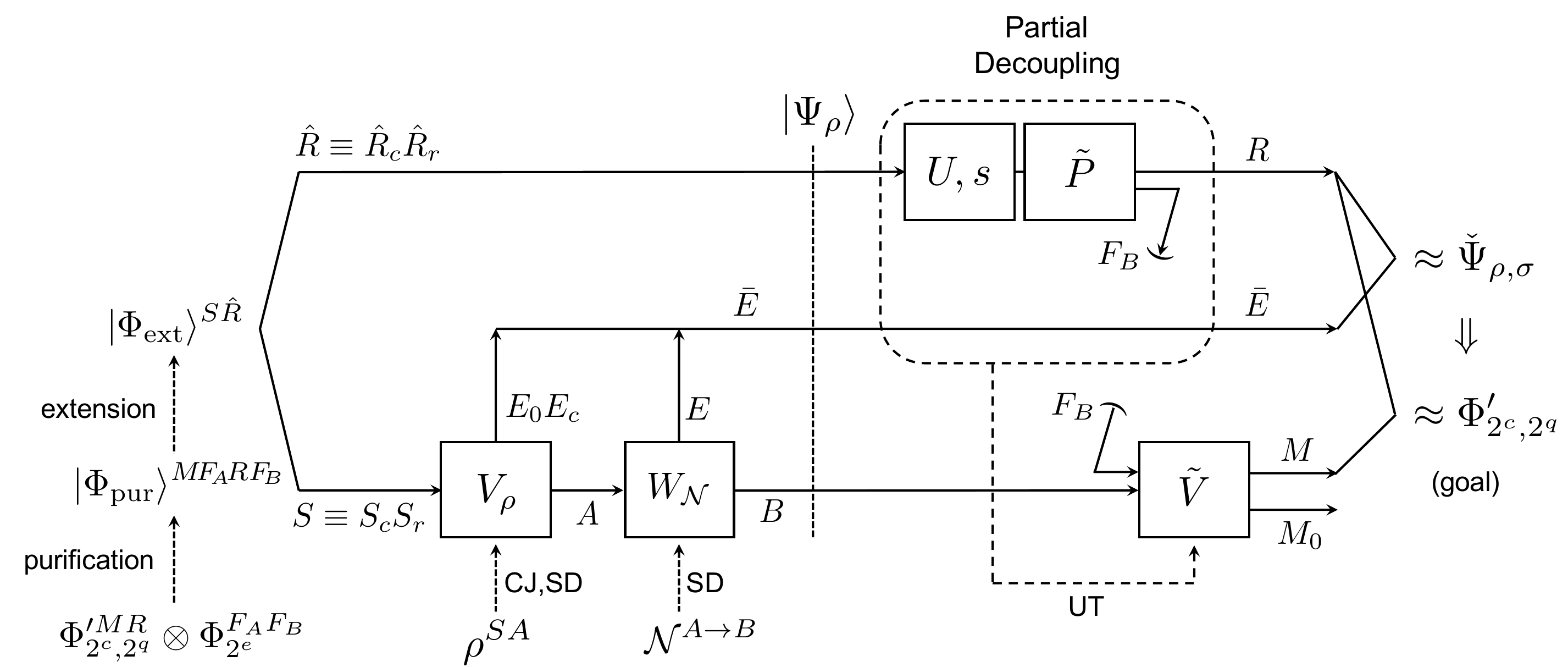}
\end{center}
\caption{
The definition of the state $|\Psi_\rho\rangle$ given by \req{narutame} and its transformation by partial decoupling are depicted.
CJ, SD and UT stand for the Choi-Jamiolkowski correspondense, the Stinespring dilation and Uhlmann's theorem, respectively.
Note that $R\equiv R_cR_q$ and $M\equiv M_cM_q$.
In composing the encoding operation, we use the fact that the actions of $U$, $s$ and $\tilde{\ca{P}}$ on system $\hat{R}$ are replaced by those of $\tilde{\ca{P}}$, $G_s^T$ and $U^T$ in this order, on system $\hat{S}$ before applying $V_\rho$. 
This trick was introduced in \cite{ADHW2009} to analyze the quantum channel capacity in an asymptotic scenario.}
\label{fig:direct}
\end{figure*}

\section{Randomized Partial Decoupling} \lsec{RPD}

In this section, we briefly review a task that we call randomized partial decoupling \cite{wakakuwa2021one} and present the direct and converse bounds for it.
This is a generalization of the decoupling theorem in the version of \cite{DBWR2010}. For the details and proofs, see the paper by the same authors \cite{wakakuwa2021one}.

Randomized partial decoupling is a task in which a bipartite quantum state $\Psi^{AR}$ is transformed by a unitary operation on $A$ and then is subject to the action of a linear CP map $\ca{T}^{A\rightarrow E}$. 
We assume that the Hilbert space $\ca{H}^A$ is decomposed into a direct-sum form as $\ca{H}^A=\bigoplus_{j=1}^J  \ca{H}_j^A$, where each $\ca{H}_j^{A}\:(j=1,\cdots,J)$ has the same dimension $r$. 
Let ${\mathcal H}^{A_c}$ be a $J$-dimensional Hilbert space with a fixed orthonormal basis $\{|j\rangle\}_{j=1}^J$ and ${\mathcal H}^{A_r}$ be an $r$-dimensional Hilbert space. 
The Hilbert space $\ca{H}^A$ is isomorphic to a tensor product Hilbert space ${\ca H}^{A_c} \otimes{\ca H}^{A_r}$, i.e., $A\cong A_cA_r$.
In terms of this decomposition, any state $\Psi^{AR}$ is written as
\alg{
\Psi^{AR}=\sum_{j,k=1}^J\outpro{j}{k}^{A_c}\otm\Psi_{jk}^{A_rR},
}
where $\Psi_{jk}^{A_rR}:=\Pi_j^A\Psi^{AR}\Pi_k^A$ with $\Pi_j$ being the projection onto $\ca{H}_j^{A}$. 
By properly choosing orthonormal bases, a maximally entangled state $\ket{\Phi}^{AA'}$ is represented by
\alg{
|\Phi\rangle^{AA'}=\left(\frac{1}{\sqrt{J}}\sum_{j=1}^J\ket{j}^{A_c}\ket{j}^{A_c'}\right)\otm\left(\frac{1}{\sqrt{r}}\sum_{\alpha=1}^r\ket{\alpha}^{A_r}\ket{\alpha}^{A_r'}\right),
\laeq{maxentdfn}
} 
where $A_c\cong A_c'$ and $A_r\cong A_r'$.

\begin{figure}[t]
\begin{center}
\includegraphics[bb={0 0 780 180}, scale=0.3]{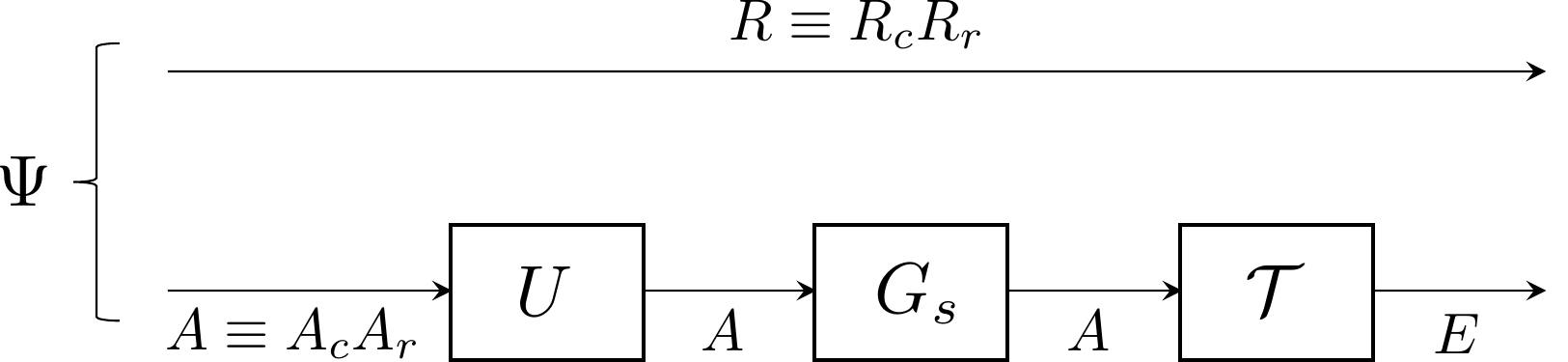}
\caption{
The procedure used in the randomized partial decoupling is depicted. For a given initial state $\Psi^{AR}$, a random unitary $U$ in the form of \req{RUrpd} and a permutation $G_s$ are first applied to $A$. Then, the system $A$ is mapped to another one $E$ by $\mathcal{T}$. 
}
\label{Fig:test}
\end{center}
\end{figure}

Consider a random unitary $U$ on $A$ in the form of
\begin{align}
U:=\sum_{j=1}^J\outpro{j}{j}^{A_c}  \otimes U_j^{A_r}.
\laeq{RUrpd}
\end{align}
where $U_j$ is independently chosen for each $j$ from the Haar measure ${\sf H}$ on the unitary group on $\ca{H}^{A_r}$.
For any state $\Psi^{AR}$, the averaged state after the action of this random unitary is given by
\begin{align}
\Psi_{\rm av}^{AR}
&:=\mbb{E}_{U_1, \dots, U_J \sim {\sf H}} [ 
U^A ( \Psi^{AR} ) U^{\dagger A}]
\laeq{qui}\\
&
=\sum_{j=1}^Jp_j \proj{j}^{A_c}\otimes \pi^{A_r}\otm\Psi_{j}^{R}. 
\end{align}
Here, $\pi^{A_r}$ is the maximally mixed state on $\ca{H}^{A_r}$, $\{p_j\}_{j=1}^J$ is a probability distribution defined by $p_j:={\rm Tr}[\bra{j}^{A_c}\Psi^{AR}\ket{j}^{A_c}]$, and $\Psi_{j}^{R}$ is a normalized state on $\ca{H}^R$ defined by $\Psi_{j}^{R}:=p_j^{-1}\Psi_{jj}^{R}$.
In the following, we denote $\mbb{E}_{U_1, \dots, U_J \sim {\sf H}}$ simply by $\mbb{E}_{U}$ when there is no ambiguity.
Let $\mbb{P}$ be the permutation group on $\{ 1,\cdots,J \}$ and  define a unitary $G_s$ for each $s\in\mbb{P}$ by
\alg{
G_s:=\sum_{j=1}^J\outpro{s(j)}{j}^{A_c}  \otimes I^{A_r}.
\label{eq:RPrpd}
}
The permutation $s$ is chosen at random according to the uniform distribution on $\mbb{P}$.
Our concern is how close the final state $\ca{T}^{A \rightarrow E} \circ \ca{G}_s^A  \circ \ca{U}^A ( \Psi^{AR} )$ is, on average over all $U$, to the averaged final state $\ca{T}^{A \rightarrow E} \circ \ca{G}_s^A ( \Psi_{\rm av}^{AR} )$, for typical choices of the permutation $s$ (see Figure \ref{Fig:test} as well).

For simplicity of analysis, we assume that $R\cong R_cR_r$, where $R_c$ is a quantum system with dimension $J$.
We also assume that $\Psi^{AR}$ is decomposed in the form of
\begin{align}
\!\!\!\!\!
\Psi^{AR}=\sum_{k,l=1}^J\outpro{k}{l}^{A_c}\!\otimes\psi_{kl}^{A_rR_r}\!\otimes\outpro{k}{l}^{R_c}\!,\!\laeq{romanof}
\end{align}
where $\psi_{kl}\in\ca{L}(\ca{H}^{A_r}\otimes\ca{H}^{R_r})$ for each $k$ and $l$.
Such states are called {\it classically coherent states} \cite{dupuis2014decoupling}.

The following theorem is the direct part of the randomized partial decoupling theorem.

\bthm{SmoothExMarkov}(Theorem 3 in \cite{wakakuwa2021one})
Consider a linear CP map $\ca{T}^{A \rightarrow E}$ and
a state $\Psi^{AR}\in\ca{S}_=(\ca{H}^{AR})$ that is decomposed as \req{romanof}.
Let $U$ and $G_s$ be random unitaries defined by (\ref{eq:RUrpd}) and (\ref{eq:RPrpd}), respectively.
Define the partial decoupling error $\Delta_{s,U}(\ca{T},\Psi)$ by
\begin{equation}
\Delta_{s,U}(\ca{T},\Psi)
:=
\left\|
\ca{T}^{A \rightarrow E} \circ \ca{G}_s^A  \bigl( \ca{U}^A ( \Psi^{AR} ) -\Psi_{\rm av}^{AR} ) \bigr)
\right\|_1,
\nn
\end{equation}
where
$
\Psi_{\rm av}^{AR}:=\mbb{E}_{U_1, \dots, U_J \sim {\sf H}} [ \ca{U}^A ( \Psi^{AR} )]
$.
Then, for any $\epsilon,\mu\geq0$,
it holds that
\begin{align}
\mbb{E}_{s,U } [\Delta_{s,U}(\ca{T},\Psi) ]
\leq  
\theta_I+\theta_{I\!I}
+4(\epsilon+\mu+\epsilon\mu).
\laeq{SmExMa}
\end{align}
The terms $\theta_I$ and $\theta_{I\!I}$ are represented by
\alg{
\!
\theta_I=
\begin{cases}
2^{-\frac{1}{2}H_I}
 &\!\!\! (J\geq2)
\\
0 &\!\!\! (J=1)
\end{cases},
\;\;
\theta_{I\!I}=
\begin{cases}
2^{-\frac{1}{2}H_{I\!I}}
 &\!\!\! (d_{A_r}\geq2)
\\
0 &\!\!\! (d_{A_r}=1)
\end{cases},
\!
}
where the exponents $H_I$ and $H_{I\!I}$ are given by
\alg{
&
H_I=
\log{(J-1)}+
H_{\rm min}^\epsilon(A|R)_{\Psi}-H_{\rm max}^\mu(A|C)_{\ca{C}(\tau)},
\\
&
H_{I\!I}=
H_{\rm min}^\epsilon(A|R)_{\ca{C}(\Psi)}-H_{\rm max}^\mu(A_r|CA_c)_{\ca{C}(\tau)},
}
respectively.
Here, $\ca{C}$ is the completely dephasing operation on $A_c$ with respect to the basis $\{|j\rangle\}_{j=1}^J$ and $\tau$ is the Choi-Jamiolkowski state of  the complementary map $\ca{T}^{A\rightarrow C}$ of $\ca{T}^{A\rightarrow E}$, i.e. $\tau = \mfk{J}(\ca{T}^{A\rightarrow C})$. 
\ethm

In \cite{wakakuwa2021one}, we also obtained a converse bound for randomized partial decoupling, which is stated by the following theorem.

\bthm{converse}(Theorem 4 in \cite{wakakuwa2021one})
Consider a linear trace-preserving CP map $\ca{T}^{A \rightarrow E}$ and
a state $\Psi^{AR}\in\ca{S}_=(\ca{H}^{AR})$ that is decomposed as \req{romanof}.
Suppose that, for $\delta>0$, there exists a normalized state in the form of
\alg{
\Omega^{ER}:=\sum_{j=1}^Jp_j\varsigma_j^E\otm\Psi_{j}^{R_r}\otm\proj{j}^{R_c},
}
such that
\alg{
\left\|
\ca{T}^{A \rightarrow E} ( \Psi^{AR} ) -\Omega^{ER}
\right\|_1
\leq
\delta.
}
Then, for any $\upsilon\in[0,1/2)$ and $\iota\in(0,1]$, it holds that
\alg{
&
\!\!
H_{\rm min}^{\lambda}(A|R)_\Psi
-H_{\rm min}^{\upsilon}(BR|C)_{\ca{T}\circ\ca{C}(\Psi)}+\log{J}
\geq
\log{\iota},
\laeq{tokiyotomare}
\\
&
\!\!
H_{\rm min}^{\lambda'}(A|R)_{\ca{C}(\Psi)}
-H_{\rm min}^{\upsilon}(BR_r|CR_c)_{\ca{T}\circ\ca{C}(\Psi)}
\nn\\
&
\quad\quad\quad\quad\quad\quad\quad\quad\quad\quad
\geq
\log{\iota}+\log{(1-2\upsilon)}.
\laeq{tokiyougoke}
}
The second terms in the L.H.S.s of \req{tokiyotomare} and \req{tokiyougoke} are for a purification $|\Psi\rangle^{ABR}$ of $\Psi^{AR}$ and the complementary channel $\ca{T}^{A \rightarrow C}$ of $\ca{T}^{A \rightarrow E}$,
with $\ca{C}$ being the completely dephasing channel on $A_c$.
The smoothing parameters $\lambda$ and $\lambda'$ are given by
\alg{
\lambda:=
&
2\sqrt{\iota+4\sqrt{20\upsilon+2\delta}}
+\sqrt{2\sqrt{20\upsilon+2\delta}}
\nn\\
&+2\sqrt{2\delta}
+2\sqrt{20\upsilon+2\delta}
+3\upsilon,
\laeq{dfnsmtlambda}
\\
\lambda':=
&
\upsilon+\sqrt{4\sqrt{\iota+2x}+2\sqrt{x}+(4\sqrt{\iota+8}+24) x
\laeq{dfnsmtlambdaII}
}
}
and $x:=\sqrt{2}\sqrt[4]{24\upsilon+2\delta}$.
\ethm

\section{Proof of The Direct Part\\(\rThm{OSDcomp})}
\lsec{direct}

We prove the direct part of the capacity theorem (\rThm{OSDcomp}) based on the direct part of the randomized partial decoupling theorem (\rThm{SmoothExMarkov}).
We follow the idea of Ref.~\cite{ADHW2009} that the problem of finding a good code for a quantum channel is equivalent to the problem of finding a good way to decouple a bipartite state constructed from the channel.

Fix an arbitrary triplet $(c,q,e)$, $\delta>0$, system $S\equiv S_cS_r$ and a state $\rho^{SA}$ that satisfy the conditions in \rThm{OSDcomp}.
We prove achievability of the triplet $(c,q,e)$ within the error $\delta$ along the following lines:
First, we construct a state $\Psi_\rho$ from the state $\rho^{SA}$, the channel $\ca{N}^{A\rightarrow B}$ and a state obtained by ``purifying'' and ``extending'' the source state and the resource state.
Second, we prove that, if a CP map achieves randomized partial decoupling of the state $\Psi_\rho$ for a particular choice of $s$ and $U$ in high precision, there exist an encoder $\ca{E}_{\rho,s,U}$ and a decoder $\ca{D}$ that accomplish the communication with a small error.
Finally, we evaluate the precision of the randomized partial decoupling based on \rThm{SmoothExMarkov}, by which we complete the proof of \rThm{OSDcomp}.

\subsection{Definitions of States and Operations}

We embed $M_c$ and $M_qF_A$ into $S_c$ and $S_r$, respectively.
This is possible because of the conditions $\dim S_c \geq 2^c$ and $\dim S_r  \geq 2^{q+e}$ in \rThm{OSDcomp}.
Similarly, we embed $R_c$ and $R_qF_B$ to larger systems $\hat{R}_c$ and $\hat{R}_r$, respectively, such that $\hat{R}_c\cong S_c$ and $\hat{R}_r\cong S_r$.
The embeddings are represented by linear isometries $P^{M_c\rightarrow S_c}$, $P^{M_qF_A \rightarrow S_r}$, $P^{R_c\rightarrow \hat{R}_c}$ and $P^{R_qF_B \rightarrow \hat{R}_r}$.
For the simplicity of notations, we denote $\hat{R}_c\hat{R}_r$ by $\hat{R}$.
In total, the systems $MF_A$ and $RF_B$ are embedded into systems $S$ and $\hat{R}$ by linear isometries
\alg{
&
P^{MF_A\rightarrow S} :=P^{M_c\rightarrow S_c}\otm P^{M_qF_A\rightarrow S_r},
\laeq{ninen0}
\\
&
P^{RF_B\rightarrow \hat{R}} :=P^{R_c\rightarrow \hat{R}_c}\otm P^{R_qF_B\rightarrow \hat{R}_r},
\laeq{ninen}
 }
respectively.
The adjoint map corresponding to these isometries are given by
\alg{
\ca{P}^{\dagger S\rightarrow MF_A}(\cdot)
&
:=(P^{MF_A\rightarrow S} )^\dagger(\cdot)(P^{MF_A\rightarrow S} ),
\\
\ca{P}^{\dagger \hat{R}\rightarrow  RF_B}(\cdot)
&
:=(P^{RF_B\rightarrow \hat{R}} )^\dagger(\cdot)(P^{RF_B\rightarrow \hat{R}} ).
}

Define a ``purified'' source-resource state by 
\alg{
&
\ket{\Phi_{\rm pur}}^{MF_ARF_B}
:=\ket{\Phi_{2^{c+q}}}^{MR}\ket{\Phi_{2^e}}^{F_AF_B}
\laeq{Phipurdfn}\\
&\quad
=
\frac{1}{\sqrt{2^c}}\sum_{j=1}^{2^c}\ket{j}^{M_c}\ket{j}^{R_c}\ket{\Phi_{2^q}}^{M_qR_q}\ket{\Phi_{2^e}}^{F_AF_B}.
\laeq{pusource}
}
Note that
\alg{
\ca{C}^{M_c}(\Phi_{\rm pur})
=
\Phi_{2^c,2^q}'^{MR}\otm\Phi_{2^e}^{F_AF_B},
\laeq{haveyou}
}
where $\ca{C}^{M_c}$ the completely dephasing operation on $M_c$ with respect to the basis $\{\ket{j}\}_{j=1}^{2^c}$.
We also introduce an ``extended''  one by
\alg{
\ket{\Phi_{\rm ext}}^{S\hat{R}}:=
\frac{1}{\sqrt{d_{S_c}}}\sum_{j=1}^{d_{S_c}}\ket{j}^{S_c}\ket{j}^{\hat{R}_c}\ket{\Phi_{d_{S_r}}}^{S_r\hat{R}_r}.
\laeq{exsource}
}
We properly choose the embedding isometries so that
\alg{
&
\ket{\Phi_{\rm pur}}^{MF_ARF_B}
\nn\\
&\quad
=
\sqrt{\frac{d_S}{2^{c+q+e}}}(P^{MF_A\rightarrow S}\otm P^{RF_B\rightarrow \hat{R}}  )^\dagger\ket{\Phi_{\rm ext}}^{S\hat{R}}
\laeq{extnonext}
}
and
\alg{
P^{M_c\rightarrow S_c}\ket{j}^{M_c}=\ket{j}^{S_c},
\quad
P^{R_c\rightarrow \hat{R}_c}\ket{j}^{R_c}=\ket{j}^{\hat{R}_c}
\laeq{UY}
}
for any $j=1,\cdots,2^c$.
Using
\alg{
\tilde{P}^{\hat{R}\rightarrow RF_B}
:=
\sqrt{\frac{d_S}{2^{c+q+e}}}\:
(P^{  RF_B\rightarrow \hat{R}})^\dagger,
\laeq{honoto}
}
the purified source-resource state and the extended one are simply related as
\alg{
P^{MF_A\rightarrow S}
\ket{\Phi_{\rm pur}}^{MF_ARF_B}
=
\tilde{P}^{\hat{R}\rightarrow RF_B}\ket{\Phi_{\rm ext}}^{S\hat{R}}.
\laeq{extnonextyy}
}

Let $\ca{E}_\rho^{S\rightarrow A}$ and $\ca{E}_{\rho_j}^{S_r\rightarrow A}$ be linear maps defined by the Choi-Jamiolkowski correspondence from $\rho^{SA}$ and $\rho_j^{S_rA}$, respectively. 
That is, $\ca{E}_\rho^{S\rightarrow A}:=\mfk{J}_S^{-1}(\rho^{SA})$ and $\ca{E}_{\rho_j}^{S_r\rightarrow A}:=\mfk{J}_{S_r}^{-1}(\rho_j^{S_rA})$.
Due to the condition that $\rho^S$ is the full-rank maximally mixed state, the two maps are completely positive and trace-preserving.
From the decomposition \req{yapzet23}, it follows that
\alg{
\ca{E}_\rho^{S\rightarrow A}(\tau)
=
\sum_{j=1}^{d_{S_c}}\ca{E}_{\rho_j}^{S_r\rightarrow A}(\bra{j}^{S_c}\tau\ket{j}^{S_c}).
\laeq{calErhodec}
}
We denote by $V_{\rho_j}^{S_r\rightarrow AE_0}$ the Stinespring dilation of $\ca{E}_{\rho_j}^{S_r\rightarrow A}$ for each $j$.
Introducing a quantum system $E_c$ with a fixed orthonormal basis $\{\ket{j}\}_{j=1}^{d_{S_c}}$, the Stinespring dilation $V_{\rho}$ of $\ca{E}_\rho^{S\rightarrow A}$ is given by
\alg{
V_\rho^{S\rightarrow AE_0E_c}
=
\sum_{j=1}^{d_{S_c}}\ket{j}^{E_c}\bra{j}^{S_c}\otm V_{\rho_j}^{S_r\rightarrow AE_0}.
\laeq{ichinen}
}
It is straightforward from \req{calErhodec} that
\alg{
\ca{E}_\rho^{S\rightarrow A}\circ\ca{C}^{S_c}
=
\ca{E}_\rho^{S\rightarrow A},
\laeq{haveyoume}
}
with $\ca{C}^{S_c}$ being the completely dephasing operation on $S_c$ with respect to the basis $\{\ket{j}\}_{j=1}^{d_{S_c}}$.

Let $W_{\ca{N}}^{A\rightarrow BE}$ be the Stinespring dilation of $\ca{N}^{A\rightarrow B}$. 
Using the extended source-resource state \req{exsource},
we define the following pure state, where $\bar{E}\equiv E_cEE_0$ (see Figure \ref{fig:direct}):
\alg{
\ket{\Psi_\rho}^{\hat{R}B\bar{E}}
:=
W_{\ca{N}}^{A\rightarrow BE}\circ
V_\rho^{S\rightarrow AE_0E_c}\ket{\Phi_{\rm ext}}^{S\hat{R}}.
\laeq{narutane}
}
Defining the state
\alg{
\ket{\rho_j}^{\hat{R}_rAE_0}:=V_{\rho_j}^{S_r\rightarrow AE_0}\ket{\Phi_{d_{S_r}}}^{S_r\hat{R}_r},
}
it follows from \req{ichinen} that
\alg{
\ket{\Psi_\rho}^{\hat{R}B\bar{E}}
=
\frac{1}{\sqrt{d_{S_c}}}\sum_{j=1}^{d_{S_c}}
\ket{j}^{\hat{R}_c}\ket{j}^{E_c}
W_{\ca{N}}^{A\rightarrow BE}\ket{\rho_j}^{\hat{R}_rAE_0}.
\laeq{narutame}
}
We trace out system $B$ to obtain the state $\Psi_\rho^{\hat{R}\bar{E}}:={\rm Tr}_B[\proj{\Psi_\rho}]$.

\begin{figure*}[t]
\begin{center}
\includegraphics[bb={0 40 1641 661}, scale=0.3]{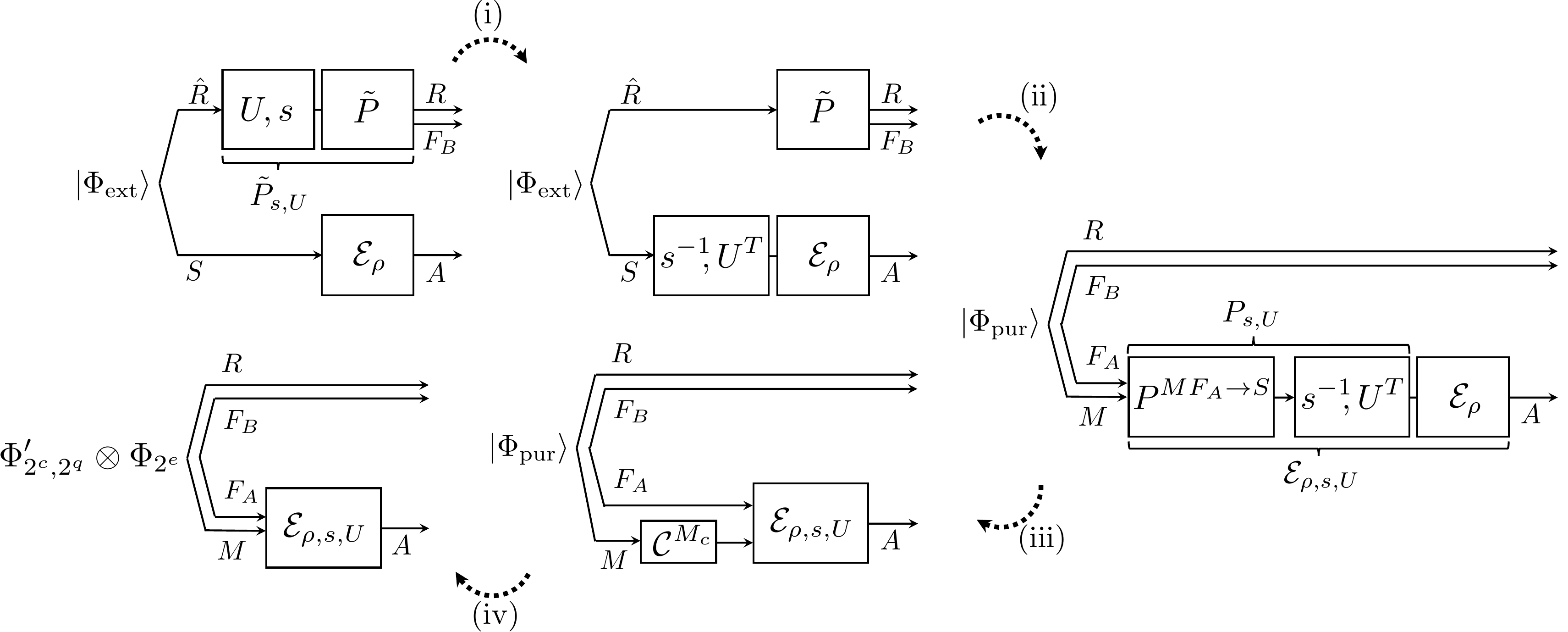}
\end{center}
\caption{
Transformation of the protocol for randomized partial decoupling of the state $\Psi_\rho$ into that for communication over the channel $\ca{N}$ is depicted. We first trace out $\bar{E}$ in Figure \ref{fig:direct} and apply these transformations to obtain Figure~\ref{fig:directprotocol}.
(i) is due to the fact that $\Phi_{\rm ext}$ is the maximally entangled state between $S$ and $\hat{R}$. 
(ii) is from Equality \req{extnonextyy}, and (iii) is because $\ca{E}_{\rho,s,U}^{MF_A\rightarrow A}=\ca{E}_{\rho,s,U}^{MF_A\rightarrow A}\circ\ca{C}^{M_c}$.
Note that $|\Phi_{\rm pur}\rangle^{\bar{M}\bar{R}}=|\Phi_{2^{c+q}}\rangle^{MR}|\Phi_{2^e}\rangle^{F_AF_B}$ as \req{Phipurdfn}.
(iv) follows from \req{haveyou}.
}
\label{fig:directtransform}
\end{figure*}

\subsection{Construction of Encoding and Decoding Operations}

We consider randomized partial decoupling of the ``bipartite'' state $\Psi_\rho^{\hat{R}\bar{E}}$ by a linear CP map ${\rm Tr}_{F_B}\circ\tilde{\ca{P}}^{\hat{R} \rightarrow RF_B}:\hat{R}\rightarrow RF_B$ (see Figure \ref{fig:direct}), where $\tilde{P}^{\hat{R}\rightarrow RF_B}$ is defined by \req{honoto}.
For a unitary $U^{\hat{R}}=\sum_{j=1}^{d_{S_c}}\proj{j}^{\hat{R}_c}\otm U_j^{\hat{R}_r}$ and a permutation $s$ on $\{ 1,\cdots,d_{S_c} \}$, define the {\it partial decoupling error} $\Delta_{s,U}$ by
\alg{
\!
\Delta_{s,U}
\!:=\!
\left\|
{\rm Tr}_{F_B}\!\circ\!\tilde{\ca{P}}^{\hat{R}\rightarrow RF_B} \! \circ \ca{G}_s^{\hat{R}} \! \circ \ca{U}^{\hat{R}} ( \Psi_\rho^{\hat{R}\bar{E}} ) 
-
\tilde{\Psi}_{\rho,s}^{R\bar{E}}
\right\|_1.
\laeq{SmExMaMaMa}
}
Here, we have defined
\alg{
\tilde{\Psi}_{\rho,s}^{R\bar{E}}
:=
{\rm Tr}_{F_B}\circ\tilde{\ca{P}}^{\hat{R}\rightarrow RF_B}  \circ \ca{G}_s^{\hat{R}} ( \Psi_{\rho,{\rm av}}^{\hat{R}\bar{E}} ),
\laeq{tukae}
}
where
\alg{
\Psi_{\rho,{\rm av}}^{\hat{R}\bar{E}}
:=
\mbb{E}_U[\ca{U}^{\hat{R}} ( \Psi_\rho^{\hat{R}\bar{E}} ) ].
\laeq{tukaa}
}
An evaluation of the partial decoupling error $\Delta_{s,U}$ will be given in the next subsection, based on the direct part of the randomized partial decoupling theorem (\rThm{SmoothExMarkov}).
We introduce the operator
\alg{
\tilde{P}_{s,U}^{\hat{R}\rightarrow RF_B}
:=
\tilde{P}^{\hat{R} \rightarrow RF_B}  G_s^{\hat{R}} U^{\hat{R}},
\laeq{dfnpsigU}
}
by which \req{SmExMaMaMa} is simply represented as
\alg{
\Delta_{s,U}
:=
\left\|
{\rm Tr}_{F_B}\!\circ\!\tilde{\ca{P}}_{s,U}^{\hat{R}\rightarrow RF_B} ( \Psi_\rho^{\hat{R}\bar{E}} ) 
-
\tilde{\Psi}_{\rho,s}^{R\bar{E}}
\right\|_1.
\laeq{SmExMaMaMa4}
}

Let $|\tilde{\Psi}_{\rho,s}\rangle^{MR\bar{E}M_0}$ be a purification of $\tilde{\Psi}_{\rho,s}^{R\bar{E}}$ with $MM_0$ being a purifying system.
Due to Uhlmann's theorem (\cite{uhlmann1976transition}; see also e.g. Chapter 9 in \cite{wildetext}) and \req{SmExMaMaMa4}, there exists a linear isometry $\tilde{V}^{BF_B\rightarrow MM_0}$ such that
\begin{eqnarray}
\left\|
\tilde{\ca{V}}^{BF_B \rightarrow MM_0} \otm
\tilde{\ca{P}}_{s,U}^{\hat{R}\rightarrow RF_B} (\Psi_\rho^{\hat{R}B\bar{E}})
-
\tilde{\Psi}_{\rho,s}^{MR\bar{E}M_0} \right\|_1 
\nn\\
\leq 2\sqrt{\Delta_{s,U}}.
\quad
\laeq{asuwo}
\end{eqnarray}
Note that $\tilde{V}^{BF_B\rightarrow MM_0}$ depends on $\rho$, $s$ and $U$ in general.
Defining
\alg{
\ca{D}^{BF_B\rightarrow M}:=({\rm Tr}_{M_0}\otm\ca{C}^{M_c})\circ\tilde{\ca{V}}^{BF_B\rightarrow MM_0},
} 
and tracing out systems $\bar{E}$ and $M_0$ in \req{asuwo},
 we obtain
\begin{eqnarray}
\left\|
\ca{D}^{BF_B\rightarrow M} \otm
\tilde{\ca{P}}_{s,U}^{\hat{R}\rightarrow RF_B}
(\Psi_\rho^{\hat{R}B})
-
\ca{C}^{M_c}( \tilde{\Psi}_{\rho,s}^{MR}) \right\|_1 
\nn\\
\leq 2\sqrt{\Delta_{s,U}}.
 \laeq{tobeforme}
\end{eqnarray}

To obtain an explicit form of the state $\tilde{\Psi}_{\rho,s}^{MR}$, we use \req{narutame} and \req{tukaa} to have
\alg{
\Psi_{\rho,{\rm av}}^{\hat{R}\bar{E}}
=
\frac{1}{d_{S_c}}\sum_{j=1}^{d_{S_c}}
\proj{j}^{\hat{R}_c}\otm\proj{j}^{E_c}
\otm\pi^{\hat{R}_r}\otm\ca{N}^{A\rightarrow E}(\rho_j^{AE_0}).
\nn
}
Combining this with \req{tukae}, and by using \req{honoto}, we obtain
\alg{
\tilde{\Psi}_{\rho,s}^{R\bar{E}}
&
=
\frac{1}{2^c}\sum_{j=1}^{2^c}
\proj{j}^{R_c}\otm\proj{s^{-1}(j)}^{E_c}
\nn\\
&\quad\quad\quad\quad\quad\quad
\otm\pi_{2^q}^{R_q}\otm\ca{N}^{A\rightarrow E}(\rho_{s^{-1}(j)}^{AE_0})
.
}
A purification of this state is given by
\begin{align}
&
\ket{\tilde{\Psi}_{\rho,s}}^{MR\bar{E}M_0}: = 
\nn\\
&\quad
\frac{1}{\sqrt{2^c}}\sum_{j=1}^{2^c}   \ket{j}^{M_c}\ket{j}^{R_c}\ket{s^{-1}(j)}^{E_c} \ket{\Phi_{2^q}}^{M_qR_q}|\varrho_{s^{-1}(j)}\rangle^{E_0EM_0},
\nn
\end{align}
with $|\varrho_j\rangle^{E_0EM_0}$ being a purification of $\ca{N}^{A\rightarrow E}(\rho_j^{AE_0})$.
Thus, we trace out $\bar{E}M_0$ to obtain $\tilde{\Psi}_{\rho,s}^{MR}=\Phi_{2^c,2^q}'^{MR}$.
Substituting this to \req{tobeforme}, and noting that $\ca{C}^{M_c}(\Phi_{2^c,2^q}'^{MR} ) =\Phi_{2^c,2^q}'^{MR}$, 
we arrive at
\begin{eqnarray}
\left\|
\ca{D}^{BF_B\rightarrow M} \otm
\tilde{\ca{P}}_{s,U}^{\hat{R}\rightarrow RF_B}
(\Psi_\rho^{\hat{R}B})
-
\Phi_{2^c,2^q}'^{MR} \right\|_1
\nn\\
 \leq 2\sqrt{\Delta_{s,U}}.
 \laeq{tobeforme2}
\end{eqnarray}

The first term in the L.H.S. of \req{tobeforme2} is calculated as follows.
Note that the state $\ket{\Phi_{\rm ext}}^{S\hat{R}}$ defined by \req{exsource} is the maximally entangled state on $S\hat{R}$.
Thus, from \req{dfnpsigU}  and \req{extnonextyy}  (see Figure \ref{fig:directtransform}), it holds that 
\alg{
&
\tilde{P}_{s,U}^{\hat{R}\rightarrow RF_B}\ket{\Phi_{\rm ext}}^{S\hat{R}}
\nn\\
&=
(G_s U)^{S^T}\otm\tilde{P}^{\hat{R}\rightarrow RF_B} 
\ket{\Phi_{\rm ext}}^{S\hat{R}}
\nn\\
&=
(G_s U)^{S^T}P^{MF_A\rightarrow S}
\ket{\Phi_{\rm pur}}^{MF_ARF_B}
\nn\\
&=
P_{s,U}^{MF_A\rightarrow S}
\ket{\Phi_{\rm pur}}^{MF_ARF_B},
\laeq{fuha}
}
where we have defined a linear isometry
\alg{
P_{s,U}^{MF_A\rightarrow S}
:=U^{S^T}G_{s^{-1}}^{S}P^{MF_A\rightarrow S}.
\laeq{hikikae}
}
Note that $G_s^{S^T}=G_{s^{-1}}^{S}$.
Combining \req{fuha} with \req{narutane}, we obtain
\alg{
&\!\!
\tilde{P}_{s,U}^{\hat{R}\rightarrow RF_B}\ket{\Psi_\rho}^{\hat{R}B\bar{E}}
\nn\\
&\!\!
=
W_{\ca{N}}^{A\rightarrow BE}\circ
V_\rho^{S\rightarrow AE_0E_c}
\circ P_{s,U}^{MF_A\rightarrow S}
\ket{\Phi_{\rm pur}}^{MF_ARF_B}\!.\!\!
\laeq{stephen}
}
We now construct an encoding operation $\ca{E}_{\rho,s,U}$ by
\alg{
\ca{E}_{\rho,s,U}^{MF_A\rightarrow A}
&
:=
\ca{E}_\rho^{S\rightarrow A} \circ \ca{P}_{s,U}^{MF_A\rightarrow S}
\laeq{dfnencrsU}
\\
&=
{\rm Tr}_{E_0E_c}\circ
\ca{V}_\rho^{S\rightarrow AE_0E_c}
\circ \ca{P}_{s,U}^{MF_A\rightarrow S}.
}
Tracing out $\bar{E} = E E_0 E_c$ in \req{stephen} yields
\alg{
\tilde{\ca{P}}_{s,U}^{\hat{R}\rightarrow RF_B}
(\Psi_\rho^{\hat{R}B})
=
\ca{N}^{A\rightarrow B}\circ\ca{E}_{\rho,s,U}^{MF_A\rightarrow A}(\Phi_{\rm pur}^{MF_ARF_B}).
\laeq{tsuji}
}
Substituting this to \req{tobeforme2}, 
we arrive at
\alg{
&
\left\|
\ca{D}^{BF_B \rightarrow M} \circ
\ca{N}^{A \rightarrow B} \circ \ca{E}_{\rho,s,U}^{MF_A\rightarrow A} (\Phi_{\rm pur}^{MF_ARF_B})
-
\Phi_{2^c,2^q}'^{MR} \right\|_1
\nn\\
&
 \leq 2\sqrt{\Delta_{s,U}}.
 \laeq{tobeforme3}
}

It remains to prove that the encoding operation defined by \req{dfnencrsU} satisfies the condition \req{encdeccls} (see Figure \ref{fig:directtransform}), i.e.,
\alg{
\ca{E}_{\rho,s,U}^{MF_A\rightarrow A}=\ca{E}_{\rho,s,U}^{MF_A\rightarrow A}\circ\ca{C}^{M_c}.
\laeq{clencrsU}
} 
From \req{ninen0}, \req{UY} and \req{hikikae}, we have
\alg{
P_{s,U}^{MF_A\rightarrow S}
\!=\!
\sum_{j=1}^{2^c}\ket{s^{-1}(j)}^{S_c}\bra{j}^{M_c}\!\otm U_{s^{-1}(j)}^{S_r^T}P^{M_qF_A\rightarrow S_r}.
\!
}
Thus, it holds that
\alg{
\ca{P}_{s,U}^{MF_A\rightarrow S}\circ\ca{C}^{M_c}
=
\ca{C}^{S_c}\circ\ca{P}_{s,U}^{MF_A\rightarrow S},
}
with $\ca{C}^{S_c}$ being the completely dephasing operation on $S_c$ with respect to the basis $\{\ket{j}\}_{j=1}^J$.
Combining this with \req{haveyoume}, we obtain
\alg{
\ca{E}_\rho^{S\rightarrow A} \circ \ca{P}_{s,U}^{MF_A\rightarrow S}\circ\ca{C}^{M_c}
=
\ca{E}_\rho^{S\rightarrow A} \circ \ca{P}_{s,U}^{MF_A\rightarrow S},
}
which implies \req{clencrsU}.
Substituting this to \req{tobeforme3}, and by using the relation \req{haveyou}, we finally arrive at
\alg{
&
\left\|
\ca{D}^{BF_B \rightarrow M} \circ
\ca{N}^{A \rightarrow B} \circ \ca{E}_{\rho,s,U}^{MF_A\rightarrow A} (\Phi_{2^c,2^q}'^{MR}\otm\Phi_{2^e}^{F_AF_B})
-
\Phi_{2^c,2^q}'^{MR} \right\|_1
\nn\\
&
 \leq 2\sqrt{\Delta_{s,U}}.
}

The deformations of expressions that we have done in this subsection are depicted in Figure~\ref{fig:directtransform}.
The figure explains the way how the communication protocol depicted in Figure \ref{fig:directprotocol} is obtained from the randomized partial decoupling protocol in Figure \ref{fig:direct}.

\subsection{Evaluation of The Errors}

To evaluate the partial decoupling error $\Delta_{s,U}$ defined by \req{SmExMaMaMa},
we apply the direct part of the randomized partial decoupling theorem (\rThm{SmoothExMarkov}) under the following correspondence:
\alg{
A_c, A_r, R_c, R_r
&
\rightarrow
\hat{R}_c, \hat{R}_r, E_c, EE_0
\\
A,R,E,C
&\rightarrow
\hat{R},\bar{E},R,F_B
\\
\Psi^{AR}
&\rightarrow\Psi_\rho^{\hat{R}\bar{E}}
\\
\ca{T}^{A\rightarrow E}
&\rightarrow {\rm Tr}_{F_B}\circ\tilde{\ca{P}}^{\hat{R}\rightarrow RF_B}
\\
\mu&=0.
}
 It follows that
there exist a unitary $U^{\hat{R}}=\sum_{j=1}^{d_{S_c}}\proj{j}^{R_c}\otm U_j^{R_qF_B}$ and a permutation $s$ such that the partial decoupling error is bounded as
\begin{align}
\Delta_{s,U}
\leq
\begin{cases}
2^{-\frac{1}{2}H_I}
+
2^{-\frac{1}{2}H_{I\!I}}
+4\epsilon
\quad\quad
(d_{S_c},d_{S_r}\geq2)
\\
2^{-\frac{1}{2}H_{I}}
+4\epsilon
\quad\quad
(q=e=0,d_{S_c}\geq2,d_{S_r}=1)
\\
2^{-\frac{1}{2}H_{I\!I}}
+4\epsilon
\quad\quad
(c=0,d_{S_c}=1,d_{S_r}\geq2).
\end{cases}
\laeq{SmExMaMa}
\end{align}
Here, the exponents $H_I$ and $H_{I\!I}$ are given by
\alg{
&
H_I=
\log{(d_{S_c}-1)}+
H_{\rm min}^\epsilon(\hat{R}|\bar{E})_{\Psi_\rho}-H_{\rm max}(\hat{R}|F_B)_{\ca{C}(\tau)},
\laeq{amaimonohoshii}\\
&
H_{I\!I}=
H_{\rm min}^\epsilon(\hat{R}|\bar{E})_{\ca{C}(\Psi_\rho)}-H_{\rm max}(\hat{R}_r|F_B\hat{R}_c)_{\ca{C}(\tau)},
\laeq{nomimonogaii}
}
and $\tau^{\hat{R}F_B}$ is the Choi-Jamiolkowski state of the complementary channel of ${\rm Tr}_{F_B}\circ\tilde{\ca{P}}^{\hat{R}\rightarrow RF_B}$. 
Using \req{pusource}-\req{extnonext}, it holds that
\alg{
\tau^{\hat{R}F_B}
&:={\rm Tr}_{F_B}\circ\tilde{\ca{P}}^{\hat{R}'\rightarrow RF_B}(\Phi_{\rm ext}^{\hat{R}\hat{R}'})
\\
&=\pi_{2^c}^{\hat{R}_c}\otm\pi_{2^q}^{\hat{R}_q}\otm\Phi_{2^e}^{\hat{F}_BF_B}.
\laeq{yotteru}
}
A simple calculation yields
\alg{
&
H_{\rm max}(\hat{R}|F_B)_{\ca{C}(\tau)}
=c+q-e,
\\
&
H_{\rm max}(\hat{R}_r|F_B\hat{R}_c)_{\ca{C}(\tau)}
=q-e.
}
Using the duality of the conditional smooth entropies, we have
\alg{
&
H_{\rm min}^\epsilon(\hat{R}|\bar{E})_{\Psi_\rho}
=-H_{\rm max}^\epsilon(\hat{R}|B)_{\Psi_\rho},
\\
&
H_{\rm min}^\epsilon(\hat{R}|\bar{E})_{\ca{C}(\Psi_\rho)}
=-H_{\rm max}^\epsilon(\hat{R}_r|BR_c)_{\Psi_\rho}.
}
Noting that $\hat{R}_c$ and $\hat{R}_r$ are isomorphic to $S_c$ and $S_r$, respectively, it follows from the definition \req{narutame} of $\Psi_\rho$ that
\alg{
\Psi_\rho^{\hat{R}B}
\cong
\ca{N}^{A\rightarrow B}(\rho^{SA}),
} 
which leads to
\alg{
&
H_{\rm max}^\epsilon(\hat{R}|B)_{\Psi_\rho}
=H_{\rm max}^\epsilon(S|B)_{\ca{N}(\rho)},
\\
&
H_{\rm max}^\epsilon(\hat{R}_r|B\hat{R}_c)_{\Psi_\rho}
=H_{\rm max}^\epsilon(S_r|BS_c)_{\ca{N}(\rho)}.
}
Substituting all these equalities to  \req{amaimonohoshii} and \req{nomimonogaii}, 
we obtain
\alg{
&
H_I=
-c-q+e-H_{\rm max}^\epsilon(S|B)_{\ca{N}(\rho)}+\log{(d_{S_c}-1)},
\laeq{amaimonohoshiiyou}
\\
&
H_{I\!I}=
-q+e-H_{\rm max}^\epsilon(S_r|BS_c)_{\ca{N}(\rho)}.
\laeq{nomimonogaiiyou}
}
We now use the conditions \req{rightonf23} and \req{rightoff23} to have
\alg{
H_I\geq -\log{\delta_1},
\quad
H_{I\!I}\geq -\log{\delta_2}.
}
Substituting these inequalities to \req{SmExMaMa}, we finally arrive at 
\alg{
\Delta_{s,U}
\leq
\begin{cases}
\sqrt{\delta_1}+\sqrt{\delta_2}+4\epsilon&\!(d_{S_c},d_{S_r}\geq2)
\\
\sqrt{\delta_1}+4\epsilon&\!(q=e=0,d_{S_c}\geq2,d_{S_r}=1)
\\
\sqrt{\delta_2}+4\epsilon&\!(c=0,d_{S_c}=1,d_{S_r}\geq2)
\end{cases}
}
and complete the proof of \rThm{OSDcomp}.
\QED

\begin{figure*}[t]
\begin{center}
\includegraphics[bb={0 0 1266 579}, scale=0.3]{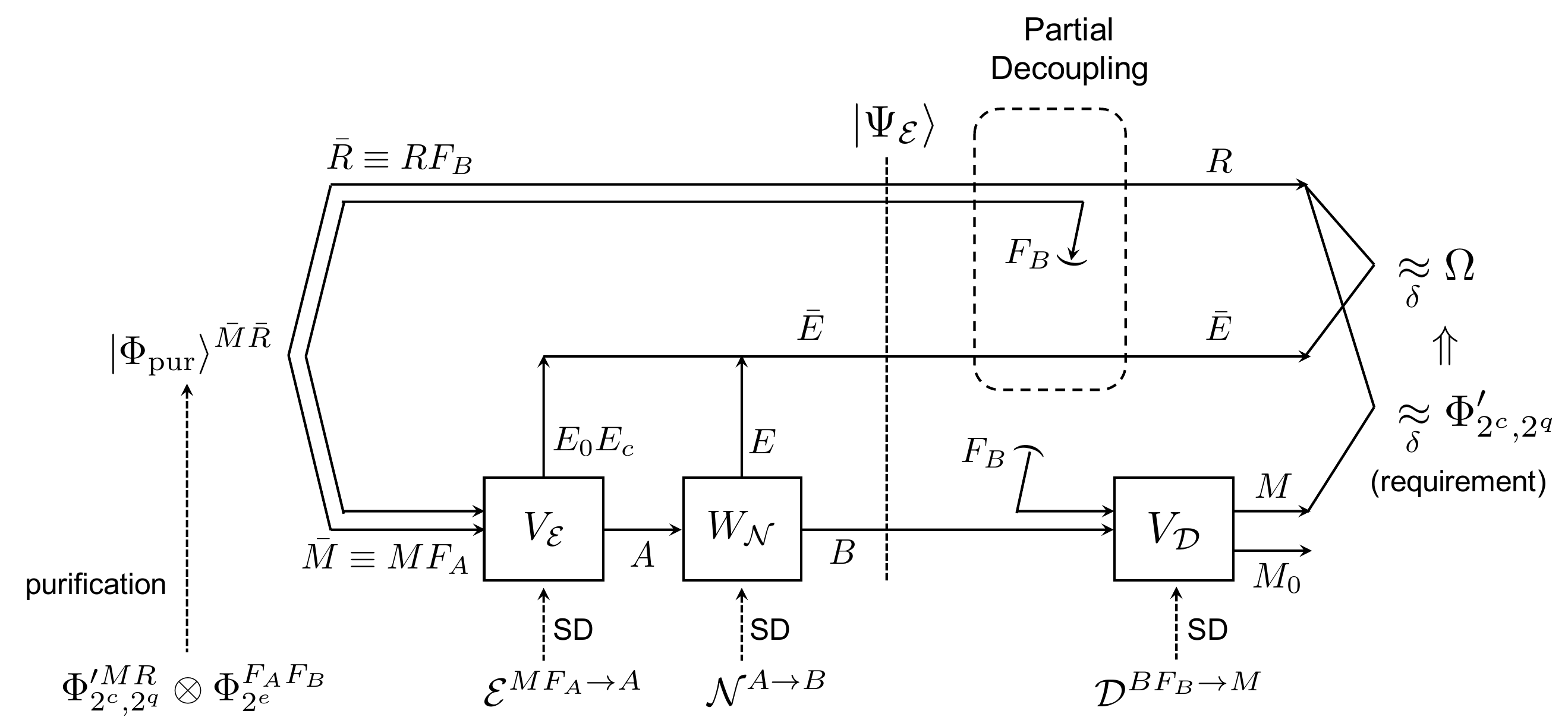}
\end{center}
\caption{
The definition of the state $\Psi_{\ca{E}}$ given by \req{ninshou} and its transformation by partial decoupling are depicted.
SD stands for the Stinespring dilation. Note that $R\equiv R_cR_q$, $M\equiv M_cM_q$ and that $|\Phi_{\rm pur}\rangle^{\bar{M}\bar{R}}=|\Phi_{2^{c+q}}\rangle^{MR}|\Phi_{2^e}\rangle^{F_AF_B}$.
}
\label{fig:converse}
\end{figure*}

\section{Proof of The Converse Part (\rThm{converseonehyb})}
\lsec{converse}

We prove the converse part of the capacity theorem (\rThm{converseonehyb}) based on the converse part of randomized partial decoupling (\rThm{converse}).
The proof proceeds as follows: 
First, we construct a state $\Psi_{\ca{E}}$ from the source state $\Phi_{2^c,2^q}'^{MR}$, the resource state $\Phi_{2^e}$, the channel $\ca{N}$ and an encoding operation $\ca{E}$.
Second, we prove that the small error condition \req{tsuzuiteku} implies that a partial-trace operation achieves partial decoupling of the state $\Psi_{\ca{E}}$.
Applying the converse part for randomized partial decoupling, we obtain a set of inequalities represented in terms of the conditional entropies of the state $\Psi_{\ca{E}}$.
Finally, we evaluate the entropies of the state to complete the proof of \rThm{converseonehyb}.

\subsection{Application of the Converse Bound for Randomized Partial Decoupling}

Suppose that a triplet $(c,q,e)$ is achievable within the error $\delta$ for the channel $\ca{N}$.
By definition, there exist an encoding operation $\ca{E}^{MF_A\rightarrow A}$ and a decoding operation $\ca{D}^{BF_B\rightarrow M}$ that satisfy the conditions 
\alg{
\left\|
 \ca{D} \circ \ca{N} \circ \ca{E} (\Phi_{2^c,2^q}'^{MR}\otm\Phi_{2^e}^{F_AF_B})  - \Phi_{2^c,2^q}'^{MR}
\right\|_1
\leq 
\delta
\laeq{tsuzuitekuu}
}
and 
\alg{
\ca{E}^{MF_A\rightarrow A}=\ca{E}^{MF_A\rightarrow A}\circ\ca{C}^{M_c}.
\laeq{encdecclss}
}
Let $\ca{V}_{\ca{D}}^{BF_B\rightarrow MM_0}$ and $\ca{W}_{\ca{N}}^{A\rightarrow BE}$ be the Stinespring dilations of $\ca{D}$ and $\ca{N}$, respectively. 
Let $E_c$ be a $2^c$-dimensional quantum system with a fixed orthonormal basis $\{\ket{j}\}_{j=1}^{2^c}$.
Due to \req{encdecclss}, a Stinespring dilation $\ca{V}_{\ca{E}}^{MF_A\rightarrow AE_0E_c}$ of $\ca{E}$ is given by
\alg{
V_{\ca{E}}^{MF_A\rightarrow AE_0E_c}
=
\sum_{j=1}^{2^c}\ket{j}^{E_c}\bra{j}^{M_c}\otm V_{\ca{E},j}^{M_qF_A\rightarrow AE_0},
\laeq{sorato}
} 
where $V_{\ca{E},j}$ is a linear isometry for each $j$.
We introduce notations $\bar{R}\equiv RF_B$, $\bar{M}\equiv MF_A$, and define a ``purified'' source-resource state $\Phi_{\rm pur}^{\bar{M}\bar{R}}$ (see Figure \ref{fig:converse}) by
\alg{
\!\!\!
\ket{\Phi_{\rm pur}}^{\bar{M}\bar{R}}\!:=\!\frac{1}{\sqrt{2^c}}\sum_{j=1}^{2^c}\ket{j}^{M_c}\ket{j}^{R_c}\ket{\Phi_{2^q}}^{M_qR_q}|\Phi_{2^e}\rangle^{F_AF_B}\!.\!\!
\laeq{kenshou}
}
Denoting $E_0EE_c$ by $\bar{E}$, we define 
a pure state $\ket{\Psi_{\ca{E}}}$ by
\alg{
\ket{\Psi_{\ca{E}}}^{\bar{R}B\bar{E}}
:=W_{\ca{N}}^{A\rightarrow BE}\circ V_{\ca{E}}^{\bar{M}\rightarrow AE_0E_c}|\Phi_{\rm pur}\rangle^{\bar{M}\bar{R}}.
\laeq{ninshou}
}
Note that $\ket{\Psi_{\ca{E}}}$ is classically coherent in $R_cE_c$, in the sense of \req{romanof}.
A purification of the state $\ca{D}\circ\ca{N} \circ \ca{E} (\Phi_{2^c,2^q}'^{MR}\otm\Phi_{2^e}^{F_AF_B})$ is then given by $\ca{V}_{\ca{D}}^{BF_B\rightarrow MM_0}(\Psi_{\ca{E}}^{\bar{R}B\bar{E}})$, with $M_0\bar{E}$ being a purifying system.
Due to the small error condition \req{tsuzuitekuu} and Uhlmann's theorem (see also \rLmm{cohpurifi} in \rApp{TechLmm}), there exists a purification $|\Omega\rangle^{R\bar{E}MM_0}$ of $\Phi_{2^c,2^q}'^{MR}$ in the form of
\alg{
\!\!\!
|\Omega\rangle
=
\frac{1}{\sqrt{2^c}}\sum_{j=1}^{2^c}\ket{j}^{M_c}\ket{j}^{R_c}\ket{j}^{E_c}\ket{\omega_j}^{E_0M_0E}\ket{\Phi_{2^q}}^{M_qR_q},
\!\!
\laeq{irut}
}
where $\ket{\omega_j}$ are normalized pure states, and satisfies
\alg{
\left\|
\ca{V}_{\ca{D}}^{BF_B\rightarrow MM_0} (\Psi_{\ca{E}}^{\bar{R}B\bar{E}}) 
 - \Omega^{R\bar{E}MM_0}
\right\|_1
\leq 
2\sqrt{\delta}. \laeq{garakuta}
}
Tracing out $MM_0$, it follows that
\alg{
\left\|
{\rm Tr}_{F_B}[\Psi_{\ca{E}}^{\bar{R}\bar{E}}]
 - \Omega^{R\bar{E}}
\right\|_1
\leq 
2\sqrt{\delta}. \laeq{kitai}
}
From \req{irut},
we have
\alg{
\Omega^{R\bar{E}}
=
\frac{1}{2^c}\sum_{j=1}^{2^c}\proj{j}^{R_c}\otm\proj{j}^{E_c}\otm\omega_j^{E_0E}\otm\pi_q^{R_q}.
}
Thus, the condition \req{kitai} implies that the map ${\rm id}^R\otm{\rm Tr}_{F_B}:\bar{R}\rightarrow R$ achieves partial decoupling of the state $\Psi_{\ca{E}}^{\bar{R}\bar{E}}$ (see Figure \ref{fig:converse}).

We apply the converse bound for randomized partial decoupling (\rThm{converse}) under the following correspondence:
\alg{
A_c,A_r,R_c,R_r
&
\rightarrow
R_c,R_qF_B,E_c,E_0E
\\
A,B,R,E,C
&\rightarrow
\bar{R},B,\bar{E},R,F_B
\\
\ket{\Psi_{\ca{E}}}^{ABR}
&\rightarrow \ket{\Psi_{\ca{E}}}^{\bar{R}B\bar{E}}
\\
\ca{T}^{A\rightarrow E}
&\rightarrow {\rm id}^R\otm{\rm Tr}_{F_B}
\\
\delta
&
\rightarrow 
2\sqrt{\delta}
\\
\upsilon&=0.
}
Noting that the complementary channel of ${\rm id}^R\otm{\rm Tr}_{F_B}$ is given by ${\rm Tr}_R\otm{\rm id}^{F_B}$,
we obtain
\alg{
&
\!\!
H_{\rm min}^{\lambda}(\bar{R}|\bar{E})_{\Psi_{\ca{E}}}
-H_{\rm min}(B\bar{E}|F_B)_{\ca{C}(\Psi_{\ca{E}})}+\log{d_{R_c}}
\nn\\
&
\quad\quad\quad\quad\quad\quad\quad\quad\quad\quad
\geq
\log{\iota},
\laeq{tokiyotomatte}
\\
&
\!\!
H_{\rm min}^{\lambda'}(\bar{R}|\bar{E})_{\ca{C}(\Psi_{\ca{E}})}
-H_{\rm min}(BE_0E|F_BE_c)_{\ca{C}(\Psi_{\ca{E}})}
\nn\\
&
\quad\quad\quad\quad\quad\quad\quad\quad\quad\quad
\geq
\log{\iota},
\laeq{tokiyougoite}
}
where $\ca{C}$ is the completely dephasing operation on $E_c$.
Substituting $\upsilon=0$ to \req{dfnsmtlambda} and \req{dfnsmtlambdaII}, the smoothing parameters $\lambda$ and $\lambda'$ are given by \req{smlam} and \req{smlamp}, respectively.

\subsection{Evaluation of Entropies}

By using the duality of the smooth conditional entropies \cite{tomamichel2010duality} (see also Lemma 27 and 29 in \cite{wakakuwa2021one}),
the first terms in \req{tokiyotomatte} and \req{tokiyougoite} are calculated to be
\alg{
&
H_{\rm min}^{\lambda}(\bar{R}|\bar{E})_{\Psi_{\ca{E}}}
=
-H_{\rm max}^{\lambda}(\bar{R}|B)_{\Psi_{\ca{E}}},
\\
&
H_{\rm min}^{\lambda'}(\bar{R}|\bar{E})_{\ca{C}(\Psi_{\ca{E}})}
=
-H_{\rm max}^{\lambda'}(R_qF_B|BR_c)_{\Psi_{\ca{E}}}.
}
To calculate the second term in \req{tokiyotomatte}, note that \req{sorato}, \req{kenshou} and \req{ninshou} imply
\alg{
&
\ca{C}^{E_c}(\Psi_{\ca{E}}^{B\bar{E}F_B})
=
\Psi_{\ca{E}}^{B\bar{E}F_B}
\\
&
\quad\quad
=
\ca{W}_{\ca{N}}^{A\rightarrow BE}\circ\ca{V}_{\ca{E}}^{MF_A\rightarrow AE_0E_c}(\Phi_{2^c,2^q}'^M\otm\Phi_{2^e}^{F_AF_B}).
}
Thus, due to the isometric invariance of the conditional max entropy (Lemma 15 in \cite{tomamichel2010duality}), we have
\alg{
&H_{\rm min}(B\bar{E}|F_B)_{\ca{C}(\Psi_{\ca{E}})}
\nn\\
&
=
H_{\rm min}(MF_A|F_B)_{{\Phi_{2^c,2^q}'}\otm\Phi_{2^e}}
\\
&
=
c+q-e.
}
Similarly, the second term in \req{tokiyougoite} is calculated to be
\alg{
&
H_{\rm min}(BE_0E|F_BE_c)_{\ca{C}(\Psi_{\ca{E}})}
\nn\\
&
=
H_{\rm min}(M_qF_A|F_BM_c)_{{\Phi_{2^c,2^q}'}\otm\Phi_{2^e}}
\\
&
=
q-e.
}
Substituting all these equalities into \req{tokiyotomatte} and \req{tokiyougoite}, we arrive at
\alg{
&
\!\!
-H_{\rm max}^{\lambda}(\bar{R}|B)_{\Psi_{\ca{E}}}
-c-q+e+\log{d_{R_c}}
\geq
\log{\iota},
\\
&
\!\!
-H_{\rm max}^{\lambda'}(R_qF_B|BR_c)_{\ca{C}(\Psi_{\ca{E}})}
-q+e
\geq
\log{\iota}.
}

Finally, we calculate the reduced state of $\Psi_{\ca{E}}$ by using  \req{sorato} and \req{ninshou} to obtain
\alg{
\Psi_{\ca{E}}^{RBF_B}
&=\ca{C}^{R_c}(\Psi_{\ca{E}}^{RBF_B})
\\
&
=
\frac{1}{2^c}\sum_{j=1}^{2^c}\proj{j}^{R_c}\otm\ca{N}^{A\rightarrow B}(\rho_j^{R_qF_BC}),
}
where
\alg{
\rho_j^{R_qF_BA}:=
\ca{E}^{MF_A\rightarrow A}(\proj{j}^{M_c}\otm\Phi_{2^q}^{M_qR_q}\otm\Phi_{2^e}^{F_AF_B}).
}
We relabel $R_c$ and $R_qF_B$ by $S_c$ and $S_r$, respectively, in which case we have $d_S\leq2^{c+q+e}$.
Noting that $\rho_j^{R_qF_B}$ is the maximally mixed state, we complete the proof of \rThm{converseonehyb}.
\QED

\section{Application to A Memoryless Channel}
\lsec{iid}

In this section, we consider an asymptotic limit of infinitely many uses of a memoryless channel.
We consider a scenario of transmitting classical and quantum messages with the assistance of shared entanglement.
First, we review a complete characterization of the achievable rate region, which was obtained by Hsieh and Wilde \cite{hsieh2010entanglement}.
Second, we present another characterization that is obtained from the direct and converse bounds for the one-shot scenario presented in \rSec{mainresults} (\rThm{OSDcomp} and \rthm{converseonehyb}).
A complete proof of the latter will be provided in \rSec{PRFasymptotic}.
Despite their seeming inconsistency, the two characterizations are equivalent.
A detailed proof for this equivalence will be provided in \rApp{AEiid}.
In \rSec{comparison}, we will apply the achievable rate region obtained in this section to special cases such as the classical capacity, the quantum capacity and the entanglement assisted classical/quantum capacities.
We will see that the known results for each case are recovered from our result.

The definition of a code for the channel $\ca{N}$ in the scenario of many uses of the channel follows from \rDfn{oneshotcode} by the correspondence $(c,q,e)\rightarrow(nC,nQ,nE)$ and $\ca{N}\rightarrow\ca{N}^{\otm n}$, where $n$ is the number of the uses of the channel.
The three-dimensional achievable rate region is defined as follows:

\bdfn{achievableRT}
A rate triplet $(C,Q,E)$ is achievable if, for any $\delta>0$, there exists $n_0\in\mbb{N}$ such that for any $n\geq n_0$, there exists an $(nC,nQ,nE,\delta)$ code for $(\ca{N}^{A\rightarrow B})^{\otm n}$.
The closure of the set of all achievable rate triplets is called the achievable rate region, and is denoted by $\ca{C}(\ca{N})$.
\edfn

\subsection{Hsieh-Wilde Theorem}

Hsieh and wilde \cite{hsieh2010entanglement} analyzed the coding problem described above and obtained a complete characterization of the achievable rate region as follows:

\bdfn{rateSN2}
Let $\ca{N}^{A\rightarrow B}$ be a quantum channel. 
Consider finite dimensional quantum systems $S_c$ and $S_r$, the former of which is equipped with a fixed orthonormal basis $\{\ket{j}\}_{j=1}^{d_{S_c}}$.
We denote $S_cS_r$ by $S$.
Consider a state $\rho^{SA}$ in the form of
\alg{
\rho^{SA}
=
\sum_{j=1}^{d_{S_c}}p_j\proj{j}^{S_c}\otm\rho_j^{S_rA},
\laeq{proogostar}
}
where $\{p_j,\rho_j\}_{j=1}^{d_{S_c}}$ is an ensemble of quantum states on $S_rA$.
Let $\Lambda(\ca{N},\rho)\in\mbb{R}^3$ be the set of all triplets $(C,Q,E)$ that satisfy
\alg{
C+2Q
&\leq I(S:B)_{\ca{N}(\rho)},
\laeq{proogo21}\\
C+Q-E
&
\leq
H(S_c)_\rho-H(S|B)_{\ca{N}(\rho)},
\laeq{proogo22}\\
Q-E
&\leq -H(S_r|BS_c)_{\ca{N}(\rho)},
\laeq{proogo23}\\
C,Q,E&\geq0,
}
and define
\alg{
\Lambda_p(\ca{N})
:=
\bigcup_{S_c,S_r,\rho}
\Lambda(\ca{N},\rho).
\laeq{llambda}
}
Here, the union is taken over all $S_c$, $S_r$ and $\rho$ that is decomposed as \req{proogostar}, 
for which we assume that $\rho_j$ is a pure state on $S_rA$ for each $j$.
The set $\Lambda_p(\ca{N}^{\otm n})$ is defined for any $n\in\mbb{N}$ along the same line.
The regularized version of $\Lambda_p$ is defined by
\alg{
\Lambda_p^{\infty}(\ca{N})
:=
\bigcup_{n=1}^\infty
\frac{1}{n}\Lambda_p(\ca{N}^{\otm n}).
\laeq{sarusarusara}
}
\edfn

\bthm{RRegionIID2}
(Theorem 1 in \cite{hsieh2010entanglement})
For any quantum channel $\ca{N}^{A\rightarrow B}$, it holds that $\ca{C}(\ca{N})=\overline{\Lambda_p^{\infty}(\ca{N})}$. 
Here, the overline represents the closure of the set.
\ethm

It was proved in \cite{hsieh2010entanglement} that, in the special cases where one or two of $C$, $Q$ or $E$ vanish,
the above theorem recovers the known results cases such as the Holevo-Schumache-Westmoreland theorem for the classical capacity \cite{holevo98,schumacher97}, the Lloyd-Shor-Devetak theorem for the quantum capacity \cite{lloyd1997capacity,shor2002quantum,devetak2005private}, the entanglement-assisted classical \cite{bennett1999entanglement,bennett2002entanglement,shor2004classical} and quantum \cite{devetak2004family,deve08} capacity theorems.
Note that, due to the chain rule of the quantum entropies, the R.H.S. of \req{proogo2} is equal to $I(S_c:B)_{\ca{N}(\rho)}
-H(S_r|BS_c)_{\ca{N}(\rho)}$.

\subsection{Alternative Expression}

We present an alternative expression for the achievable rate region that is obtained from the one-shot result presented in \rSec{mainresults}.

\bdfn{rateSN}
Let $\ca{N}^{A\rightarrow B}$ be a quantum channel.
Consider finite dimensional quantum systems $S_c$ and $S_r$. Let $S$ be $S_cS_r$, and $\{\ket{j}\}_{j=1}^{d_{S_c}}$ be a fixed orthonormal basis of $S_c$.
Consider a state $\rho^{SA}$ in the form of
\alg{
\rho^{SA}
=
\sum_{j=1}^{d_{S_c}}p_j\proj{j}^{S_c}\otm\rho_j^{S_rA},
\laeq{proogo}
}
where $\{p_j,\rho_j\}_{j=1}^{d_{S_c}}$ is an ensemble of quantum states on $S_rA$.
Let $\Theta(\ca{N},\rho)\subset\mbb{R}^3$ be the set of all triplets $(C,Q,E)$ that satisfy
\alg{
Q+E
&\leq
 H(S_r|S_c)_\rho,
\laeq{proogo1}\\
C+Q-E
&
\leq
H(S_c)_\rho-H(S|B)_{\ca{N}(\rho)},
\laeq{proogo2}\\
Q-E
&\leq -H(S_r|BS_c)_{\ca{N}(\rho)},
\laeq{proogo3}\\
C,Q,E&\geq0,
\laeq{proogo4}
}
and define
\alg{
\Theta_{\pi}(\ca{N})
:=
\bigcup_{S_c,S_r,\rho}
\Theta(\ca{N},\rho).
\laeq{ttheta}
}
Here, the union is taken over all finite dimensional systems $S_c$, $S_r$ and all states $\rho$ that is decomposed as \req{proogo}, 
for which we assume that $\rho^{S}$ is the full-rank maximally mixed state on $S$.
The set $\Theta_{\pi}(\ca{N}^{\otm n})$ is defined for any $n\in\mbb{N}$ along the same line.
The regularized version of $\Theta_{\pi}$ is defined by
\alg{
\Theta_{\pi}^{\infty}(\ca{N})
:=
\bigcup_{n=1}^\infty
\frac{1}{n}\Theta_{\pi}(\ca{N}^{\otm n}).
\laeq{sarusaru}
}
\edfn

\bthm{RRegionIID1}
For any quantum channel $\ca{N}^{A\rightarrow B}$, it holds that $\ca{C}(\ca{N})=\overline{\Theta_{\pi}^{\infty}(\ca{N})}$.
Here, the overline represents the closure of the set.
\ethm

\noindent
A proof of \rThm{RRegionIID1} will be provided in \rSec{PRFasymptotic}.

In \rSec{comparison}, we will apply \rThm{RRegionIID1} to the special cases where one or two of $C$, $Q$ or $E$ vanish.
We prove that the known results for those special cases such as the Holevo-Schumache-Westmoreland theorem for the classical capacity \cite{holevo98,schumacher97}, the Lloyd-Shor-Devetak theorem for the quantum capacity \cite{lloyd1997capacity,shor2002quantum,devetak2005private} and the entanglement-assisted capacity theorems \cite{bennett1999entanglement,bennett2002entanglement,shor2004classical,devetak2004family,deve08} are recovered from \rThm{RRegionIID1}, in the same way as those are obtained from \rThm{RRegionIID2}.

\subsection{Equivalence of The Two Expressions}
\lsec{altexpr}

The region defined by \rDfn{rateSN} may look different from the one defined by \rDfn{rateSN2}.
But the two regions are actually the same, i.e., it holds that
\alg{
\overline{\Theta_{\pi}^{\infty}(\ca{N})}
=
\overline{\Lambda_p^{\infty}(\ca{N})}.
\laeq{SfeqR}
}
The point is that we take the union over $S_c$, $S_r$ and $\rho$ both in \req{ttheta} and in \req{llambda}.
Taking the union eliminate the seeming mismatch between the conditions \req{proogo21} and \req{proogo1}.


To be more precise, we introduce two alternative expressions for the achievable rate region.
We define
\alg{
\Theta(\ca{N})
:=
\bigcup_{S_c,S_r,\rho}
\Theta(\ca{N},\rho),
}
where $\Theta(\ca{N},\rho)\in\mbb{R}^3$ is given by \rDfn{rateSN}. 
Here, the union is taken over all $S$ and $\rho$ that is in the form of \req{proogo}.
We do not require the condition that $\rho^S$ is the full-rank maximally mixed state on $S$.
We also define
\alg{
\Lambda(\ca{N})
:=
\bigcup_{S_c,S_r,\rho}
\Lambda(\ca{N},\rho),
}
where $\Lambda(\ca{N},\rho)$ is defined in \rDfn{rateSN2}.
The union is taken over all $S$ and $\rho$ that is decomposed as \req{proogostar}, 
but we do not require that $\{\rho_j^{S_rA}\}_j$ are pure states.
Note that the only difference between $\Theta(\ca{N},\rho)$ and $\Lambda(\ca{N},\rho)$ is in one of the inequality conditions for $(C,Q,E)$. That is, we require $Q+E\leq H(S_r|S_c)_\rho$ for $\Theta(\ca{N},\rho)$ and $C+2Q\leq I(S:B)_{\ca{N}(\rho)}$ for $\Lambda(\ca{N},\rho)$ (see Inequalities \req{proogo1} and \req{proogo21}).
The two sets are regularized into
\alg{
&
\Theta^{\infty}(\ca{N})
:=
\bigcup_{n=1}^\infty
\frac{1}{n}\Theta(\ca{N}^{\otm n}),
\\
&
\Lambda^{\infty}(\ca{N})
:=
\bigcup_{n=1}^\infty
\frac{1}{n}\Lambda(\ca{N}^{\otm n}).
\laeq{sarusaru2}
}
In appenxid \rApp{AEiid}, we prove the following proposition that implies \req{SfeqR}:

\bprp{SfeqR2}
For any quantum channel $\ca{N}$, it holds that
\alg{
\overline{\Theta_{\pi}^{\infty}(\ca{N})}
=
\overline{\Theta^{\infty}(\ca{N})}
=
\overline{\Lambda^{\infty}(\ca{N})}
=\overline{\Lambda_p^{\infty}(\ca{N})}.
\laeq{SfeqR2}
}
\eprp

The first equality in \req{SfeqR2} implies that the condition that the reduced state of the state $\rho$ on system $S_cS_r$ is the completely mixed state does not impose any restriction on the achievable rate region.
I.e., the rate region does not change when we drop this condition.
In \rSec{comparison}, we will invoke this fact to show that the known capacity theorems in special cases are obtained from \rThm{RRegionIID1}.

\subsection{Remark On The Non-Optimality of Time-Sharing Strategies}

Ref.~\cite{devetak2005capacity} addressed the task of simultaneously transmitting classical and quantum messages through a noisy quantum channel without the assistance of shared entanglement, i.e. the case where $E=0$ in \rDfn{achievableRT}.
A strategy for this task is to use $n\lambda$ copies of the channel to transmit classical messages and the remaining $n(1-\lambda)$ copies for quantum messages ($0\leq\lambda\leq1$), which is often referred to as the time sharing strategy.
It achieves the rate pair
\alg{
(C,Q)=(\lambda C(\ca{N}),(1-\lambda)Q(\ca{N})),
\laeq{timesharing}
}
with $C(\ca{N})$ and $Q(\ca{N})$ being the classical and quantum capacities of the channel.
Ref.~\cite{devetak2005capacity} proved that the time sharing strategy is not optimal in general: For a certain channel $\ca{N}$, there exists an achievable rate pair $(C,Q)$ that cannot be represented as \req{timesharing}.
In contrast, the time sharing strategy is optimal when the entanglement resource is freely available.
This is because $t>0$ qubits of quantum communication is converted to $2t$ bits of classical communication and vice versa, by means of superdense coding and quantum teleportation.

In the case where a {\it limited} amount of shared entanglement is available, it is not clear whether or not the time sharing strategy is optimal. 
This problem may be of independent interest because it would be closely related to the non-additivity of the classical capacity of a quantum channel under a limited amount of entanglement assistance \cite{zhu2017superadditivity}.
We leave this problem as an open question.

\section{Proof of \rThm{RRegionIID1}}
\lsec{PRFasymptotic}

In this section, we provide a proof for \rThm{RRegionIID1}.

\subsection{Proof of The Direct Part}

We prove the direct part of \rThm{RRegionIID1}, i.e.
\alg{
\ca{C}(\ca{N})
\supseteq
\overline{\Theta_{\pi}^{\infty}(\ca{N})}.
\laeq{motorcycle}
}
Since $\ca{C}(\ca{N})$ is a closed set, 
it suffices to prove that, for any $n$, a rate triplet $(C,Q,E)$ is achievable if $(C,Q,E)$ is an inner point of $\frac{1}{n}\Theta_{\pi}(\ca{N}^{\otm n})$.
We only consider the case where $n=1$.
It is straightforward to generalize the proof for $n\geq 2$. 
Fix an arbitrary state $\rho$ that satisfies the condition \req{proogo}, in addition to the condition that $\rho^S$ is the full-rank maximally mixed state on $S$. Fix an arbitrary triplet $(C,Q,E)$ that is an inner point of $\Theta(\ca{N},\rho)$.
Then, there exists $\nu>0$ such that
\alg{
Q+E
&\leq
 H(S_r|S_c)_\rho,
\laeq{proogol1}\\
C+Q-E
&
\leq
H(S_c)_\rho-H(S|B)_{\ca{N}(\rho)}-2\nu,
\laeq{proogol2}\\
Q-E
&\leq -H(S_r|BS_c)_{\ca{N}(\rho)}-2\nu.
\laeq{proogol3}
}
Fix arbitrary $\epsilon,\delta>0$ and choose sufficiently large $m$.
Due to the fully-quantum asymptotic equipartition property (\cite{tomamichel2009fully}: see also \rLmm{fqaep} in Appendix \rapp{PSE}), it holds that
\alg{
&
mH(S|B)_{\ca{N}(\rho)}
\geq
H_{\rm max}^{\epsilon}(S^m|B^m)_{\ca{N}(\rho)^{\otm m}}
-m\nu,
\nn\\
&
mH(S_r|BS_c)_{\ca{N}(\rho)}
\geq
H_{\rm max}^{\epsilon}(S_r^m|B^mS_c^m)_{\ca{N}(\rho)^{\otm m}}
-m\nu.
\nn
}
Combining this with \req{proogol1}-\req{proogol3}, and noting that $H(S_r|S_c)_\rho\leq\log{\dim{S_r}}$ and $H(S_c)_\rho=\log{ d_{S_c}}$, we obtain
\alg{
m(Q+E)
&\leq
m\log{\dim{S_r}},
\laeq{majen0}\\
m(C+Q-E)
&
\leq
m\log{d_{S_c}}
-H_{\rm max}^{\epsilon}(S^m|B^m)_{\ca{N}(\rho)^{\otm m}}
\nn\\
&
\quad\quad\quad\quad\quad\quad\quad\quad\quad
-m\nu,
\laeq{majen1}\\
m(Q-E)
&\leq
 -H_{\rm max}^{\epsilon}(S_r^m|B^mS_c^m)_{\ca{N}(\rho)^{\otm m}}
\nn\\
&
\quad\quad\quad\quad\quad\quad\quad\quad\quad
-m\nu.
\laeq{majen2}
}
We choose $m$ sufficiently large so that we have $-m\nu\leq\log{\delta}$. 
Denoting $mC,mQ,mE$ by $c,q,e$, respectively, it follows that
\alg{
q+e
&
\leq
\log{\dim{S_r^m}},
\laeq{tcard}\\
c+q-e
&
\leq
-H_{\rm max}^{\epsilon}(S^m|B^m)_{\ca{N}^{\otm m}(\rho^{\otm m})}
\nn\\
&\quad\quad\quad\quad\quad\quad\quad
+\log{d_{S_c}^m}
+\log{\delta},
\laeq{madema1}\\
q-e
&
\leq
-H_{\rm max}^{\epsilon}(S_r^m|B^mS_c^m)_{\ca{N}^{\otm m}(\rho^{\otm m})}
+\log{\delta}.
\laeq{madema2}
}

We separately consider the cases $d_{S_c}^m\geq\max\{2^c,2\}$ and $d_{S_c}^m<\max\{2^c,2\}$.
For the former case, we simply proceed with \req{madema1} to obtain
\alg{
c+q-e
&
\leq
-H_{\rm max}^{\epsilon}(S^m|B^m)_{\ca{N}(\rho)^{\otm m}}
\nn\\
&\quad\quad\quad\quad
+\log{(d_{S_c}^m-1)}
+\log{2\delta},
\laeq{madema1-2}
}
where we have used $d_{S_c}^m/(d_{S_c}^m-1)\leq2$.
Combining this with \req{tcard} and \req{madema2}, it follows from \rThm{OSDcomp} that there exists a $(c,q,e,\delta')$ code for the channel $\ca{N}^{\otm m}$, where $\delta':=2\sqrt{\sqrt{2\delta}+\sqrt{\delta}+4\epsilon}$.
For the latter case, we introduce a system $S_c'$ such that $d_{S_c}^md_{S_c'}\geq\max\{2^c,2\}$.
Denoting $S_c'S_c^m$ by $\hat{S}_c$ and $\hat{S}_cS_r^m=S_c'S^m$ by $\hat{S}$, we define the state
\alg{
\hat{\rho}_m^{\hat{S}A^m}
:=
\frac{1}{d_{S_c'}}\sum_{j'=1}^{d_{S_c'}}\proj{j'}^{S_c'}\otm\rho^{SA}.
}
Due to the property of the smooth max entropy for product states (see \rLmm{Hmaxprod} in \rApp{PSE}), we have
\alg{
&
H_{\rm max}^{\epsilon}(S^m|B^m)_{\ca{N}^{\otm m}(\rho^{\otm m})}
\geq
H_{\rm max}^{\epsilon}(\hat{S}|B^m)_{\ca{N}^{\otm m}(\hat{\rho}_m)}
\nn\\
&
\quad\quad\quad\quad\quad\quad\quad\quad\quad\quad\quad\quad\quad\quad\quad\quad\quad\quad
-\log{d_{\hat{S}_c}},
\nn\\
&
H_{\rm max}^{\epsilon}(S_r^m|B^mS_c^m)_{\ca{N}^{\otm m}(\rho^{\otm m})}
=
H_{\rm max}^{\epsilon}(S_r^m|B^m\hat{S}_c)_{\ca{N}^{\otm m}(\hat{\rho}_m)}.
\nn
}
Thus, Inequalities \req{madema1} and \req{madema2} yield
\alg{
c+q-e
&
\leq
-H_{\rm max}^{\epsilon}(\hat{S}|B^m)_{\ca{N}^{\otm m}(\hat{\rho}_m)}
\nn\\
&\quad\quad\quad\quad\quad\quad\quad
+\log{(d_{\hat{S}_c}-1)}
+\log{2\delta},
\laeq{madema1-3}\\
q-e
&
\leq
-H_{\rm max}^{\epsilon}(S_r^m|B^m\hat{S}_c)_{\ca{N}^{\otm m}(\hat{\rho}_m)}
+\log{\delta},
\laeq{madema2-3}
}
where we have used $d_{\hat{S}_c}/(d_{\hat{S}_c}-1)\leq2$ in the first line.
These two inequalities together with \req{tcard} imply that there exists a $(c,q,e,\delta')$ code for the channel $\ca{N}^{\otm m}$.
Since $\epsilon$ and $\delta$ can be arbitrarily small in both cases, we complete the proof of the direct part \req{motorcycle}.
\QED

\subsection{Proof of The Converse Part}

The converse part of \rThm{RRegionIID1} is given by
\alg{
\ca{C}(\ca{N})
\subseteq
\overline{\Theta_{\pi}^{\infty}(\ca{N})},
}
and is proved as follows.
Suppose that a rate triplet $(C,Q,E)$ is an inner point of $\ca{C}(\ca{N})$.
By definition, for any $\delta>0$ and sufficiently large $n$, there exist a $(nC,nQ,nE,\delta)$ code for the channel $\ca{N}^{\otm n}$.
Due to \rThm{converseonehyb}, there exists a quantum system $S\equiv S_cS_r$ and a state in the form of
\alg{
\rho_n^{SA^n}=\frac{1}{d_{S_c}}\sum_{j=1}^{d_{S_c}}\proj{j}^{S_c}\otm\rho_j^{S_rA^n},
\laeq{yapuyapu}
 }
 satisfying $d_{S}\leq2^{n(C+Q+E)}$ and $\rho^{S}=\pi^{S}$,
such that for any $\iota\in(0,1]$, it holds that
\alg{
n(Q+E)
&
\leq
\log{\dim{S_r}}
,
\\
n(C+Q-E)
&
\leq
\log{d_{S_c}}
-H_{\rm max}^{\lambda}(S|B^n)
-
\log{\iota},
\\
n(Q-E)
&
\leq
-H_{\rm max}^{\lambda'}(S_r|B^nS_c)-\log{\iota}.
}
The entropies are for the state $\ca{N}^{\otm n}(\rho_n)$, and the smoothing parameters $\lambda$ and $\lambda'$ are given by \req{smlam} and \req{smlamp}.
Note that we have $\rho^{S}=\ca{N}(\rho)^{S}$ since $\ca{N}$ acts only on $A$.
From the condition $\rho^{S}=\pi^{S}$, we have $\log{\dim{S_r}}=H(S_r|S_c)$.
Using the relation between the smooth max entropy and the von Neumann entropy (\rLmm{contHmax} in \rApp{PSE}), we also have
\alg{
&
H_{\rm max}^{\lambda}(S|B^n)
\geq
H(S|B^n)
-
\eta(\lambda)\log{d_{S}},
\\
&
H_{\rm max}^{\lambda'}(S_r|B^nS_c)
\geq
H(S_r|B^nS_c)
-
\eta(\lambda')\log{d_{S}},
}
where $\eta$ is a function that satisfies $\lim_{x\rightarrow0}\eta(x)=0$.
Combining these inequalities, we arrive at
\alg{
Q+E
&
\leq
\frac{1}{n}H(S_r|S_c)
,
\nn\\
C+Q-E
&
\leq
\frac{1}{n}(\log{d_{S_c}}-H(S|B^n)
+
\eta(\lambda)\log{d_{S}}
-
\log{\iota}),
\nn
\\
Q-E
&
\leq
\frac{1}{n}(
-H(S_r|B^nS_c)
+
\eta(\lambda')\log{d_{S}}-\log{\iota}).
\nn
}
By taking the limit of $n\rightarrow\infty$ and $\iota,\delta\rightarrow0$, we obtain $(C,Q,E)\in{\Theta_{\pi}^{\infty}(\ca{N})}$ and complete the proof. 
\QED

\section{Comparison to Previous Results in Special Cases}
\lsec{comparison}

In this section, we apply the results presented in \rSec{mainresults} and \rSec{iid} to the special cases where one or two of $C$, $Q$ and $E$ is/are equal to zero and compare the results with those in the previous literature. 
That is, we investigate the classical capacity, the quantum capacity, the entanglement-assisted classical capacity, the entanglement-assisted quantum capacity and the simultaneous capacity for classical and quantum information without the entanglement assistance, and discuss the relation with the existing results.
We consider a quantum channel $\ca{N}$ with the input system $A$ and the output system $B$.

For the one-shot scenario, we will invoke \rThm{OSDcomp}, \rThm{converseonehyb}, \rCrl{OSDcompunlim} and \rCrl{converseonehybunlim}.
The smoothing parameters $\epsilon(\delta,\delta')$, $\lambda(\delta,\iota)$ and $\lambda'(\delta,\iota)$ are set to be $\epsilon(\delta,\delta'):=\delta^2/16-\sqrt{\delta'}/4$, \req{smlam} and \req{smlamp}, respectively. 
When we write $\widehat{\sup}$, the supremum is taken over all finite dimensional quantum systems $S\equiv S_cS_r$ and all classical-quantum states $\rho^{S_cS_rA}$ between $S_c$ and $S_r$ such that $\rho^{S_cS_r}$ is the full-rank maximally mixed state. 
For each of the special cases, we compare our results with several results in the previous literature.
We, however, do not provide quantitative comparisons because all the results, including ours, involve complex optimizations, which makes quantitative analyses intractable.
All the one-shot results presented or reviewed in this section shall coincide with the known coding theorems in the asymptotic limit of infinitely many uses of a memoryless channel.

For the asymptotic scenario, we invoke \rThm{RRegionIID1} and prove that the known capacity theorems proved in the previous literature are recovered.
As mentioned in the discussion below \rPrp{SfeqR2}, and as proved in \rApp{ageha} in detail, the restriction in \rDfn{rateSN} that the reduced state of $\rho$ on $S$ is the completely mixed state does not impose any restriction on the achievable rate region.
Therefore, in the following subsections, we assume that this restriction is removed.
It is convenient to note that the R.H.S.s of \req{proogo1}-\req{proogo3} are evaluated as
\alg{
H(S_r|S_c)_\rho
&\geq\max\{0,-H(S_r|BS_c)_\rho\},
\laeq{orenji} \\
 H(S_c)_\rho-H(S|B)_{\ca{N}(\rho)}
&
 \geq
 -H(S_r|BS_c)_{\ca{N}(\rho)}.
}

\subsection{Classical Capacity without Entanglement Assistance}

The $\delta$-classical capacity $c(\ca{N},\delta)$ of a channel $\ca{N}$ is defined as the supremum of $c$ such that the triplet $(c,q=0,e=0)$ is achievable within the error $\delta$ for the channel $\ca{N}$.
Lower and upper bounds on the $\delta$-classical capacity are obtained from \rThm{OSDcomp} and \rThm{converseonehyb} by letting $q=e=0$.
We have
\alg{
&
\widehat{\sup_{S_c,\rho}}
\sup_{\delta'\in(0,\delta^4\!/\!16]}
\left[
\log{d_{S_c}}
-H_{\rm max}^{\epsilon(\delta,\delta')}(S_c|B)_{\ca{N}(\rho)}
+\log{2\delta'}
\right]
\nn\\
&\quad\quad\quad\quad\quad\quad\quad
\leq 
c(\ca{N},\delta)
\leq
\nn\\
&
\widehat{\sup_{S_c,S_r,\rho}} 
\inf_{\iota\in(0,1]}
\left[
\log{d_{S_c}}
\!-\!H_{\rm max}^{\lambda(\delta,\iota)}(S_cS_r|B)_{\ca{N}(\rho)}
\!-\!\log{\iota}
\right],
\laeq{kasina1}
}
where the map $\ca{C}$ is the completely dephasing operation on $S$ with respect to a fixed orthonormal basis.

Dupuis et al.~\cite{dupuis2014decoupling} addressed the one-shot classical capacity of a channel $\ca{N}'^{A\rightarrow B}$ with classical input and quantum output (the classical-quantum channel), namely a channel that satisfies $\ca{N}'^{A\rightarrow B}=\ca{N}'^{A\rightarrow B}\circ\ca{C}^A$ with $\ca{C}$ being the completely dephasing map on $A$ with respect to a fixed orthonormal basis.
Based on the `dequantizing theorem' (Theorem 3.1 in \cite{dupuis2014decoupling}),
they obtained a lower bound that holds for any $0<\delta<1$:
\alg{
c(\ca{N}',\delta)
\geq
&
\sup_{\varepsilon\in[0,\delta^2/8)]}
\left[
\log{(d_A-1)}
-
H_{\rm max}^\varepsilon(A|B)_\omega
\right.
\nn\\
&\quad\quad\quad\quad\quad\quad
\left.
+2\log{(\delta^2-8\varepsilon)}-1
\right],
\laeq{ypppodo}
}
where $\omega$ is the Choi-Jamio\l kowski state of $\ca{N}'$.
Renes {\it et al.} \cite{renes2011noisy} addressed the same task and obtained both lower and upper bounds (see Theorem 1 in \cite{renes2011noisy}).
The channel is modeled by an input alphabet $\ca{X}$ and the set of output quantum states $\{\theta_x\}_{x\in\ca{X}}$.
They obtained lower and upper bounds that yield
\alg{
&
\max_{P_X}
\left[
H_{\rm min}^{\delta/8}(X)_{\vartheta}
-H_{\rm max}^{\delta/8}(X|B)_{\vartheta}
+4\log{\delta}-16
\right]
\nn\\
&\quad
\leq 
c(\ca{N},\delta)
\leq
\max_{P_X}
\left[
H_{\rm min}(X)_{\vartheta}
-H_{\rm max}^{\sqrt{2\delta}}(X|B)_{\vartheta}
\right]
\laeq{kasina2}
}
for any $\delta>0$,
where the maximization is taken over all probability distribution $P_X$ on alphabet $\ca{X}$ and the entropies are for the state $\vartheta=\sum_{x\in\ca{X}}p_x\proj{x}^X\otm\theta_x^B$.
From both results, by taking supremum over all sets of output states of the form $\{\ca{N}(\rho_x)\}_{x\in\ca{X}}$, we obtain one-shot lower bounds that are similar to \req{ypppodo}.
The upper bounds in \req{kasina1} and \req{kasina2} are different in that \req{kasina1} includes system $S_r$.
We shall see below that this seeming mismatch vanishes in the asymptotic scenario.

Wang {\it et al.} \cite{wang2012one} addressed the same task in terms of the hypothesis testing approach and obtained lower and upper bounds that yield
\alg{
&
\sup_{P_X}D_H^{\delta/2}(\vartheta^{XB}\|\vartheta^X\otm\vartheta^B)+\log{\epsilon}-4
\nn\\
&\leq
c(\ca{N},\delta)
\leq
\sup_{P_X}D_H^{\delta}(\vartheta^{XB}\|\vartheta^X\otm\vartheta^B).
}
Here, $D_H$ is the hypothesis testing relative entropy and $\vartheta$ is defined in the same way as above.
The hypothesis testing approach to this problem was further improved by Datta {\it et al.} \cite{datta2013smooth} and Matthews {\it et al.} \cite{matthews2014finite}.
In particular, the result in Ref.~\cite{datta2013smooth} for the one-shot scenario recovers the strong converse rate \cite{winter1999coding,ogawa1999strong} when applied to the asymptotic scenario.

In the asymptotic limit, we obtain from \req{proogo2} and \req{sarusaru} that
\alg{
&C(\ca{N})
\leq
\lim_{n\rightarrow\infty}\frac{1}{n}\sup_{S_c,S_r,\rho}[H(S_c)_\rho-H(S|B^n)_{\ca{N}^{\otm n}(\rho)}]
\laeq{tenen}\\
&
\!\!\!\!
=
\lim_{n\rightarrow\infty}\frac{1}{n}\sup_{S_c,S_r,\rho}[
I(S_c:B^n)_{\ca{N}^{\otm n}(\rho)}
-H(S_r|B^nS_c)_{\ca{N}^{\otm n}(\rho)}
],
\nn
}
where the supremum is taken over all finite-dimensional quantum systems $S_c$, $S_r$ and all classical-quantum states $\rho$ between $S_c$ and $S_rA$.
Due to the argument in \cite{wildetext} (see the argument after Theorem 24.2.2 therein), we may, without loss of generality, assume that $S_r$ is a trivial (one-dimensional) system.
Hence, if $C$ is below the R.H.S. of \req{tenen}, both the conditions \req{proogo1} and \req{proogo3} are satisfied as well.
Therefore the inequality \req{tenen} is achievable and we recover the Holevo-Schumacher-Westmoreland theorem \cite{holevo98,schumacher97}, which yields
\alg{
C(\ca{N})
=
\lim_{n\rightarrow\infty}\frac{1}{n}\sup_{S_c,\rho}[
I(S_c:B^n)_{\ca{N}^{\otm n}(\rho)}
].
}

\subsection{Quantum Capacity without Entanglement Assistance}

The $\delta$-quantum capacity $q(\ca{N},\delta)$ of a channel $\ca{N}$ is defined as the supremum of $c$ such that the triplet $(c=0,q,e=0)$ is achievable within the error $\delta$ for the channel $\ca{N}$.
Lower and upper bounds on the $\delta$-quantum capacity are obtained from \rThm{OSDcomp} and \rThm{converseonehyb} by letting $c=e=0$, which yields
\alg{
&
\widehat{\sup_{S_r,\rho}}
\sup_{\delta'\in(0,\delta^4\!/\!16]}
\left[
-H_{\rm max}^{\epsilon(\delta,\delta')}(S_r|B)_{\ca{N}(\rho)}
+\log{\delta'}
\right]
\nn\\
&\quad\quad\quad\quad\quad
\leq
q(\ca{N},\delta)
\leq
\nn\\
&
\widehat{\sup_{S_r,\rho}}
\inf_{\iota\in(0,1]}
\left[
-H_{\rm max}^{\lambda'(\delta,\iota)}(S_r|B)_{\ca{N}(\rho)}
-\log{\iota}
\right].
\laeq{harahetta1}
}

Datta {\it et al.} \cite{datta2011apex} and Buscemi {\it et al.} \cite{buscemi2010quantum} addressed this task based on the decoupling approach.
Ref.~\cite{datta2011apex} defines the one-shot $\delta$-error quantum capacity $q_{\delta}^{(1)}(\ca{N})$ in terms of the entanglement transmission fidelity (see Definition 11 therein) and obtained the following lower and upper bounds that holds for any $0<\delta\leq1$:
\alg{
&
\max_{\ca{M}\subseteq\ca{H}^A}\left[
-H_{\rm max}^{\varepsilon}(S_r|B)_{\ca{N}(\Phi_{\ca{M}})}
\right]
+2\log{\left(\varepsilon+\sqrt{4\sqrt{\varepsilon}-4\varepsilon}\right)}
\nn\\
&\quad\quad
\leq q_{\delta}^{(1)}(\ca{N})
\leq
\max_{\ca{M}\subseteq\ca{H}^A}\left[
-H_{\rm max}^{\delta}(S_r|B)_{\ca{N}(\Phi_{\ca{M}})}
\right].
\laeq{harahetta2}
}
Here, the maximization is taken over all subspaces $\ca{M}\subseteq\ca{H}^A$, $|\Phi_{\ca{M}}\rangle\in\ca{M}\otimes\ca{H}^{S_r}$ is the maximally entangled state with Schmidt rank $\dim{\ca{M}}$, and $\varepsilon>0$ is chosen so that $\delta=2(5(\varepsilon+\sqrt{4\sqrt{\varepsilon}-4\varepsilon}))^{1/2}+2\sqrt{\varepsilon}$.
A comparison of this result with the one in \cite{buscemi2010quantum} was discussed in Section 3.3 of \cite{datta2011apex}.
Besides the values of the smoothing parameters, the difference between \req{harahetta1} and \req{harahetta2} is in whether the supremum over the states is limited to the maximally entangled state or not.
Tomamichel {\it et al.} \cite{tomamichel2016quantum} addressed the same task from the hypothesis-testing approach.

In the asymptotic limit, we have
\alg{
Q(\ca{N})=\lim_{n\rightarrow\infty}\frac{1}{n}\sup_{S_c,S_r,\rho}[-H(S_r|B^nS_c)_{\ca{N}^{\otm n}(\rho)}],
}
where the supremum is taken over all finite-dimensional quantum systems $S_c$, $S_r$ and a classical-quantum state $\rho$ between $S_c$ and $S_rA$.
Since the von Neumann entropy conditioned by a classical system is equal to the entropy averaged over the values of the classical system, without loss of generality we can assume that $S_c$ is a trivial (one-dimensional) system. 
Thus, we recover the Lloyd-Shor-Devetak theorem \cite{lloyd1997capacity,shor2002quantum,devetak2005private}, which yields
\alg{
Q(\ca{N})
=\lim_{n\rightarrow\infty}\frac{1}{n}\sup_{S_r,\rho}[-H(S_r|B^n)_{\ca{N}^{\otm n}(\rho)}].
\laeq{daitata}
}
One can also obtain \req{daitata} directly from \req{harahetta1}.

\subsection{Classical Capacity with Free Entanglement Assistance}

The entanglement-assisted $\delta$-classical capacity $c_e(\ca{N},\delta)$ is defined as the supremum of $c$ such that there exists $e>0$ and the triplet $(c,q=0,e)$ is achievable within the error $\delta$ for the channel $\ca{N}$.
Lower and upper bounds on the entanglement-assisted $\delta$-classical capacity is obtained from \rCrl{OSDcompunlim} and \rCrl{converseonehybunlim} by letting $q=0$, which yields
\alg{
&\widehat{\sup_{S,\rho}}
\sup_{\delta'\in(0,\delta^4\!/\!16]}
\left[
\log{d_{S}}
-H_{\rm max}^{\epsilon(\delta,\delta')}(S|B)_{\ca{N}(\rho)}
+\log{\delta'}
\right]
\nn\\
&\quad\quad\quad\quad\quad\quad
\leq 
c_e(\ca{N},\delta)
\leq
\nn\\
&\widehat{\sup_{S,\rho}}
\inf_{\iota\in(0,1]}
\left[
\log{d_{S}}
-H_{\rm max}^{\lambda'(\delta,\iota)}(S|B)_{\ca{N}(\rho)}
-\log{\iota}
\right].
}

Datta {\it et al.} \cite{datta2012one} investigated this task and obtained lower and upper bounds based on the decoupling approach.
Their proof is based on the fact that the classical capacity is twice as large as the quantum capacity if entanglement resource is freely available.
They defined the one-shot entanglement-assisted classical capacity $c_{{\rm ea},\epsilon}^{(1)}(\ca{N})$ in terms of the worst-case error probability (see Definition 13 in \cite{datta2012one}) and obtained lower and upper bounds thereof. 
They proved that for any $0<\varepsilon<1$ and $\varepsilon''$ such that $\varepsilon=2\sqrt[4]{2\sqrt{27\varepsilon''}+27\varepsilon''}$, it holds that
\alg{
&
\!\!\!\!\!\!
\max_{\varrho\in\ca{S}(\ca{H}^A)}\left[H_{\rm min}^{\varepsilon''}(S)_{\ca{N}(\phi)}-H_{\rm max}^{\varepsilon''}(S|B)_{\ca{N}(\phi)}\right]+4\log{\varepsilon''}-2
\nn\\
&\quad\quad\quad\quad\quad\quad\quad
\leq 
c_{{\rm ea},\epsilon}^{(1)}(\ca{N})
\leq
\nn\\
&
\max_{\varrho\in\ca{S}(\ca{H}^A)}\left[H_{\rm min}^{4\varepsilon}(S)_{\ca{N}(\phi)}-H_{\rm max}^{8\varepsilon+4\sqrt{2\sqrt{\varepsilon}}}(S|B)_{\ca{N}(\phi)}\right]
\nn\\
&\quad\quad\quad\quad\quad\quad\quad\quad\quad\quad\quad\quad\quad\quad\quad\quad
+\log{\frac{1}{2\sqrt{2}\varepsilon}},
}
where $\phi^{SA}$ is a purification of $\varrho\in\ca{S}(\ca{H}^A)$.

Datta {\it et al.} also addressed this task from the hypothesis testing approach and obtained a lower bound in terms of the average error probability \cite{datta2016second}.
Namely, they proved that for any $0<\delta<1$ and $0<\epsilon<\delta/2$, it holds that
\begin{eqnarray}
c_e(\ca{N},\delta)
\geq
D_H^{\delta-2\epsilon}(\ca{N}^{A'\rightarrow B}(\vartheta^{AA'})\|\ca{N}^{A'\rightarrow B}(\kappa^{AA'}))
\nn
\\-\log{\frac{1-\delta}{\epsilon^2}},
\end{eqnarray}
where $\vartheta^{AA'}$ is an arbitrary pure state that can be represented as a superposition of the maximally entangled state in orthogonal subspaces and $\kappa$ is the probabilistic mixture of the maximally mixed state in those subspaces.
Matthews {\it et al.} \cite{matthews2014finite} obtained a converse bound from the hypothesis testing approach, which yields
\alg{
c_e(\ca{N},\delta)
\leq
\max_{\rho\in\ca{S}(\ca{H}^{A'})}\min_{\sigma\in\ca{S}(\ca{H}^B)}D_H^\delta(\ca{N}^{A'\rightarrow B}(\phi^{AA'})\|\rho^A\otm\sigma^B),
}
where $\phi$ is a purification of $\rho$.
These two bounds are more general than ours, in that they do not assume that the resource state is the maximally entangled state.
The hypothesis testing approach to the entanglement-assisted quantum capacity has further been developed based on the convex splitting and position based coding \cite{qi2018applications,anshu2018building,anshu2019near}.

In the asymptotic limit, the conditions \req{proogo1}-\req{proogo4} are equivalent to
\alg{
\begin{cases}
C
\leq
H(S_c)_\rho-H(S|B)_{\ca{N}(\rho)}+E,
\\
H(S_r|BS_c)_{\ca{N}(\rho)}^+
\leq
E
\leq
 H(S_r|S_c)_\rho,
 \end{cases}
\laeq{proogo3p}
}
where
\alg{
H(S_r|BS_c)_{\ca{N}(\rho)}^+
:=
\max\{H(S_r|BS_c)_{\ca{N}(\rho)},0\}.
}
The two-dimensional achievable rate region in this scenario was originally proved by Shor \cite{shor2004classical}, which is represented by the condition
\alg{
\begin{cases}
C\leq I(S_cS_r:B)_{\ca{N}(\varrho)}
\\
E\geq H(S_r|S_c)_{\varrho}
\end{cases}.
\laeq{yoroko}
}
The regions \req{proogo3p} and \req{yoroko} coincide with each other by taking the union over all classical-quantum states $\rho$ and $\varrho'$ between $S_c$ and $S_rA$.
To show this, observe that the nontrivial vertices of the region \req{proogo3p} are given by
\alg{
(C_1,E_1)&=(I(S_cS_r:B)_{\ca{N}(\rho)},H(S_r|S_c)_{\rho}),
\\
(C_2,E_2)&=(I(S_c:B)_{\ca{N}(\rho)},H(S_r|BS_c)_{\ca{N}(\rho)}^+).
}
In both cases, there exists a state $\varrho$ such that $C_i\leq I(S_cS_r:B)_{\ca{N}(\varrho)}$ and $E\geq H(S_r|S_c)_{\varrho}$.
For $(C_2,E_2)$, we may choose $\varrho=\proj{0}^{S_r}\otm\rho^{S_cA}$.
Conversely, let $(C^*,E^*)$ be the nontrivial vertex of the region defined by \req{yoroko}, i.e., let $C^*$ and $E^*$ be equal to the R.H.S.s of \req{yoroko}.
It is straightforward to verify that $(C^*,E^*)$ satisfies the condition \req{proogo3p}.
Thus, by taking the union over all $\rho$ in \req{proogo3p} and by regularizing it, we obtain the two-dimensional achievable rate region proved by Shor \cite{shor2004classical}.

As proved in \cite{shor2004classical}, the single-letter formula for the entanglement-assisted classical capacity by Bennet {\it et al.} \cite{bennett1999entanglement,bennett2002entanglement} is obtained from \req{yoroko}.
An alternative proof is obtained from \rThm{RRegionIID1} as follows.
One can see if a pair $(C,E)$ satisfies the conditions \req{proogo1}-\req{proogo3} (which is equivalent to \req{proogo3p} when $Q=0$), it holds that
\alg{
C
\leq
I(S_r:B|S_c)_{\ca{N}(\rho)}.
}
Conversely, if the above inequality holds, we have
\alg{
C-H(S_c)_\rho+H(S|B)_{\ca{N}(\rho)}
&\leq
H(S_r|S_c)_\rho.
}
Since the R.H.S. of the above inequality is nonnegative, there exists $E\geq0$ that satisfies the conditions \req{proogo1}-\req{proogo4}.
Hence, we have
\alg{
C_e(\ca{N})
=
\lim_{n\rightarrow\infty}\frac{1}{n}\sup_{S_c,S_r,\rho}I(S_r:B|S_c)_{\ca{N}^{\otm n}(\rho)},
}
where the supremum is taken over all finite-dimensional quantum systems $S_c$, $S_r$ and a classical-quantum state $\rho$ between $S_c$ and $S_rA^n$.
Since the von Neumann entropy conditioned by a classical system is equal to the entropy averaged over the values of the classical system, without loss of generality we can assume that $S_c$ is a trivial (one-dimensional) system. 
Thus, we have
\alg{
C_e(\ca{N})
=
\lim_{n\rightarrow\infty}\frac{1}{n}\sup_{S_r,\rho}I(S_r:B)_{\ca{N}^{\otm n}(\rho)}.
}
It is known that the mutual information of quantum channels is additive, and that it is sufficient to take the supremum over all pure states (see e.g.~Section 12.4 in \cite{wildetext}).
Thus, we arrive at the entanglement-assisted classical capacity theorem by Bennet {\it et al.} \cite{bennett1999entanglement,bennett2002entanglement}, which yields
\alg{
C_e(\ca{N})
=
\sup_{|\phi\rangle^{AA'}}I(S_r:B)_{\ca{N}(\phi)}.
\laeq{orenki}
}

\subsection{Quantum Capacity with Free Entanglement Assistance}

The entanglement-assisted $\delta$-quantum capacity $q_e(\ca{N},\delta)$ is defined as the supremum of $q$ such that there exists $e>0$ and the triplet $(c=0,q,e)$ is achievable within the error $\delta$ for the channel $\ca{N}$.
Lower and upper bounds on the entanglement-assisted $\delta$-quantum capacity is obtained from \rCrl{OSDcompunlim} and \rCrl{converseonehybunlim} by letting $c=0$, which yields
\alg{
&\frac{1}{2}
\widehat{\sup_{S,\rho}}
\sup_{\delta'\in(0,\delta^4\!/\!16]}
\left[
\log{d_{S}}
-H_{\rm max}^{\epsilon(\delta,\delta')}(S|B)_{\ca{N}(\rho)}
+\log{\delta'}
\right]
\nn\\
&\quad\quad\quad\quad\quad\quad\quad
\leq 
q_e(\ca{N},\delta)
\leq
\nn\\
&
\frac{1}{2}
\widehat{\sup_{S,\rho}}
\inf_{\iota\in(0,1]}
\left[
\log{d_{S_r}}
-H_{\rm max}^{\lambda'(\delta,\iota)}(S|B)_{\ca{N}(\rho)}
-\log{\iota}
\right].
}

A similar result was obtained in Datta et al.~\cite{datta2012one} based on the decoupling approach.
They define the one-shot entanglement-assisted quantum capacity $q_{{\rm ea},\epsilon}^{(1)}$ in terms of the entanglement transmission fidelity (see Definition 6 and 7 therein) and obtained lower and upper bounds thereof. 
They proved that for any $0<\varepsilon<1$ and $\varepsilon'$ such that $\varepsilon=2\sqrt{2\sqrt{27\varepsilon'}+27\varepsilon'}$, it holds that
\alg{
&
\max_{\varrho\in\ca{S}(\ca{H}^A)}\frac{1}{2}\left[H_{\rm min}^{\varepsilon'}(S)_{\ca{N}(\phi)}-H_{\rm max}^{\varepsilon'}(S|B)_{\ca{N}(\phi)}\right]+2\log{\varepsilon'}
\nn\\
&\quad\quad\quad\quad\quad\quad\quad
\leq 
q_{{\rm ea},\varepsilon}^{(1)}(\ca{N})
\leq
\nn\\
&
\max_{\varrho\in\ca{S}(\ca{H}^A)}\frac{1}{2}\left[H_{\rm min}^{\varepsilon}(S)_{\ca{N}(\phi)}-H_{\rm max}^{2\varepsilon+4\sqrt[4]{\varepsilon}}(S|B)_{\ca{N}(\phi)}\right]+\log{\frac{\sqrt{2}}{\varepsilon}},
}
where $\phi^{SA}$ is a purification of $\varrho\in\ca{S}(\ca{H}^A)$.
They also proved that similar bounds hold for the entanglement-assisted quantum capacity defined in terms of the minimum output fidelity (see Theorem 11 in \cite{datta2012one}).

In the asymptotic limit, one can see that if there exists $(Q,E)$ satisfying the conditions \req{proogo1} and \req{proogo3}, it holds that
\alg{
\begin{cases}
Q
&
\leq
\frac{1}{2}I(S_r:B|S_c)_{\ca{N}(\rho)},
\\
Q-E
&\leq
-H(S_r|BS_c)_{\ca{N}(\rho)}
\end{cases}
\laeq{darkw}
}
Conversely, if both of the above inequality hold, there exists $\Delta\geq0$ such that
\alg{
Q+E
&
\leq
H(S_r|S_c)_{\ca{N}(\rho)}+\Delta,
\\
Q-E
&\leq
-H(S_r|BS_c)_{\ca{N}(\rho)}-\Delta.
}
Indeed, letting $\Delta=Q+E-H(S_r|S_c)_{\ca{N}(\rho)}$ yields
\alg{
-H(S_r|BS_c)_{\ca{N}(\rho)}-\Delta
&
=I(S_r:B|S_c)_{\ca{N}(\rho)}-Q-E
\nn\\
&
\geq
Q-E,
}
where the second line follows from the first condition in \req{darkw}.
For any $m\in\mbb{N}$, let $S_m$ be an $2^{\lfloor m\Delta\rfloor}$-dimensional quantum system, $\pi_m$ be the maximally mixed state thereon, and $\varrho_m\equiv\pi_m\otm\rho$. Denoting $S_mS_r^m\equiv \hat{S}_r$, we obtain
\alg{
Q+E
&
\leq
\frac{1}{m}[H(\hat{S}_r|S_c^m)_{\ca{N}^{\otm m}(\varrho_m)}+1],
\\
Q-E
&\leq
-\frac{1}{m}H(\hat{S}_r|B^nS_c^n)_{\ca{N}^{\otm n}(\varrho_m)}.
}
Taking the limit of $m$ to infinity, we observe that the pair $(Q,E)$ belongs to the region defined by \rDfn{rateSN} for $C=0$.
Thus, we obtain the two-dimensional rate region obtained by Devetak {\it et al.} \cite{deve08} (Theorem 7.7 in \cite{deve08}).
I.e., the region obtained from the set of all pairs $(Q,E)$ satisfying
\alg{
\begin{cases}
Q
&
\leq
\frac{1}{2}I(A:B)_{\ca{N}(\phi)}
\\
Q-E
&\leq
-H(A|B)_{\ca{N}(\phi)}
\end{cases},
}
by taking the union over all pure states $|\phi\rangle^{AA'}$ and by regularizing it.
Note that the region defined in \cite{deve08} is convex because of regularization.
Hence, the conditioning by a classical system $S_c$ does not change the rate region.

\subsection{Simultaneous Capacity of Classical and Quantum Information without Entanglement Assistance}

We consider the one-shot capacity region for simultaneously transmitting classical and quantum messages through the channel without entanglement assistance. 
The $\delta$-simultaneous capacity region of a channel $\ca{N}$ is defined as the set of all pairs $(c,q)\in\mbb{R}_\geq^2$ such that the triplet $(c,q,e=0)$ is achievable within the error $\delta$ for the channel $\ca{N}$. 
We denote the achievable rate region by $\Gamma_\delta(\ca{N})$, and assume that $\delta\in(0,2]$. 
For an arbitrary system $S\equiv S_cS_r$ and an arbitrary state $\rho^{SA}$ that is diagonal in $S_c$ with respect to a fixed orthonormal basis, let $\Gamma_{\delta,\delta',\epsilon}^{\rm in}(\ca{N},\rho)$ be the set of all pairs $(c,q)\in\mbb{R}_\geq^2$ that satisfy
\alg{
\begin{cases}
c+q
&
\leq
-H_{\rm max}^{\epsilon(\delta,\delta')}(S|B)_{\ca{N}(\rho)}
+\log{(d_{S_c}-1)}
+\log{\delta'},
\\
q
&
\leq
-H_{\rm max}^{\epsilon(\delta,\delta')}(S_r|BS_c)_{\ca{N}(\rho)}
+\log{\delta'(1-2\epsilon(\delta,\delta'))}
\end{cases}
\nn
}
and let $\Gamma_{\delta,\iota}^{\rm out}(\ca{N},\rho)$ be the one that satisfy
\alg{
\begin{cases}
c+q
&
\leq
-H_{\rm max}^{\lambda(\delta,\iota)}(S|B)_{\ca{N}(\rho)}
+\log{d_{S_c}}-
\log{\iota},
\\
q
&
\leq
-H_{\rm max}^{\lambda'(\delta,\iota)}(S_r|BS_c)_{\ca{N}(\rho)}-\log{\iota}.
\end{cases}
\laeq{soroIIii}
}
It follows from \rThm{OSDcomp} and \rThm{converseonehyb} that
\alg{
\bigcup_{S,\rho}
\bigcup_{\delta'}
\Gamma_{\delta,\delta'}^{\rm in}(\ca{N},\rho)
\subseteq
\Gamma_\delta(\ca{N})
\subseteq
\bigcup_{S,\rho}
\bigcap_{\iota}
\Gamma_{\delta,\iota}^{\rm out}(\ca{N},\rho).
\laeq{cqregionone}
}
Here, the union over $\delta'$ and the intersection over $\iota$ are taken in the intervals $\delta'\in(0,\delta^4/16]$ and $\iota\in(0,1]$, respectively.
The union over $\rho$ is taken over all classical-quantum states $\rho$ between $S_c$ and $S_rA$ such that $\rho^S$ is the full-rank maximally mixed.

The same communication task has been analyzed by Salek et al.~\cite{salek2019one} for the one-shot scenario based on the hypothesis-testing approach. 
They first proved the inner and outer bounds on the two-dimensional capacity region for simultaneously transmitting public and private messages (see Theorem 2 and 3 therein).
Then they apply the argument by Devetak {\it et al.} \cite{devetak2005private}, which states that the private classical capacity of a quantum channel is equal to the quantum capacity.
The outer bound is given as the union, over all states $\rho$, of the set of rate pairs $(c,q)$ satisfying
\alg{
\begin{cases}
c&\leq I_H^\delta(X:B)_\varrho
\\
q&\leq I_H^{\sqrt{\delta}}(Y:B|X)_\varrho-I_{\rm max}^{\sqrt{2\delta'}}(Y:E|X)_\varrho
\end{cases}.
\laeq{soroII}
}
Here, $I_H$ and $I_{\rm max}$ are the hypothesis testing mutual information and the max mutual information, and $\delta$ and $\delta'$ are the error parameters for the classical message and the quantum message, respectively.
The mutual informations are for the state of the form $\varrho=\sum_{x,y}p(x,y)\proj{x}\otm\proj{y}\otm\ca{N}^{A\rightarrow B,E}(\rho_{x,y})$, where $\ca{N}^{A\rightarrow E}$ is the complementary channel of $\ca{N}^{A\rightarrow B}$.
The inner bound is given in a similar form, with the additional terms on the R.H.S.s of the two inequality that depends only on the smoothing parameters and the errors.

One may think that our bound \req{cqregionone} is not consistent with those obtained \req{soroII}.
The condition \req{soroII} defines a rectangle, while \req{soroIIii} defines a trapezoid.
Note, however, that the actual bounds are obtained by taking the union over states $\rho$ in both cases, which resolves this seeming mismatch.

In the asymptotic limit, we obtain from the conditions \req{proogo1}-\req{proogo3} that
\alg{
\begin{cases}
C+Q
&
\leq
H(S_c)_\rho-H(S|B)_{\ca{N}(\rho)},
\\
Q
&\leq -H(S_r|BS_c)_{\ca{N}(\rho)},
\end{cases}
}
which is equivalent to the conditions obtained by Hsieh and Wilde in \cite{hsieh2010entanglement} (see Theorem 5).
Originally, Devetak and Shor \cite{devetak2005capacity} addressed this problem and obtained the achievable rate region which is characterized by the following conditions:
\alg{
\begin{cases}
C
&
\leq
I(S_c:B)_{\ca{N}(\rho)},
\\
Q
&\leq -H(S_r|BS_c)_{\ca{N}(\rho)}.
\end{cases}
}
As discussed in \cite{hsieh2010entanglement}, by taking the union over all state $\rho$ and by regularizing it, the above regions coincide with each other.

\section{Conclusion}
\lsec{cncl}

In this paper, we analyzed the task of simultaneously transmitting classical and quantum messages through a noisy quantum channel assisted by a limited amount of shared entanglement in the one-shot scenario. 
We have derived direct and converse bounds for the one-shot achievable rate region.  
To our knowledge, this is the first time that trade-off relations that apply simultaneously to all the three resources of classical communication, quantum communication, and shared entanglement has been obtained in the one-shot channel coding scenario.
We then applied the one-shot result to the asymptotic scenario of infinitely many uses of a memoryless channel.
We obtained a complete characterization of the triple resource trade-offs, which coincides with the prior result by Hsieh and Wilde \cite{hsieh2010entanglement}. 
Numerical calculations of the achievable rate region for specific qubit channels are given in \cite{nakata2021one}. 
In \cite{wakakuwa2020one}, we analyzed quantum state redistribution for a classical-quantum hybrid source in the one-shot scenario in terms of randomized partial decoupling. 
To investigate relations between these two tasks is left as future work.

To investigate quantum channel coding problems in the one-shot scenario, an approach from hypothesis testing (see e.g.~\cite{mosonyi2009generalized,wang2012one,tomamichel2016quantum,matthews2014finite,datta2013smooth,datta2016second,salek2019one,wilde2017converse}) is complementary to the one based on decoupling.
The hypothesis-testing approach has an advantage over the decoupling approach in that it recovers the optimal asymptotic results not only in the i.i.d. setting but also in the non-i.i.d. (information spectrum) setting.
This approach has been widely used for quantum channel coding problems e.g.~to investigate the strong converse bounds and the second-order asymptotics.
Along this line, the concepts of convex splitting \cite{anshu2017quantum} and position-based coding \cite{anshu2018building} have recently proved useful in various settings of one-shot quantum channel coding \cite{qi2018applications,salek2019one,anshu2018building,anshu2019near,salek2018one,wilde2017position,gagatsos2020covert,sen2021inner,wang2022towards,anshu2022efficient,salek2020single,salek2019publicness}.
One future direction is to apply the hypothesis-testing approach to the problem addressed in this paper and obtain triple-resource trade-off bounds that are asymptotically tight in the non-i.i.d. setting.

\section*{Acknowledgement}

The authors thank Hayata Yamasaki and Min-Hsiu Hsieh for useful discussions.
E. W. is supported by JSPS KAKENHI Grant No.~18J01329. Y. N. is supported by JST, PRESTO Grant No.~JPMJPR1865, Japan.

\appendices

\section{Technical Lemmas}
\lapp{TechLmm}

We introduce some technical lemmas that are used in the main text.

\blmm{cqerroreq}
Consider two states $\rho$ and $\sigma$ in the form of 
\alg{
&
\rho=\frac{1}{K}\sum_{k=1}^K\proj{k}^X\otm\proj{k}^Y\otm\rho_k^A,
\\
&
\sigma=\frac{1}{K}\sum_{k=1}^K\proj{k}^X\!\otm\!\left(\sum_{k'=1}^Kp(k'|k)\proj{k'}^Y\!\otm\!\sigma_{kk'}^A\right)\!,\!
}
where $\{\ket{k}\}_k$ is an orthonormal basis of $\ca{H}^X$ and $\ca{H}^Y$, $\{p(k'|k)\}_{k'=1}^K$ is a conditional probability distribution, and $\rho_k$ and $\sigma_{kk'}$ are normalized states on $A$ for each $k$ and $k'$.
Suppose that we have
\alg{
&
\frac{1}{K}\sum_{k=1}^K(1-p(k|k))\leq\frac{\delta}{3},
\\
&
\frac{1}{K}\sum_{k=1}^K
\left\|\rho_k-\sum_{k'=1}^Kp(k'|k)\sigma_{kk'}\right\|_1
\leq
\frac{\delta}{3}
}
for $\delta>0$.
Then, it holds that
\alg{
\left\|\rho-\sigma\right\|_1\leq\delta.
}
\elmm

\bprf
Using the properties of the trace distance (see e.g.~ Section 9.1 in \cite{wildetext}), we have
\alg{
&
\left\|\rho-\sigma\right\|_1
\nn\\
&
=
\frac{1}{K}\sum_{k=1}^K
\left\|\proj{k}^Y\otm\rho_k^A-\sum_{k'=1}^Kp(k'|k)\proj{k'}^Y\!\otm\!\sigma_{kk'}^A\right\|_1
\nn\\
&
=
\frac{1}{K}\sum_{k=1}^K
\left\|\rho_k^A-p(k|k)\sigma_{kk}^A\right\|_1
\nn\\
&\quad\quad\quad\quad\quad\quad
+
\frac{1}{K}\sum_{k=1}^K\sum_{k'\neq k}p(k'|k)\left\|\sigma_{kk'}^A\right\|_1
\\
&
\leq
\frac{1}{K}\sum_{k=1}^K
\left\|\rho_k^A-\sum_{k'=1}^Kp(k'|k)\sigma_{kk'}^A\right\|_1
\nn\\
&\quad\quad\quad\quad\quad\quad
+
\frac{2}{K}\sum_{k=1}^K\sum_{k'\neq k}p(k'|k)\left\|\sigma_{kk'}^A\right\|_1
\laeq{acchii}\\
&
=
\frac{1}{K}\sum_{k=1}^K
\left\|\rho_k^A-\sum_{k'=1}^Kp(k'|k)\sigma_{kk'}^A\right\|_1
+
\frac{2}{K}\sum_{k=1}^K(1-p(k|k))
\nn\\
&
\leq
\delta,
}
where Inequality \req{acchii} follows due to the triangle inequality.
\QED
\eprf

\blmm{cohpurifi}
Consider a state $\rho$ on $XA$ and a pure state $\ket{\phi}$ on $XYAB$ that take the forms of
\alg{
&
\rho
=
\sum_kp_k\proj{k}^X\otm\varrho_k^A,
\\
&
\ket{\phi}
=
\sum_k\sqrt{p_k}\ket{k}^X\ket{k}^Y\ket{\varphi_k}^{AB},
}
where $\{p_k\}_k$ is a probability distribution, $\varrho_k$ is a state on $A$ and $\ket{\varphi_k}$ is a pure state on $AB$ for each $k$.
Suppose that 
\alg{
\left\|\phi^{XA}-\rho^{XA}\right\|_1\leq\delta.
}
Then, there exists a purification $\ket{\Psi}^{XYAB}$ of $\rho$ that takes the form of 
\alg{
\ket{\Psi}
=
\sum_k\sqrt{p_k}\ket{k}^X\ket{k}^Y\ket{\psi_k}^{AB},
}
and satisfies
$\|\proj{\Psi}-\proj{\phi}\|_1\leq2\sqrt{\delta}$.
\elmm

\bprf
Any purification $\ket{\Psi'}$ of $\rho$ is represented as
\alg{
\ket{\Psi'}
=
\sum_k\sqrt{p_k}\ket{k}^X\ket{\psi_k'}^{YAB},
}
where $\ket{\psi_k'}$ is a purification of $\varrho_k$ for each $k$.
It follows that
\alg{
|\inpro{\Psi'}{\phi}|
&
=
\left|\sum_kp_k\bra{\psi_k'}^{YAB}\ket{k}^Y\ket{\psi_k}^{AB}\right|
\\
&
\leq
\sum_kp_k\left|\bra{\psi_k'}^{YAB}\ket{k}^Y\ket{\psi_k}^{AB}\right|
\\
&
\leq
\sum_kp_k\max_{\psi_k''}|\inpro{\psi_k''}{\psi_k}|
\\
&
=
\sum_kp_k|\inpro{\psi_k^*}{\psi_k}|,
\laeq{riigon}
}
where we have defined
\alg{
\ket{\psi_k^*}
:=
\argmax_{\ket{\psi_k''}}|\inpro{\psi_k''}{\psi_k}|.
} 
The maximization in the fourth line is taken over all purifications $\ket{\psi_k''}^{AB}$ of $\varrho_k^A$.
We consider a state
\alg{
\ket{\Psi}
=
\sum_k\sqrt{p_k}\ket{k}^X\ket{k}^Y\ket{\psi_k^*}^{AB}.
}
Due to \req{riigon}, it holds that
\alg{
F(\rho^{XA},\phi^{XA})
=
\max_{|\Psi'\rangle}|\inpro{\Psi'}{\phi}|
=
|\inpro{\Psi}{\phi}|
=
F(\ket{\Psi},\ket{\phi}),
\nn
}
where $F$ is the fidelity defined by 
$
F(\sigma,\tau):=\|\sqrt{\sigma}\sqrt{\tau}\|_1
$.
The maximization is taken over all purifications of $\rho^{XA}$
and the first equality follows from Uhlmann's theorem \cite{uhlmann1976transition}.
By using the relation between the trace distance and the fidelity (see e.g.~Section 9.2.3 in \cite{nielsentext}), we obtain
\alg{
&
1-F(\phi^{XA},\rho^{XA})
\leq
\left\|\phi^{XA}-\rho^{XA}\right\|_1
\leq\delta,
\\
&
\left\|\proj{\Psi}-\proj{\phi}\right\|_1
\leq
2\sqrt{1-F(\ket{\Psi},\ket{\phi})}.
}
Combining these all together, we complete the proof.
\QED
\eprf

\section{Properties of Entropies}
\lapp{PSE}

In this section, we summarize properties of quantum entropies that are used in the proofs of the main results.
Note that the set of positive semidefinite operators, normalized states and subnormalized states are defined by
\begin{align}
&
\ca{P}(\ca{H}) = \{\rho \in {\rm Her}(\ca{H}) : \rho \geq 0 \},
\\
&
\ca{S}_=(\ca{H}) = \{\rho \in \ca{P}(\ca{H}) : \tr [\rho]=1 \},
\\
&
\ca{S}_{\leq}(\ca{H}) = \{\rho \in \ca{P}(\ca{H}) : \tr [\rho] \leq 1 \}.
\end{align}

\begin{lmm}
\label{lmm:duality}
(Definition 14, Equality (6) and Lemma 16 in \cite{tomamichel2010duality})
For any subnormalized pure state $|\psi\rangle$ on system $ABC$, and for any $\epsilon>0$, 
$H_{\rm max}^\epsilon(A|B)_\psi=
-
H_{\rm min}^\epsilon(A|C)_\psi$.
\end{lmm}

\blmm{DPIsmoothmax}
(Corollary of Theorem 18 in  \cite{tomamichel2010duality})
For any state $\rho^{AB}\in\ca{S}_=(\ca{H}^{AB})$, any CPTP map $\ca{E}^{A\rightarrow B}$ and any $\epsilon\geq0$, it holds that
\alg{
H_{\rm max}^{\epsilon}(A|B)_\rho
\leq
H_{\rm max}^{\epsilon}(A|C)_{\ca{E}(\rho)}.
}
\elmm

\begin{lmm}\label{lmm:SE2}
(Corollary of Lemma 20 in \cite{tomamichel2010duality})
For any $\rho^{AB}\in\ca{S}_\leq(\ca{H}^{AB})$, it holds that
\alg{
-\log{d_A}
\leq
H_{\rm max}(A|B)_\rho
-
\log{{\rm Tr}[\rho^{AB}]}
\leq
\log{d_A}.
}
\elmm

\begin{lmm}\label{lmm:SE3}
[Lemma A.2 in \cite{DBWR2010}]
For any $\rho^{AB}\in\ca{S}_=(\ca{H}^{AB})$, $\sigma^{CD}\in\ca{S}_=(\ca{H}^{CD})$ and any $\epsilon,\epsilon'\geq0$, it holds that
\alg{
\!\!
H_{\rm min}^{\epsilon+\epsilon'}(AC|BD)_{\rho\otm\sigma}
\geq
H_{\rm min}^{\epsilon}(A|B)_{\rho}
\!+\!
H_{\rm min}^{\epsilon'}(C|D)_{\sigma}.
\!\!\!
}
\elmm

\blmm{fqaep}
(Theorem 1 in \cite{tomamichel2009fully})
For any $\rho\in\ca{S}_=(\ca{H}^{AB})$ and $0<\epsilon<1$, it holds that
\alg{
\lim_{n\rightarrow\infty}\frac{1}{n}H_{\rm max}^\epsilon(A^n|B^n)_{\rho^{\otm n}}
=
H(A|B)_\rho.
}
\elmm

\blmm{maxvNub}
(Corollary of Lemma 2 in \cite{tomamichel2009fully})
For any $\rho\in\ca{S}_=(\ca{H}^{AB})$, it holds that
\alg{
H_{\rm max}(A|B)_\rho\geq H(A|B)_\rho.
}
\elmm

\blmm{Hmaxprod}
For any $\rho^{AB}\in\ca{S}_=(\ca{H}^{AB})$ and $\xi^C\in\ca{S}_=(\ca{H}^{C})$, it holds that
\alg{
H_{\rm max}^{\epsilon}(A|BC)_{\rho\otm\xi}
&
=
H_{\rm max}^{\epsilon}(A|B)_{\rho},
\laeq{Hmaxprodeq}
\\
H_{\rm max}^{\epsilon}(AC|B)_{\rho\otm\xi}
&
\leq
H_{\rm max}^{\epsilon}(A|B)_{\rho}
+
\log{d_C}.
\laeq{Hmaxprodineq}
}
\elmm

\bprf
To prove Equality \req{Hmaxprodeq}, define an operation $\ca{E}_\xi^{B\rightarrow BC}$ by $\ca{E}_\xi(\tau^B)=\tau^B\otm\xi^C$.
Due to the monotonicity of the smooth max entropy (\rLmm{DPIsmoothmax}) under $\ca{E}_\xi^{B\rightarrow BC}$ and the partial trace operation, it holds that
\alg{
&
H_{\rm max}^{\epsilon}(A|B)_{\rho}
\leq
H_{\rm max}^{\epsilon}(A|BC)_{\ca{E}_\xi(\rho)}
\nn\\
&\quad
=
H_{\rm max}^{\epsilon}(A|BC)_{\rho\otm\xi}
\leq
H_{\rm max}^{\epsilon}(A|B)_{\rho},
}
which implies \req{Hmaxprodeq}.
To prove \req{Hmaxprodineq}, note that \rLmm{SE3} and the duality relation (\rLmm{duality}) imply, for any $\eta\in\ca{S}_=(\ca{H}^D)$,
\alg{
&
H_{\rm max}^{\epsilon+\epsilon'}(AC|BD)_{\rho\otm\xi\otm\eta}
\nn\\
&
\quad
\leq
H_{\rm max}^{\epsilon}(A|B)_{\rho}
+
H_{\rm max}^{\epsilon'}(C|D)_{\xi\otm\eta}.
}
We particularly choose $\epsilon'=0$.
Due to Inequality \req{Hmaxprodeq}, the L.H.S. is equal to $H_{\rm max}^{\epsilon}(AC|B)_{\rho\otm\xi}$.
From \rLmm{SE2}, the second term in the R.H.S. is bounded as
$
H_{\rm max}(C|D)_{\xi\otm\eta}
\leq
\log{d_C},
$
which completes the proof.
\QED
\eprf

\blmm{tomerare}
For any $\rho^{AB}\in\ca{S}_=(\ca{H}^{AB})$ and $\epsilon\geq0$, it holds that
\alg{
-\log{d_A}
\leq
H_{\rm max}^\epsilon(A|B)_\rho
-
\log{(1-2\epsilon)}.
}
\elmm

\bprf
Let $\hat{\rho}^{AB}\in\ca{B}^\epsilon(\rho)$ be such that $H_{\rm max}^\epsilon(A|B)_\rho=H_{\rm max}(A|B)_{\hat\rho}$.
Due to \rLmm{SE2}, it holds that
\alg{
-\log{d_A}
\leq
H_{\rm max}(A|B)_{\hat{\rho}}
-
\log{{\rm Tr}[\hat{\rho}^{AB}]}.
\laeq{katachino}
}
Using the triangle inequality for the trace distance, we have
\alg{
{\rm Tr}[\hat{\rho}^{AB}]
=\|\hat{\rho}^{AB}\|_1
\geq
\|\rho^{AB}\|_1-\|\rho^{AB}-\hat{\rho}^{AB}\|_1
\geq
1-2\epsilon,
\nn
}
where the last inequality follows from the relation between the trace distance and the purified distance \req{relTDPD}.
Substituting this to \req{katachino}, we complete the proof.
\QED
\eprf

\blmm{contHmax}
For any $0\leq\epsilon<1$ and any state $\rho\in\ca{S}_=(\ca{H}^{AB})$, it holds that
\alg{
H_{\rm max}^\epsilon(A|B)_\rho
\geq
H(A|B)_\rho
-
\eta(\epsilon)\log{d_A},
\laeq{contHmax}
}
where $\eta$ is a function that satisfies $\lim_{x\rightarrow0}\eta(x)=0$ and is independent of the dimensions of the systems.
\elmm

\bprf
Let $\hat{\rho}\in\ca{B}^\epsilon(\rho)$ be a subnormalized state such that $H_{\rm max}^\epsilon(A|B)_\rho=H_{\rm max}^\epsilon(A|B)_{\hat{\rho}}$.
From \rLmm{maxvNub}, it holds that
\alg{
H_{\rm max}(A|B)_{\hat{\rho}/{\rm Tr}[\hat{\rho}]}\geq H(A|B)_{\hat{\rho}/{\rm Tr}[\hat{\rho}]}.
}
Thus, Inequality \req{contHmax} follows due to the Alicki-Fannes inequality (\!\cite{alicki04}, see also Inequality (89) in \cite{wakakuwa2017coding}).
\QED
\eprf

\blmm{entpinbo}
Let $\{\Pi_m\}_{m=1}^M$ be a complete set of orthogonal projectors on a finite dimensional Hilbert space $\ca{H}^A$, and let $X$ be a quantum system with a fixed orthonormal basis $\{\ket{m}\}_{m=1}^M$.
Consider a map $\ca{E}:A\rightarrow XA$ defined by
\alg{
\ca{E}(\cdot):=\sum_{m=1}^M\proj{m}^X\otm\Pi_m(\cdot)\Pi_m^A.
}
For any state $\rho$ on system $AB$, it holds that
\alg{
&
H(A|B)_\rho
\leq
H(XA|B)_{\ca{E}(\rho)},
\\
&
H(A|BX)_{\ca{E}(\rho)}
\leq
H(A|B)_\rho
+\log{M}.
}
\elmm

\bprf
The first inequality follows from the isometric invariance of the conditional quantum entropy and its monotonicity under the completely dephasing operation (see e.g. Corollary 11.9.4 in \cite{wildetext}).
Note that the map $\ca{E}$ is represented as $\ca{E}=\ca{C}^X\circ\ca{V}^{A\rightarrow XA}$, where $V$ is a linear isometry defined by $V:=\sum_{m=1}^M\ket{m}^X\otm\Pi_m^A$ and $\ca{C}$ is the completely dephasing operation on $X$ with respect to the basis $\{\ket{m}\}_{m=1}^M$.
To prove the second inequality, let $X'$ be a $M$-dimensional Hilbert space with a fixed orthonormal basis $\{\ket{m}\}_{m=1}^M$.
Define a linear isometry $V:\ca{H}^A\rightarrow \ca{H}^X\otm\ca{H}^{X'}\otm\ca{H}^A$ by
\alg{
V:=\sum_{m=1}^M\ket{m}^{X}\otm\ket{m}^{X'}\otm\Pi_m^A.
}
A Stinespring dilation of the map $\ca{E}$ is given by $\ca{E}={\rm Tr}_{X'}\circ\ca{V}$.
It holds that
\alg{
&
H(A|B)_\rho
\\
&
=
H(XX'A|B)_{\ca{V}(\rho)}
\\
&
\geq
H(X|B)_{\ca{V}(\rho)}+
H(A|BX)_{\ca{V}(\rho)}
\nn\\
&\quad\quad\quad\quad+
H(X'|ABX)_{\ca{V}(\rho)}
\\
&
\geq
H(A|BX)_{\ca{V}(\rho)}-\log{\dim{X}}
\\
&
=
H(A|BX)_{\ca{E}(\rho)}-\log{M},
}
where the first line follows from the isometric invariance of the conditional quantum entropy, the second line from the chain rule, the third line due to $H(X|B)_{\ca{V}(\rho)}\geq0$ and $H(X'|ABX)_{\ca{V}(\rho)}\geq-\log{\dim{X}}$, and the last line from  $\ca{V}(\rho)^{AB}=\ca{E}(\rho)$.
\QED
\eprf

\blmm{fano}
Let $\{p_x,\rho_x\}_{x\in\ca{X}}$ be an ensemble of states on system $A$, and suppose that there exists a POVM $\{M_x\}_{x\in\ca{X}}$ such that
\alg{
\sum_{x\in\ca{X}}p_x{\rm Tr}[M_x\rho_x]
\geq
1-\epsilon.
\laeq{caution}
}
Then, for the state
\alg{
\rho^{XA}:=\sum_{x\in\ca{X}}p_x\proj{x}^X\otm\rho_x^A,
}
it holds that
\alg{
H(X|A)_\rho\leq\eta(\epsilon)\log{|\ca{X}|},
}
where $\eta$ is a function that satisfies $\lim_{\epsilon\rightarrow0}\eta(\epsilon)=0$ and is independent of the dimensions of the systems.
\elmm

\bprf
Let $\hat{X}$ be a quantum system with a fixed orthonormal basis $\{\ket{x}\}_{x\in\ca{X}}$, and define a CPTP map $\ca{M}:A\rightarrow \hat{X}$ by $\ca{M}(\cdot):={\rm Tr}[M_x(\cdot)]\proj{x}^{\hat{X}}$.
It follows that
\alg{
\tilde{\rho}^{X\hat{X}}
&:={\rm id}^X\otm\ca{M}(\rho^{XA})
\\
&=
\sum_{x,x'\in\ca{X}}p_x\proj{x}^X\otm p_{x'|x}\proj{x'}^{\hat{X}},
}
where $\{p_{x'|x}\}_{x'\in\ca{X}}$ is a conditional probability distribution defined by $p_{x'|x}={\rm Tr}[M_{x'}\rho_x]$.
Thus, due to the monotonicity of the conditional quantum entropy, we have
\alg{
H(X|A)_\rho
\leq
H(X|\hat{X})_{\tilde{\rho}}.
\laeq{kaikai}
}
With a slight abuse of notation, let $(X,\hat{X})$ be a pair of random variables that takes values in $\ca{X}\times\ca{X}$ according to a joint probability distribution $P\equiv\{p(x,\hat{x})\}$, where $p(x,\hat{x}):=p_xp_{x'|x}$.
Due to the condition \req{caution}, it holds that
\alg{
\!
P(X\neq\hat{X})
=
\sum_{\substack{x,x'\in\ca{X}\\x\neq x'}}p(x,\hat{x})
=
\sum_{x\in\ca{X}}p_x(1-p_{x|x})
\leq
\epsilon.
\!
}
Thus, due to Fano's inequality (see e.g.~Theorem 2.10.1 in \cite{cover05}), it follows that
\alg{
H(X|\hat{X})_{\tilde{\rho}}
=
H(X|\hat{X})_{P}
\leq
\eta(\epsilon)\log{|\ca{X}|}.
}
Combining this with \req{kaikai}, we complete the proof.
\QED
\eprf

\section{Method of Types and Type Subspaces}
\lapp{typesub}

In this section, we briefly review the definitions and properties of types and type subspaces.
For the details, see e.g. Section 13.7 and 14.3 in \cite{wildetext}.
The properties of type subspaces presented in this section will be used in \rApp{AEiid} to prove the equivalence between the achievable rate region for the asymptotic limit given by \rThm{RRegionIID1} in the main text and the one obtained in Ref.~\cite{hsieh2010entanglement}

Let $\ca{X}$ be a finite alphabet and $n\in\mbb{N}$.
A probability distribution $\{t(x)\}_{x\in\ca{X}}$ is called a {\it type of length $n$} if $nt(x)\in\mbb{N}$ for all $x\in\ca{X}$.
Let $x^n\equiv x_1\cdots x_n$ be a sequence of variables of length $n$ such that $x_i\in\ca{X}$ for each $i$. 
The sequence $x^n$ is said to be of type $t$ if
\alg{
\frac{1}{n}N(x|x^n)=t(x)
}
for all $x\in\ca{X}$, where  $N(x|x^n)$ is the number of the symbol $x\in\ca{X}$ that appears in the sequence $x^n$.
The {\it type class} corresponding to the type $t$ of length $n$, which we denote by $T_t^n$, is the set of all sequences whose type is $t$.
Let $\mathfrak{T}(\ca{X},n)$ be the set of all types of length $n$ on the alphabet $\ca{X}$.
It holds that
\alg{
T_t^n\cap T_{t'}^n=\emptyset\;(t\neq t'),
\quad
\ca{X}^n=\bigcup_{t\in\mathfrak{T}(\ca{X},n)}T_t^n.
\laeq{kaii}
}
The size of $\mathfrak{T}(\ca{X},n)$ is bounded above by a polynomial function of $n$ as
\alg{
|\mathfrak{T}(\ca{X},n)|
\leq (n+1)^{|\ca{X}|}.
\laeq{shinju}
}
By definition, any two sequences in the same type class are transformed with each other by permuting the elements. That is, for any $t\in\mathfrak{T}(\ca{X},n)$ and $x^n,x'^n\in T_t^n$, there exists a permutation $s$ such that $x_i'=x_{s(i)}$ for any $1\leq i\leq n$.

Let $\ca{H}$ be a Hilbert space with a fixed orthonormal basis $\{\ket{x}\}_{x\in\ca{X}}$.
For any $n\in\mbb{N}$, the {\it type subspace} corresponding to the type $t$ of length $n$ is defined by 
\alg{
\ca{H}_t^n:={\rm span}\{\ket{x^n} :  x^n\in T_t^n\}
\subseteq\ca{H}^{\otm n},
}
where $\ket{x^n}=\ket{x_1}\cdots\ket{x_n}$,
and the {\it type projector} is defined by
\alg{
\Pi_t^n:=\sum_{ x^n\in T_t^n}\proj{x^n}.
}
It follows from \req{kaii} that
\alg{
\Pi_t^n\Pi_{t'}^n=0\;(t\neq t'),
\quad
I=\sum_{t\in\mathfrak{T}(\ca{X},n)}\Pi_t^n.
}
Consider a state $\rho\in\ca{S}_=(\ca{H})$ and suppose that the eigen decomposition of $\rho$ is given by $\rho=\sum_{x\in\ca{X}}p_x\proj{x}$. 
By definition, it holds that
\alg{
\Pi_t^n\rho^{\otm n}\Pi_t^n
=
\sum_{ x^n\in T_t^n}p_{x^n}\proj{x^n}
=
q_t\Pi_t^n,
}
where
\alg{
q_t
:=
\prod_{x\in\ca{X}}p_x^{nt(x)}.
}
The projectors $\{\Pi_t^n\}_{t\in\mathfrak{T}(\ca{X},n)}$ are called the {\it type projectors corresponding to the state $\rho^{\otm n}$.}

Consider an ensemble $\{p_j,\rho_j\}_{j=1}^J$, where $J\in\mbb{N}$, $\rho_j\in\ca{S}_=(\ca{H})$, and fix arbitrary $n\in\mbb{N}$.
For any type $t$ of length $n$ over $[J]:=\{1,\cdots,J\}$, define a sequence of length $n$ by
\begin{eqnarray}
{\bm j}_t
:=
\underbrace{11\cdots1}_{nt(1)}
\underbrace{22\cdots2}_{nt(2)}
\cdots 
\underbrace{JJ\cdots J}_{nt(J)}.
\laeq{dfnbfjt}
\end{eqnarray}
For each $j\in[J]$, 
let $\{\Pi_{\nu_j}\}_{\nu_j\in\mathfrak{T}(\ca{X},nt(j))}$ be the set of type projectors corresponding to the state ${\rho_j}^{\otm nt(j)}$. 
We define a projector on ${\mathcal H}^{\otimes n}$ by
\alg{
\Pi_{\vec{\nu}|{\bm j}_t}
:=
\Pi_{\nu_1}\otm\cdots\otm\Pi_{\nu_J}
}
for each $\vec{\nu}:=\nu_1\cdots\nu_J\in\bigotimes_{j=1}^J\mathfrak{T}(\ca{X},nt_j)$.
Using \req{shinju}, the number of conditional type projectors is bounded above by
\alg{
\prod_{j=1}^J
|\mathfrak{T}(\ca{X},nt(j))|
\leq
\prod_{j=1}^J
(nt(j)+1)^{|\ca{X}|}
\leq
(n+1)^{2{\rm dim}\ca{H}}.
}

As mentioned above, for any $t\in\mathfrak{T}(J,n)$ and ${\bm j}\in T_t^n$, there exists a permutation $s$ such that $j_i=j_{t,s(i)}$ for each $1\leq i\leq n$,
where $j_{t,s(i)}$ is the $s(i)$-th element of the sequence ${\bm j}_t$ defined by \req{dfnbfjt}.
Let $P_s$ be a unitary that acts on $\ca{H}^{\otm n}$ as
\alg{
P_s(\ket{\varphi_1}\otm\cdots\otm\ket{\varphi_n})
=
\ket{\varphi_{s(1)}}\otm\cdots\otm\ket{\varphi_{s(n)}}.
}
We define the set of {\it conditional type projectors} $\{\Pi_{\vec{\nu}|{\bm j}}\}_{\vec{\nu}}$ on ${\mathcal H}^{\otimes n}$ by
\alg{
\Pi_{\vec{\nu}|{\bm j}}
=
P_s\Pi_{\vec{\nu}|{\bm j}_t}P_s^\dagger.
}
By definition, it holds that
\alg{
{\rm Tr}[\Pi_{\vec{\nu}|{\bm j}}\rho_{{\bm j}}]
=
{\rm Tr}[P_s\Pi_{\vec{\nu}|{\bm j}}P_s^\dagger P_s\rho_{{\bm j}}P_s^\dagger]
=
{\rm Tr}[\Pi_{\vec{\nu}|{\bm j}_t}\rho_{{\bm j}_t}].
}
Note that the number of conditional type projectors for the sequences ${\bm j}$ is equal to that for ${\bm j}'$ if the two sequences belong to the same type class.

\section{Proof of \rPrp{SfeqR2}}
\lapp{AEiid}

In this appendix, we prove \rPrp{SfeqR2} in \rSec{iid}.
For the simplicity of notations, we denote $d_{S_c}$ by $J$.

\subsection{Proof of $\overline{\Theta_{\pi}^{\infty}(\ca{N})}=\overline{\Theta^{\infty}(\ca{N})}$}
\lapp{ageha}

By definition, it is straightforward to verify that $\overline{\Theta_{\pi}^{\infty}(\ca{N})}\subseteq\overline{\Theta^{\infty}(\ca{N})}$.
Thus, 
it suffices to prove the converse relation $\overline{\Theta_{\pi}^{\infty}(\ca{N})}\supseteq\overline{\Theta^{\infty}(\ca{N})}$.
We prove this by showing that
$\overline{\Theta_{\pi}^{\infty}(\ca{N})}\supseteq\frac{1}{n}\Theta(\ca{N}^{\otm n},\rho)$ for any $n$ and any state $\rho$
in the form of
\alg{
\rho^{SA^n}
=
\sum_{j=1}^Jp_j\proj{j}^{S_c}\otm\rho_j^{S_rA^n},
}
where we do not require that $\rho^{S}=\pi^{S}$.
We only consider the case where $n=1$. It is straightforward to generalize the proof for $n\geq2$.

\subsubsection{Construction of States}

Fix an arbitrary $\epsilon>0$ and choose sufficiently large $m$.
Let $\mathfrak{T}(J,m)$ be the set of all types of length $m$ over $[J]:=\{1,\cdots,J\}$, and $T_t^n\subset [J]^{\times m}$ be the type class corresponding to a type $t\in\mathfrak{T}(J,m)$ (see \rApp{typesub} for the definitions and properties of types and type subspaces). 
For any $t\in\mathfrak{T}(J,m)$ and $\vec{j}\in T_t^n$, 
let $\{\Pi_{\varsigma|\vec{j}}\}_{\varsigma=1}^{\theta(\vec{j})}$
be the set of conditional type projectors on $(\ca{H}^{S_r})^{\otm m}$ with respect to $\rho_{\vec{j}}^{S_r^m}$.
Here, $\theta(\vec{j})$ is the number of the conditional type subspaces.
Note that $\theta(\vec{j})=\theta(\vec{j}')$ if $\vec{j}$ and $\vec{j}'$ belong to the same type class.
Thus, we will denote $\theta(\vec{j})$ also as $\theta(t)$ if $\vec{j}\in T_t^n$. 
Define probability distributions $\{p_t\}_{t\in\mathfrak{T}(J,m)}$ and $\{p_{\varsigma|\vec{j}}\}_{\varsigma=1}^{\theta(\vec{j})}$ for each $\vec{j}\in J^{\times m}$ by
\alg{
p_t:=\sum_{\vec{j}\in T_t^n}p_{\vec{j}},
\quad
p_{\varsigma|\vec{j}}
:=
{\rm Tr}[\Pi_{\varsigma|\vec{j}}\rho_{\vec{j}}^{S_r^m}].
} 
Due to the properties of the conditional type projectors, it holds that $p_{\varsigma|\vec{j}}=p_{\varsigma|\vec{j}'}$ for any $\vec{j}$ and $\vec{j}'$ in the same type set $t$, which we denote by $p_{\varsigma|t}$. 
We define a probability distribution $\{p_{\varsigma,t}\}_{\varsigma\in[\theta(t)],t\in\mathfrak{T}(J,m)}$ by 
\alg{
p_{\varsigma,t}=p_{t}\cdot p_{\varsigma|t}.
\laeq{rrt}
}

Let $Y$ and $Y'$ be quantum systems with dimensions $|\mathfrak{T}(J,m)|$ and $\theta_*:=\max_{t\in\mathfrak{T}(J,m)}|\theta(t)|$, respectively.
Consider states
\alg{
\rho_{\varsigma,\vec{j}}^{S_r^mA^m}
&
:=
p_{\varsigma|\vec{j}}^{-1}\Pi_{\varsigma|\vec{j}}^{S_r^m}\rho_{\vec{j}}^{S_r^mA^m}\Pi_{\varsigma|\vec{j}}^{S_r^m},
\laeq{cofeemilk}\\
\rho_{\varsigma,t}^{S^mA^m}
&
:=
\frac{1}{| T_t^n|}
\sum_{\vec{j}\in T_t^n}\proj{\vec{j}}^{S_c^m}
\otm
\rho_{\varsigma,\vec{j}}^{S_r^mA^m}
\laeq{110}
}
and define
\alg{
&
\rho_m^{YY'S^mA^m}
:=
\sum_{t\in\mathfrak{T}(J,m)}
p_t
\proj{t}^Y
\otm
\frac{1}{| T_t^n|}
\sum_{\vec{j}\in T_t^n}\proj{\vec{j}}^{S_c^m}
\nn\\
&
\quad\quad\quad\quad\quad\quad\quad
\otm
\sum_{\varsigma=1}^{\theta(\vec{j})}p_{\varsigma|\vec{j}}
\proj{\varsigma}^{Y'}\otm
\rho_{\varsigma,\vec{j}}^{S_r^mA^m}.
\laeq{111}
}
By the definition of the type subspaces, it is straightforward to verify that $\rho_{\varsigma,t}^{S^m}$ has a flat distribution on its support, that is,
\alg{
\rho_{\varsigma,t}^{S^m}
=
\frac{1}{| T_t^n|}
\sum_{\vec{j}\in T_t^n}\proj{\vec{j}}^{S_c^m}
\otm
\frac{\Pi_{\varsigma,\vec{j}}^{S_r^m}}{{\rm Tr}[\Pi_{\varsigma,\vec{j}}]}.
\laeq{flDs}
}
By using \req{rrt} and \req{110}, the state \req{111} is rewritten into
\alg{
\rho_m^{YY'S^mA^m}
=
\sum_{t\in\mathfrak{T}(J,m)}\sum_{\varsigma=1}^{\theta(t)}
p_{\varsigma,t}\proj{\varsigma,t}^{YY'}
\otm
\rho_{\varsigma,t}^{S^mA^m}.
\nn
}
For the later convenience, we introduce a map $\ca{E}_{\vec{j}}:S_r^m\rightarrow Y'S_r^m$ by 
\alg{
\ca{E}_{\vec{j}}(\cdot)=
\sum_{\varsigma=1}^{\theta(\vec{j})}
\proj{\varsigma}^{Y'}
\otm
\Pi_{\varsigma|\vec{j}}^{S_r^m}(\cdot)\Pi_{\varsigma|\vec{j}}^{S_r^m}
}
for each $\vec{j}$, which leads to
\alg{
\sum_{\varsigma=1}^{\theta(\vec{j})}p_{\varsigma|\vec{j}}
\proj{\varsigma}^{Y'}\otm
\rho_{\varsigma,\vec{j}}^{S_r^mA^m}
=
\ca{E}_{\vec{j}}(\rho_{\vec{j}}^{S_r^mA^m}).
\laeq{yasa}
}
A useful relation which follows from the properties of the conditional type projectors is that for any $\vec{j}$ and $\vec{j}'$ in the same type set, there exits a permutation $s$ such that
\alg{
\rho_{\varsigma,\vec{j}'}^{S_r^m}
=
\ca{P}_s^{S_r^m}
(\rho_{\varsigma,\vec{j}}^{S_r^m})
\laeq{koori}
}
and
\alg{
&
(\ca{N}^{A\rightarrow B})^{\otm m}(\rho_{\varsigma,\vec{j}'}^{S_r^mA^m})
\nn\\
&
\quad\quad
=
\ca{P}_s^{S_r^m}
\otm
\ca{P}_s^{B^m}
\circ(\ca{N}^{A\rightarrow B})^{\otm m}
(\rho_{\varsigma,\vec{j}}^{S_r^mA^m}).
\laeq{mizu}
}
These properties will be used in the following subsections to calculate the entropies of the states.

\subsubsection{Calculation of Entropies}

Let us calculate entropies and mutual informations of the state defined above.
We use the definition of the state $\rho_m$ given by \req{111} and the fact that $(\rho^{S_cA})^{\otm m}=\rho_m^{S_c^mA^m}$.
Due to the properties of the quantum mutual information, we have
\alg{
&
mI(S_c:B)_{\ca{N}(\rho)}
\\
&
=
I(S_c^m:B^m)_{\ca{N}^{\otm m}(\rho^{\otm m})}
\\
&=
I(S_c^m:B^m)_{\ca{N}^{\otm m}(\rho_m)}
\\
&\leq
I(YY'S_c^m:B^m)_{\ca{N}^{\otm m}(\rho_m)}
\\
&
=
I(YY':B^m)_{\ca{N}^{\otm m}(\rho_m)}
\nn\\
&\quad\quad\quad\quad
+
I(S_c^m:B^m|YY')_{\ca{N}^{\otm m}(\rho_m)}
\\
&\leq
I(S_c^m:B^m|YY')_{\ca{N}^{\otm m}(\rho_m)}
+
H(YY')
\\
&
\leq
\sum_{t\in\mathfrak{T}(J,m)}
\sum_{\varsigma=1}^{\theta(t)}
p_{\varsigma,t}I(S_c^m:B^m)_{\ca{N}^{\otm m}(\rho_{\varsigma,t})}
\nn\\
&\quad\quad\quad\quad\quad\quad\quad\quad\quad\quad\quad\quad
+\log{\theta_*|\mathfrak{T}(J,m)|}.
\laeq{noritake1}
}
We also have
\alg{
&
mH(S_r|BS_c)_{\ca{N}(\rho)}
\\
&
=
H(S_r^m|B^mS_c^m)_{\ca{N}^{\otm m}(\rho^{\otm m})}
\\
&
=
\sum_{\vec{j}\in[J]^{\times m}}p_{\vec{j}}
H(S_r^m|B^m)_{\ca{N}^{\otm m}(\rho_{\vec{j}})}
\\
&
\geq
\sum_{\vec{j}\in[J]^{\times m}}p_{\vec{j}}
[
H(S_r^m|B^mY')_{\ca{N}^{\otm m}\otm\ca{E}_j(\rho_{\vec{j}})}-\log{\theta(\vec{j})}
]
\laeq{norimaki2}\\
&
\geq
\sum_{\vec{j}\in[J]^{\times m}}p_{\vec{j}}
H(S_r^m|B^mY')_{\ca{N}^{\otm m}\otm\ca{E}_j(\rho_{\vec{j}})}
-\log{\theta_*},
\laeq{street2}
}
where \req{norimaki2} follows from \rLmm{entpinbo} in \rApp{PSE}.
Using \req{yasa} and \req{110},
the first term in \req{street2} is calculated as
\alg{
&
\sum_{\vec{j}\in[J]^{\times m}}p_{\vec{j}}
H(S_r^m|B^mY')_{\ca{N}^{\otm m}\otm\ca{E}_j(\rho_{\vec{j}})}
\\
&
=
\sum_{\vec{j}\in[J]^{\times m}}p_{\vec{j}}
\sum_{\varsigma=1}^{\theta(\vec{j})}p_{\varsigma|\vec{j}}
H(S_r^m|B^m)_{\ca{N}^{\otm m}(\rho_{\varsigma,\vec{j}})}
\\
&
=
\sum_{t\in\mathfrak{T}(J,m)}
\sum_{\vec{j}\in T_t^n}
\frac{p_t}{| T_t^n|}
\sum_{\varsigma=1}^{\theta(t)}p_{\varsigma|t}
H(S_r^m|B^m)_{\ca{N}^{\otm m}(\rho_{\varsigma,\vec{j}})}
\\
&
=
\sum_{t\in\mathfrak{T}(J,m)}
\sum_{\varsigma=1}^{\theta(t)}p_{\varsigma,t}
\!\cdot\!
\frac{1}{| T_t^n|}
\sum_{\vec{j}\in T_t^n}
H(S_r^m|B^m)_{\ca{N}^{\otm m}(\rho_{\varsigma,\vec{j}})}
\\
&
=
\sum_{t\in\mathfrak{T}(J,m)}
\sum_{\varsigma=1}^{\theta(t)}p_{\varsigma,t}
\cdot
H(S_r^m|B^mS_c^m)_{\ca{N}^{\otm m}(\rho_{\varsigma,t})}.
\laeq{kagami}
}
Here, the last line follows from the fact that the entropies of the state $\ca{N}^{\otm m}(\rho_{\varsigma,\vec{j}})$ depends only on $\varsigma$ and the type of $\vec{j}$, because of the local unitary equivalence \req{mizu}.
Similarly, we have
\alg{
&
mH(S_r|S_c)_\rho
=
H(S_r^m|S_c^m)_{\rho^{\otm m}}
=
\sum_{\vec{j}\in[J]^{\times m}}p_{\vec{j}}H(S_r^m)_{\rho_{\vec{j}}}
\nn\\
&
\leq
\sum_{\vec{j}\in[J]^{\times m}}p_{\vec{j}}
H(S_r^mY')_{\ca{E}_{\vec{j}}(\rho_{\vec{j}})}
\laeq{norimaki}\\
&
=
\sum_{\vec{j}\in[J]^{\times m}}p_{\vec{j}}
[H(Y')_{\ca{E}_{\vec{j}}(\rho_{\vec{j}})}
+
H(S_r^m|Y')_{\ca{E}_{\vec{j}}(\rho_{\vec{j}})}
]
\\
&
\leq
\sum_{\vec{j}\in[J]^{\times m}}p_{\vec{j}}
[\log{\theta(\vec{j})}
+
H(S_r^m|Y')_{\ca{E}_{\vec{j}}(\rho_{\vec{j}})}
]
\\
&
\leq
\sum_{\vec{j}\in[J]^{\times m}}p_{\vec{j}}
H(S_r^m|Y')_{\ca{E}_{\vec{j}}(\rho_{\vec{j}})}
+
\log{\theta_*}
\\
&
=
\sum_{t\in\mathfrak{T}(J,m)}
\sum_{\varsigma=1}^{\theta(t)}p_{\varsigma,t}
\cdot
H(S_r^m|S_c^m)_{\rho_{\varsigma,t}}
+
\log{\theta_*}.
\laeq{kagami2}
}
The second line follows from \rLmm{entpinbo} in \rApp{PSE}, and the last line from the similar argument as in \req{kagami}, for which we use the local unitary equivalence \req{koori}.
The cardinalities of the type sets $\mathfrak{T}(J,m)$ and $\theta_*$ are bounded from above by
\alg{
|\mathfrak{T}(J,m)|
\leq
(m+1)^J,
\quad
\theta_*
\leq
(m+1)^{2d_A}.
\laeq{punisher}
}

Consider an arbitrary inner point $(C,Q,E)$ of $\Theta(\ca{N},\rho)$ and choose sufficiently large $m$.
Combining Inequalities \req{noritake1}, \req{street2}, \req{kagami}, \req{kagami2} and \req{punisher} with the conditions \req{proogo1}-\req{proogo3},
 it follows that
\alg{
Q+E
&
\leq
\frac{1}{m}\sum_{\varsigma,t}p_{\varsigma,t}H(S_r^m|S_c^m)_{\rho_{\varsigma,t}},
\\
C+Q-E
&\leq
\frac{1}{m}\sum_{\varsigma,t}p_{\varsigma,t}
[H(S_c^m)_{\rho_{\varsigma,t}}
\nn\\
&\quad\quad\quad\quad\quad\quad
-
H(S^m|B^m)_{\ca{N}^{\otm m}(\rho_{\varsigma,t})}
],
\laeq{kore}\\
Q-E
&\leq
-\frac{1}{m}\sum_{\varsigma,t}p_{\varsigma,t}H(S_r^m|B^mS_c^m)_{\rho_{\varsigma,t}}.
}
Here, we have used the fact that
the chain rule of quantum enrtopies implies 
\alg{
&
I(S_c:B)_{\ca{N}(\rho)}-H(S_r|BS_c)_{\ca{N}(\rho)}
\nn\\
&\quad
=
H(S_c)_{\rho}-H(S|B)_{\ca{N}(\rho)}
}
and
\alg{
&
I(S_c^m:B^m)_{\ca{N}^{\otm m}(\rho_{\varsigma,t})}
-
H(S_r^m|B^mS_c^m)_{\ca{N}^{\otm m}(\rho_{\varsigma,t})}
\nn\\
&=
H(S_c^m)_{\rho_{\varsigma,t}}-H(S^m|B^m)_{\ca{N}^{\otm m}(\rho_{\varsigma,t})}.
}
Thus, we arrive at
\alg{
(C,Q,E)
&
\in
\sum_{\varsigma,t}p_{\varsigma,t}\cdot\frac{1}{m}\Theta_{\pi}(\ca{N}^{\otm m},\rho_{\varsigma,t})
\\
&
\subset
{\rm conv}\frac{1}{m}\Theta_{\pi}(\ca{N}^{\otm m})
\\
&
\subset
{\rm conv}\Theta_{\pi}^{\infty}(\ca{N})
\subseteq
\overline{\Theta_{\pi}^{\infty}(\ca{N})},
}
where the last line follows from the convexity of $\overline{\Theta_{\pi}^{\infty}(\ca{N})}$ (see the next subsection).
This completes the proof of $\frac{1}{n}\Theta(\ca{N}^{\otm n},\rho)\subseteq\overline{\Theta_{\pi}^{\infty}(\ca{N})}$ for $n=1$.
The proofs for $n\geq2$ are obtained along the same line.
\QED

\begin{figure*}[t]
\centerline{
\subfigure[]{\includegraphics[bb={0 30 557 486},scale=0.28]{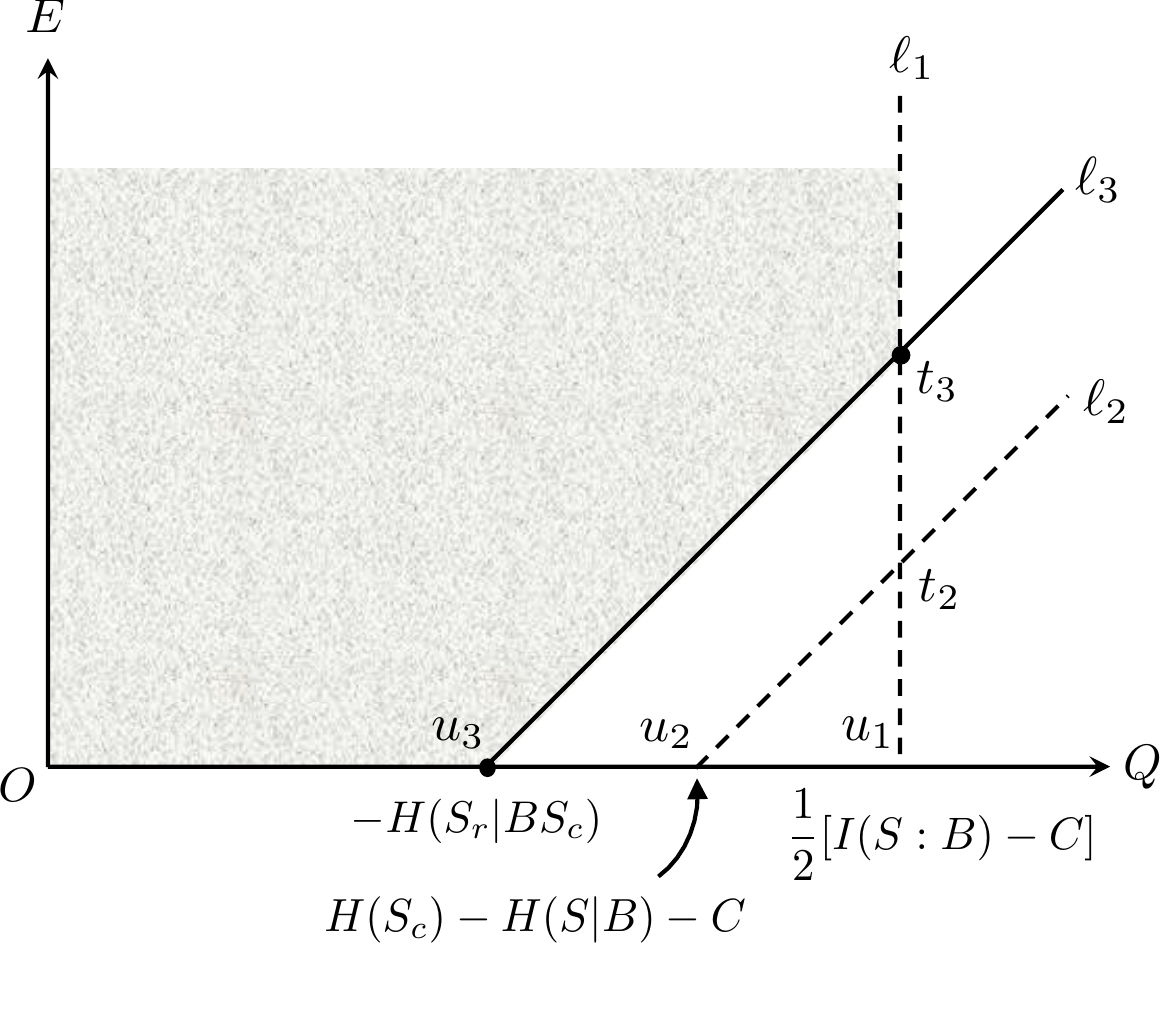}}
\quad
\subfigure[]{\includegraphics[bb={0 32 557 486},scale=0.28]{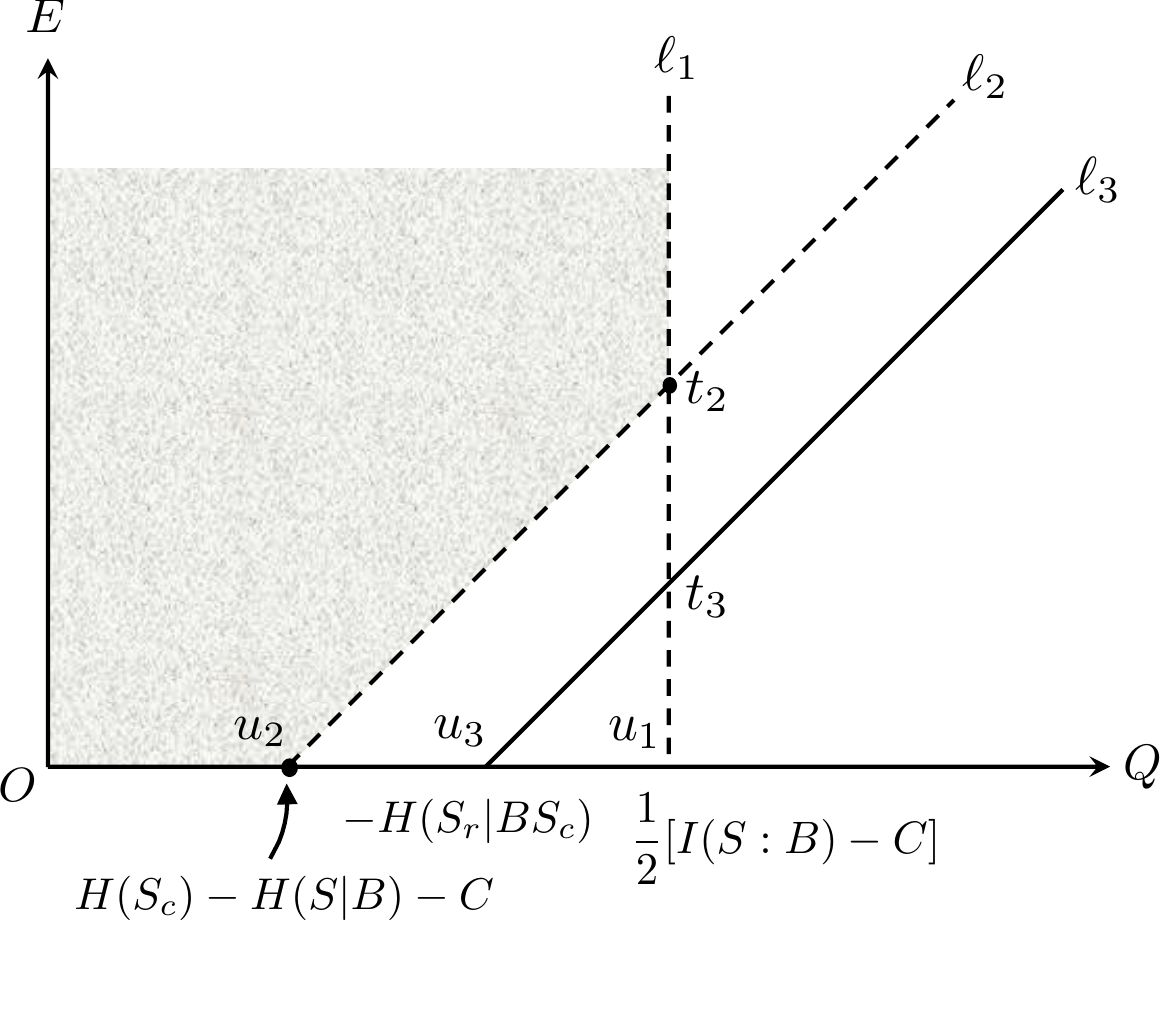}}
\quad
\subfigure[]{\includegraphics[bb={0 0 591 456},scale=0.28]{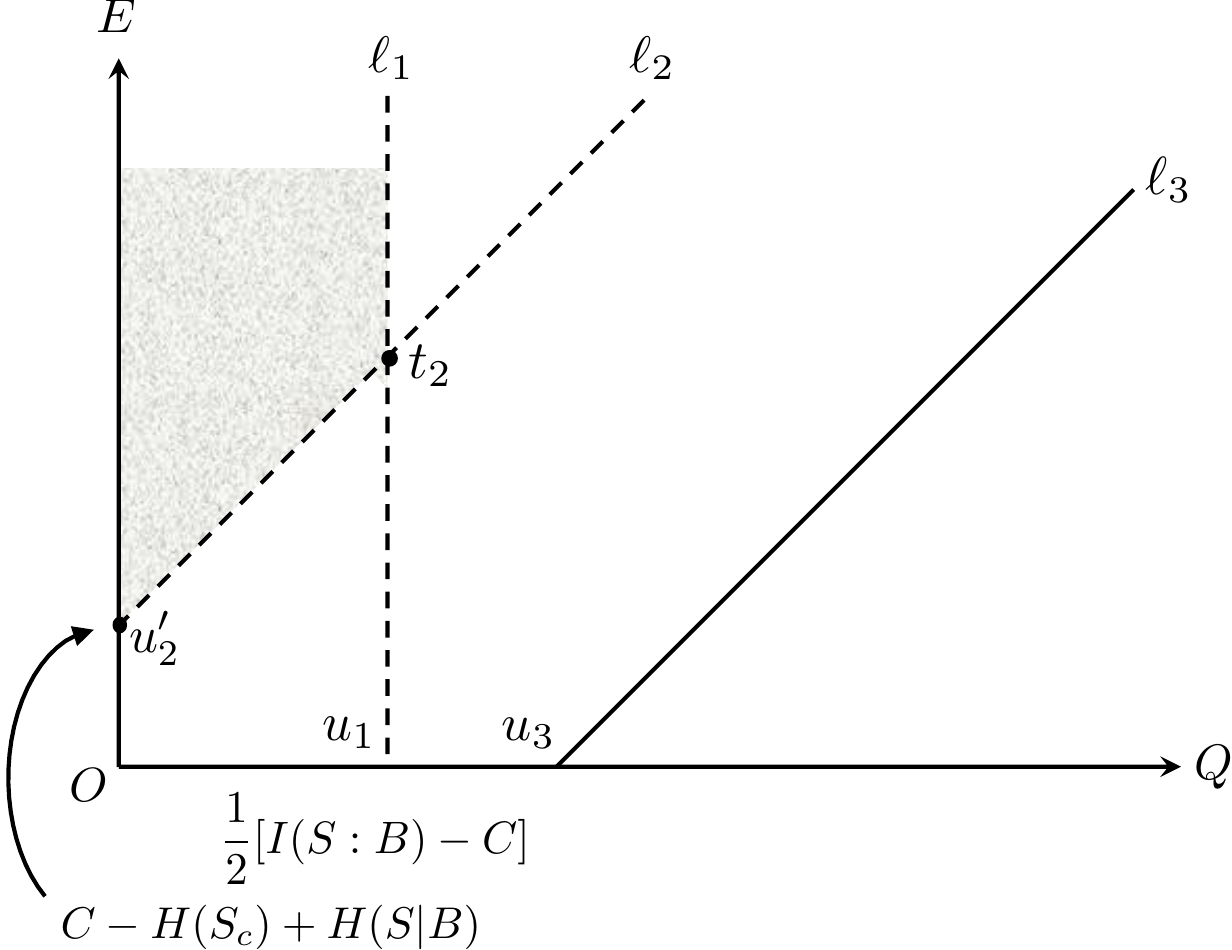}}
}
\caption{
The two dimensional regions of $(Q,E)$ satisfying inequalities \req{proogo21}-\req{proogo23} in the case of $-H(S_r|BS_c)_{\ca{N}(\rho)}>0$ are depicted.
The figures (a), (b) and (c) are for $0\leq C\leq I(S_c:B)$, $I(S_c:B)\leq C\leq I(S_c:B)-H(S_r|BS_c)$ and $I(S_c:B)-H(S_r|BS_c)\leq C\leq I(S:B)$, respectively.
Lines $\ell_1$, $\ell_2$ and $\ell_3$ represent the boundaries represented by inequalities \req{proogo21}-\req{proogo23}. 
The points of intersection of the three lines and the axes are denoted by $u_1$, $u_2$, $u_3$, $t_2$, $t_2'$ and $t_3$. 
At $C=0$, both $u_1$ and $u_2$ are on the right than $u_3$.
Thus $u_3$ and $t_3$ are vertices of the region that yield $P_1^+$ and $P_2$, respectively.
The two points $u_1$ and $u_2$ approaches to the origin as $C$ increases.
At $C=I(S_c:B)$, the point $u_2$ coincides $u_3$, in which case the points $u_2=u_3$ and $t_2=t_3$ are the vertices $P_3^+$ and $P_4$.
In $C=H(S_c)-H(S|B)$, the point $u_2$ meets the origin, and yields $P_5$.
Finally, the point $u_1$ coincides the origin at $C=I(S:B)$,  in which case the point of $u_2'=t_2$ yields $P_6$.
}
\label{fig:rateregion1}
\end{figure*}

\begin{figure}[h]
\centerline{
\subfigure[]{\includegraphics[bb={0 -10 422 623},scale=0.28]{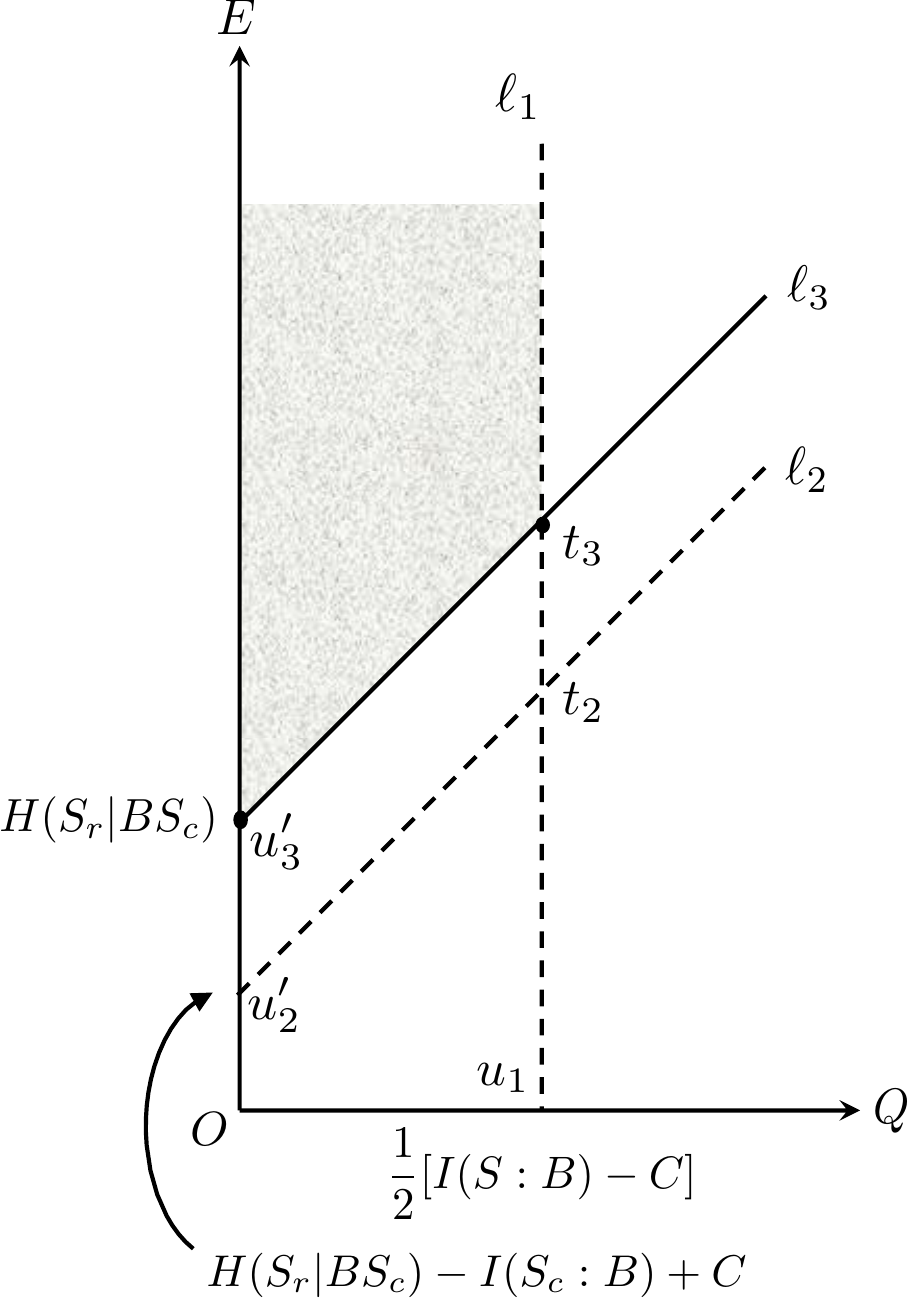}}
\quad
\subfigure[]{\includegraphics[bb={0 -10 380 623},scale=0.28]{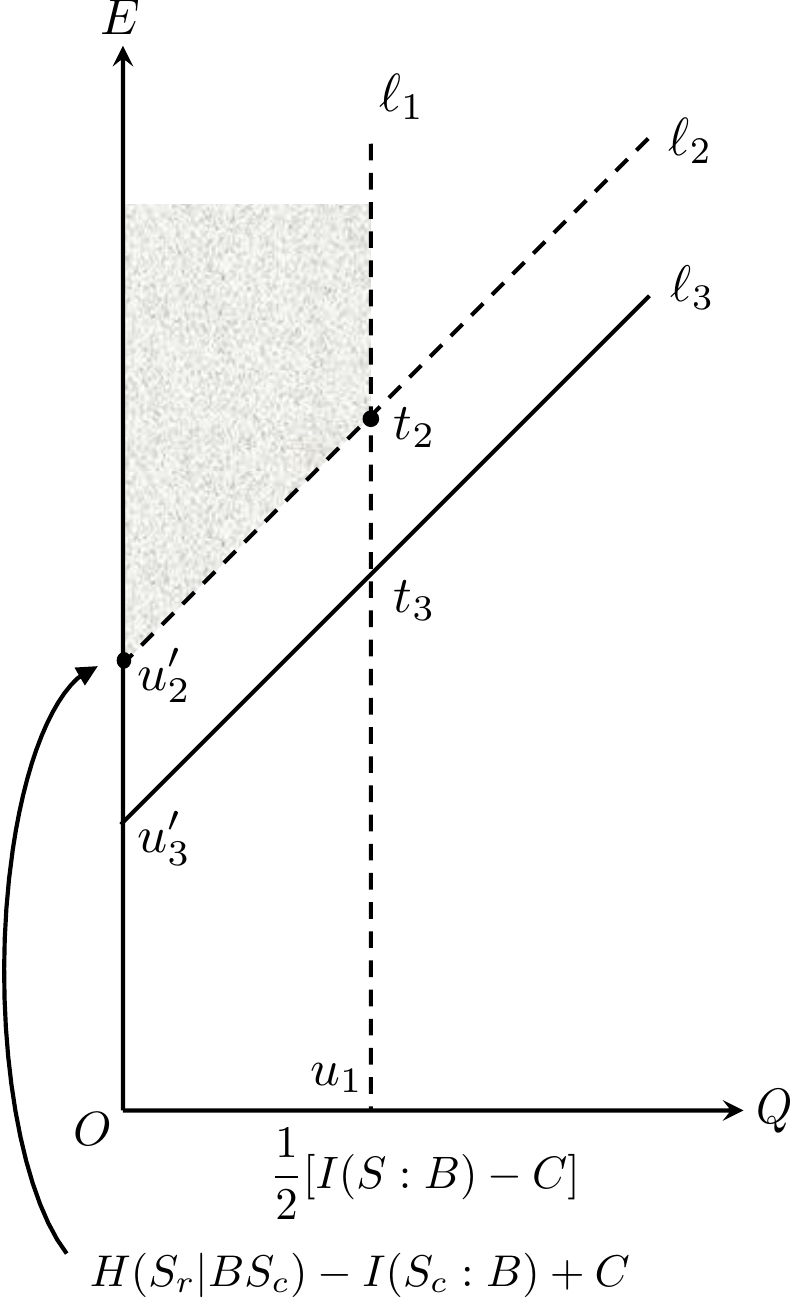}}
}
\caption{
The two dimensional regions of $(Q,E)$ satisfying inequalities \req{proogo21}-\req{proogo23} in the case of $-H(S_r|BS_c)_{\ca{N}(\rho)}<0$ are depicted.
The figure (a) is for $0\leq C\leq I(S_c\!:\!B)$, and (b) is for $I(S_c\!:\!B)\leq C\leq I(S\!:\!B)$.
Lines $\ell_1$, $\ell_2$ and $\ell_3$ represent the boundaries represented by inequalities \req{proogo21}-\req{proogo23}, respectively, and the crossing points are denoted by $u_1$,  $u_2'$, $u_3'$, $t_2$ and $t_3$.
At $C=0$, the point $u_2'$ is below $u_3'$,
in which case $u_3'$ and $t_3$ are vertices of the region that yield $P_1^-$ and $P_2$, respectively.
The point $u_2'$ approaches $u_3'$ as $C$ increases, and coincides it at $C=I(S_c:B)$.
In this case, the points $u_2'=u_3'$ and $t_2=t_3$ correspond to the vertices $P_3^-$ and $P_4$.
At $C=I(S:B)$,  the point $u_1$ coincides the origin, where the point $u_2'=t_2$ yields $P_6$.
}
\label{fig:rateregion2}
\end{figure}

\subsubsection{Proof of ${\rm conv}\Theta_{\pi}^{\infty}(\ca{N})\subseteq\overline{\Theta_{\pi}^{\infty}(\ca{N})}$}

We prove that the convex hull of $\Theta_{\pi}^{\infty}(\ca{N})$ is a subset of $\overline{\Theta_{\pi}^{\infty}(\ca{N})}$.
Fix arbitrary $\lambda_1,\lambda_2>0$ such that $\lambda_1+\lambda_2=1$, and suppose that $(C^{(i)},Q^{(i)},E^{(i)})\in\Theta_{\pi}^{\infty}(\ca{N})$ for $i=1,2$.
By definition, for any sufficiently large $n$, there exist quantum systems $S^{(i)}\equiv S_c^{(i)}S_r^{(i)}$ and a quantum state $\rho_i$ on $S^{(i)}C^{n\lambda_i}$ such that $\rho_i^{S^{(i)}}$ is the maximally mixed state, and it holds that
\alg{
n_i(Q^{(i)}\!+\!E^{(i)})
&\leq
 H(S_r^{(i)}|S_c^{(i)})_{\rho_i},
\laeq{hashigo}\\
n_i(C^{(i)}\!+\!Q^{(i)}\!-\!E^{(i)})
&
\leq
H(S_c^{(i)})_{\rho_i}
\nn\\
&\quad
-H(S^{(i)}|B^{n\lambda_i})_{\ca{N}^{\otm n_i}(\rho_i)},
\\
n_i(Q^{(i)}\!-\!E^{(i)})
&\leq
 -H(S_r^{(i)}|B^{n\lambda_i}S_c^{(i)})_{\ca{N}^{\otm n_i}(\rho_i)}
\laeq{hashigo2}
}
for $i=1,2$, where $n_1:=\lfloor n\lambda_1\rfloor$ and $n_2:=\lceil n\lambda_i\rceil$.
Define $S_c\equiv S_c^{(1)}S_c^{(2)}$, $S_r \equiv  S_r^{(1)}S_r^{(2)}$, $S\equiv S_cS_r$ and consider a quantum state $\bar{\rho}$ on $SA^n$ defined by
\alg{
\bar{\rho}^{\:SA^n}:=
\rho_1^{\:S^{(1)}\!A^{n_1}}
\otm
\rho_2^{\:S^{(2)}\!A^{n_2}}.
}
It is straightforward to verify that the state is diagonal on $S_c$ with respect to a fixed basis, and that $\bar{\rho}^{S}$ is the full-rank maximally mixed state on $S$. 
Due to the additivity of the conditional quantum entropy, we have
\alg{
H(S_r|S_c)_{\bar{\rho}}
=
H(S_r^{(1)}|S_c^{(1)})_{\rho_1}
+
H(S_r^{(2)}|S_c^{(2)})_{\rho_2}
}
and so forth. 
Define 
\alg{
(C_n,Q_n,E_n)\equiv\sum_{i=1,2}n_i(C^{(i)},Q^{(i)},E^{(i)}).
}
It follows from \req{hashigo}-\req{hashigo2} that
\alg{
Q_n+E_n
&\leq
 H(S_r|S_c)_{\bar{\rho}}
\\
C_n+Q_n-E_n
&
\leq
H(S_c)_{\bar{\rho}}-H(S|B^n)_{\ca{N}^{\otm n}(\bar{\rho})}
\\
Q_n-E_n
&\leq
 -H(S_r|B^nS_c)_{\ca{N}^{\otm n}(\bar{\rho})},
}
which implies $(C_n,Q_n,E_n)\in\Theta_{\pi}(\ca{N}^{\otm n})$.
Furthermore, it is straightforward to verify that
\alg{
\lim_{n\rightarrow\infty}\frac{1}{n}(C_n,Q_n,E_n)
&
=
(\bar{C},\bar{Q},\bar{E})
\\
&
:=
\sum_{i=1,2}\lambda_i(C^{(i)},Q^{(i)},E^{(i)}).
}
This implies $(\bar{C},\bar{Q},\bar{E})\in\overline{\Theta_{\pi}^\infty(\ca{N})}$, and completes the proof.
\QED

\subsection{Proof of $\overline{\Theta^{\infty}(\ca{N})}=\overline{\Lambda^{\infty}(\ca{N})}$}

Suppose that a triplet $(C,Q,E)$ belongs to $\Theta(\ca{N},\rho)$, which is defined by Inequalities \req{proogo1}-\req{proogo4}. Noting that $\rho^{S}=\ca{N}(\rho)^{S}$, Inequalities \req{proogo1} and \req{proogo2} implies \req{proogo21}.
Thus, we have $\Theta(\ca{N},\rho)\subseteq\Lambda(\ca{N},\rho)$, which leads to $\overline{\Theta^{\infty}(\ca{N})}\subseteq\overline{\Lambda^{\infty}(\ca{N})}$.

To prove the converse relation $\overline{\Theta^{\infty}(\ca{N})}\supseteq\overline{\Lambda^{\infty}(\ca{N})}$, we show that $\Theta^{\infty}(\ca{N})\supseteq\Lambda(\ca{N},\rho)$ for any state $\rho$. 
Note that $\Lambda(\ca{N},\rho)$ is a convex polytope such that $(C,Q,E+\Delta E)\in\Lambda(\ca{N},\rho)$ for any $(C,Q,E)\in\Lambda(\ca{N},\rho)$ and any $\Delta E>0$ (see \rDfn{rateSN2}).
Thus, it suffices to prove that (i) all vertices of $\Lambda(\ca{N},\rho)$ belongs to $\Theta^{\infty}(\ca{N})$, and that (ii) if $(C,Q,E)\in\Theta^{\infty}(\ca{N})$, then $(C,Q,E+\Delta E)\in\Theta^{\infty}(\ca{N})$ for any $\Delta E>0$.

\subsubsection{Vertices of $\Lambda(\ca{N},\rho)$}

Consider the following points in $\mbb{R}^3$,
where all entropies and mutual informations are for the state $\ca{N}(\rho)$:
\alg{
&P_0:=(0,\;0,\;0)
\nn\\
&
P_1^+:=(0,\;-H(S_r|BS_c),\;0)
\nn\\
&
P_1^-:=(0,\;0,\;H(S_r|BS_c))
\nn\\
&
P_2:=\left(0,\;\frac{1}{2}I(S:B),
\frac{1}{2}I(S\!:\!B)+H(S_r|BS_c)\right)
\nn\\
&
P_3^+:=\left(I(S_c:B),\;-H(S_r|BS_c),\;0\right)
\nn\\
&
P_3^-:=\left(I(S_c:B),\;0,\;H(S_r|BS_c)\right)
\nn\\
&
P_4:=\left(I(S_c\!:\!B),\frac{1}{2}I(S_r\!:\!D|S_c),
\frac{1}{2}I(S_r\!:\!D|S_c)\!+\!H(S_r|BS_c)\right)
\nn\\
&
P_5:=\left(H(S_c)-H(S|B),\;0,\;0\right)
\nn\\
&
P_6:=\left(I(S\!:\!B),\;0,\;H(S_r|S_c)\right)
\nn
}
The vertices of $\Lambda(\ca{N},\rho)$ are $P_0,P_1^+,P_2,P_3^+,P_4,P_5,P_6$ in the case of $-H(S_r|BS_c)_{\ca{N}(\rho)}>0$ and $P_1^-,P_2,P_3^-,P_4,P_6$ when $-H(S_r|BS_c)_{\ca{N}(\rho)}<0$ (Figures \ref{fig:rateregion1} and \ref{fig:rateregion2}: see also Section VI in \cite{hsieh2010entanglement}).
Note that, by the chain rule of the mutual information, it holds that
\alg{
&
H(S_c)_\rho
-
H(S|B)_{\ca{N}(\rho)}
\nn\\
&
=
I(S_c:B)_{\ca{N}(\rho)}-H(S_r|BS_c)_{\ca{N}(\rho)}
\\
&
=
I(S:B)_{\ca{N}(\rho)}-H(S_r|S_c)_{\ca{N}(\rho)}.
}
By a simple calculation, it is straightforward to verify that all of the above points except $P_2$ belong to $\Theta(\ca{N},\rho)$, and consequently to $\Theta^{\infty}(\ca{N})$.

\subsubsection{Proof of $P_2\in\Theta^{\infty}(\ca{N})$}

Consider the point $P_2$ represented by the coordinate $(C_2,Q_2,E_2)$, where
\alg{
&
C_2=0,
\quad
Q_2=\frac{1}{2}I(S:B)_{\ca{N}(\rho)},
\\
&
E_2=
\frac{1}{2}I(S:B)_{\ca{N}(\rho)}+H(S_r|S_cB)_{\ca{N}(\rho)}.
}
A simple calculation yields
\alg{
Q_2+E_2
&=H(S_r|S_c)_{\ca{N}(\rho)}+I(S_c:B)_{\ca{N}(\rho)},
\laeq{trop1}\\
C_2+Q_2-E_2
&=
-H(S_r|S_cB)_{\ca{N}(\rho)},
\laeq{trop2}\\
Q_2-E_2
&=
-H(S_r|S_cB)_{\ca{N}(\rho)}.
\laeq{trop3}
}
Fix arbitrary $\epsilon,\delta>0$ and choose sufficiently large $n$. 
Due to the data compression theorem for classical information source with quantum side information (Theorem 1 in \cite{devetak2003classical}),
 there exist a countable set $\ca{Y}_{n,\delta}$ satisfying
 \alg{
|\ca{Y}_{n,\delta}|\leq 2^{n(H(S_c|B)_{\ca{N}(\rho)}+\delta)},
}
a function $f:[J]^{\times n}\rightarrow \ca{Y}_{n,\delta}$ and 
 for each $y\in\ca{Y}_{n,\delta}$, there exists a measurement $\{M_{\vec{j}}^y\}_{\vec{j}\in[J]^{\times n}}$ on $B^n$  that satisfies
\alg{
\sum_{\vec{j}\in[J]^{\times n}}
p_{\vec{j}}
{\rm Tr}\left[
M_{\vec{j}}^y(\ca{N}^{A\rightarrow B})^{\otm n}(\rho_{\vec{j}}^{A^n})
\right]
\geq
1-\epsilon.
\laeq{rect}
}
We introduce a $|\ca{Y}_{n,\delta}|$-dimensional quantum system $Y$ and define a state
\alg{
\rho_n^{YS^nA^n}
:=
\sum_{\vec{j}\in [J]^{\times n}}p_{\vec{j}}
\proj{f(\vec{j})}^Y
\otm
\proj{\vec{j}}^{S_c^n}
\otm
\rho_{\vec{j}}^{S_r^nA^n}.
\nn
}
We denote the system $S_c^nS_r^n$ by $\hat{S}_r$.
It is straightforward to verify that
\alg{
\rho_n^{YS^nA^n}
=
(\rho^{SA})^{\otm n}.
\laeq{ican}
}

Using the properties of quantum entropies and \req{ican}, we have
\alg{
&
H(\hat{S}_r|Y)_{\rho_n}
\nn\\
&
=
H(S_c^nS_r^n|Y)_{\rho_n}
\\
&
=
H(S_c^nY)_{\rho_n}
-
H(Y)_{\rho_n}
+
H(S_r^n|S_c^nY)_{\rho_n}
\\
&=
H(S_c^n)_{\rho_n}
-
H(Y)_{\rho_n}
+
H(S_r^n|S_c^n)_{\rho_n}
\\
&\geq
H(S_c^n)_{\rho_n}
-
|\ca{Y}_{n,\delta}|
+
H(S_r^n|S_c^n)_{\rho_n}
\\
&=
nH(S_c)_{\rho}
-
|\ca{Y}_{n,\delta}|
+
nH(S_r|S_c)_{\rho}
\\
&
\geq
nI(S_c:B)_{\ca{N}(\rho)}
+
nH(S_r|S_c)_{\rho}-n\delta,
}
where $\eta$ is a function that satisfies $\lim_{\epsilon\rightarrow0}\eta(\epsilon)=0$ and is independent of the dimensions of the systems.
From \req{rect}, \req{ican} and \rLmm{fano} in \rApp{PSE}, we also have
\alg{
&
H(\hat{S}_r|YB^n)_{\ca{N}^{\otm n}(\rho_n)}
\nn\\
&=
H(S_c^nS_r^n|YB^n)_{\ca{N}^{\otm n}(\rho_n)}
\\
&
=
H(S_c^n|YB^n)_{\ca{N}^{\otm n}(\rho_n)}
+
H(S_r^n|S_c^nYB^n)_{\ca{N}^{\otm n}(\rho_n)}
\\
&
=
H(S_c^n|YB^n)_{\ca{N}^{\otm n}(\rho_n)}
+
H(S_r^n|S_c^nB^n)_{\ca{N}^{\otm n}(\rho_n)}
\\
&
=
H(S_c^n|YB^n)_{\ca{N}^{\otm n}(\rho_n)}
+
nH(S_r|S_cB)_{\ca{N}(\rho)}
\\
&
\leq
nH(S_r|S_cB)_{\ca{N}(\rho)}
+
2n\eta(\epsilon)\log{d_{S_c}}.
}
In addition, a simple calculation using the chain rule yields
\alg{
&
H(Y)_{\rho_n}
-
H(Y\hat{S}_r|B^n)_{\ca{N}^{\otm n}(\rho_n)}
\nn\\
&=
-H(\hat{S}_r|YB^n)_{\ca{N}^{\otm n}(\rho_n)}
+
I(Y:B^n)_{\ca{N}^{\otm n}(\rho_n)}
\\
&\geq
-H(\hat{S}_r|YB^n)_{\ca{N}^{\otm n}(\rho_n)}.
}
Combining these relations with \req{trop1}-\req{trop3}, we arrive at
\alg{
Q_2+E_2
&\leq
\frac{1}{n}H(\hat{S}_r|Y)_{\ca{N}^{\otm n}(\rho_n)}
+
\delta,
\\
C_2+Q_2-E_2
&\leq
\frac{1}{n}[H(Y)_{\rho_n}
-H(Y\hat{S}_r|B^n)_{\ca{N}^{\otm n}(\rho_n)}]
\nn\\
&
\quad\quad\quad\quad\quad\quad\quad\quad\quad
+
2\eta(\epsilon)\log{d_{S_c}},
\\
Q_2-E_2
&\leq
-\frac{1}{n}H(\hat{S}_r|YB^n)_{\ca{N}^{\otm n}(\rho_n)}
\nn\\
&
\quad\quad\quad\quad\quad\quad\quad\quad
+
2\eta(\epsilon)\log{d_{S_c}}.
}
Since these relations hold for any small $\epsilon,\delta>0$ and sufficiently large $n$,
we obtain  $P_2\in\Theta^{\infty}(\ca{N})$.
\QED

\subsubsection{Proof of $(C,Q,E+\Delta E)\in\Theta^{\infty}(\ca{N})$ for any $\Delta E>0$ and $(C,Q,E)\in\Theta^{\infty}(\ca{N})$}

We complete the proof of $\Theta^{\infty}(\ca{N})\supseteq\Lambda(\ca{N},\rho)$ by showing that, 
if $(C,Q,E)\in\Theta^{\infty}(\ca{N})$, then $(C,Q,E+\Delta E)\in\Theta^{\infty}(\ca{N})$ for any $\Delta E>0$.
It suffices to prove that for any $n\in\mbb{N}$ and $\Delta E>0$, it holds that $(C,Q,E+\Delta E)\in\Theta^{\infty}(\ca{N})$ if $(C,Q,E)\in\frac{1}{n}\Theta(\ca{N}^{\otm n})$.
We only consider the case where $n=1$.
It is straightforward to generalize the proof for $n\geq2$.

Consider a triplet $(C,Q,E)\in\Theta(\ca{N})$, and fix arbitrary $\Delta E>0$ and $m\in\mbb{N}$.
By definition, there exist finite dimensional quantum systems $S_c$, $S_r$ and a state in the form of
\alg{
\rho^{SA}=\sum_{j=1}^Jp_j\proj{j}^{S_c}\otm\rho_j^{S_rA},
}
such that
\alg{
Q+E
&\leq H(S_r|S_c)_\rho,
\laeq{prooga1}\\
C+Q-E
&
\leq
H(S_c)_\rho-H(S|B)_{\ca{N}(\rho)},
\laeq{prooga2}\\
Q-E
&\leq -H(S_r|BS_c)_{\ca{N}(\rho)}.
\laeq{prooga3}
}
Define $\Delta E_m:=\lfloor m\Delta E\rfloor$, and
let $S_r'$ be a quantum system with dimension $2^{\Delta E_m}$.
Consider a state
\alg{
\rho_m^{S^mS_r'A^m}=(\rho^{SA})^{\otm m}\otm\pi^{S_r'},
}
where $\pi$ is the full-rank maximally mixed state on $S_r'$.
Relabelling $S_c^m$ by $\hat{S}_c$, $S_r^mS_r'$ by $\hat{S}_r$ and $\hat{S}_c\hat{S}_r$ by $\hat{S}$,
the above state is represented as
\alg{
\rho_m^{\hat{S}A^m}
:=
\sum_{\vec{j}\in[J]^{\times m}}p_{\vec{j}}\proj{\vec{j}}^{\hat{S}_c}\otm\hat{\rho}_{\vec{j}}^{\hat{S}_rA^m},
}
where
\alg{
\hat{\rho}_{\vec{j}}^{\hat{S}_rA^m}
:=
\rho_{\vec{j}}^{S_r^mA^m}\otm\pi^{S_r'}
}
and
\alg{
p_{\vec{j}}
:=
p_{j_1}\cdots p_{j_m},
\quad
\rho_{\vec{j}}
:=
\rho_{j_1}\otm\cdots\otm\rho_{j_m}
}
for $\vec{j}=j_1\cdots j_m$.
Noting that
\alg{
H(\hat{S}_r|\hat{S}_c)_{\rho_m}
=
mH(S_r|S_c)_\rho
+
\Delta E_m
}
 and so forth, it follows from \req{prooga1}-\req{prooga3} that
\alg{
m(Q+E)+\Delta E_m
&\leq
 H(\hat{S}_r|\hat{S}_c)_{\rho_m},
\\
m(C+Q-E)-\Delta E_m
&
\leq
H(\hat{S}_c)_\rho
\nn\\
&\quad\quad
-H(\hat{S}|B^m)_{\ca{N}^{\otm m}(\rho_m)},
\\
m(Q-E)-\Delta E_m
&\leq
 -H(\hat{S}_r|B^m\hat{S}_c)_{\ca{N}^{\otm m}(\rho_m)}.
}
This implies $(C,Q,E+\Delta E_m/m)\in\Theta^{\infty}(\ca{N})$.
Noting that $\lim_{m\rightarrow\infty}(\Delta E_m/m)=\Delta E$, this implies $(C,Q,E+\Delta E)\in\overline{\Theta^{\infty}(\ca{N})}$ and completes the proof.
\QED

\subsection{Proof of $\overline{\Lambda^{\infty}(\ca{N})}=\overline{\Lambda_p^{\infty}(\ca{N})}$}

It is straightforward to verify that $\overline{\Lambda^{\infty}(\ca{N})}\supseteq\overline{\Lambda_p^{\infty}(\ca{N})}$. Thus, we prove $\overline{\Lambda^{\infty}(\ca{N})}\subseteq\overline{\Lambda_p^{\infty}(\ca{N})}$ by showing that $\Lambda(\ca{N}^{\otm n})\subseteq\Lambda_p(\ca{N}^{\otm n})$ for any $n$. We only consider the case $n=1$. It is straightforward to generalize the proof for $n\geq2$.

Fix an arbitrary state $\rho$ in the form of \req{proogostar}, and suppose that $(C,Q,E)\in\Lambda(\ca{N},\rho)$. For each $j$, let $\{q_{k|j},\ket{\phi_{j,k}}\}_k$ be an ensemble of pure states on $S_rA$ such that $\rho_j=\sum_kq_{k|j}\proj{\phi_{j,k}}$. 
We denote $p_jq_{k|j}$ by $p_{jk}$.
Let $Y$ be a finite dimensional quantum system with a fixed orthonormal basis $\{\ket{k}\}_k$, 
and define a state $\tilde{\rho}^{YSA}$ by
\alg{
\tilde{\rho}^{YSA}
:=
\sum_{j=1}^J\sum_kp_{jk}\proj{k}^Y\otm\proj{j}^{S_c}\otm\proj{\phi_{j,k}}^{S_rA}.
\nn
}
It is straightforward to verify that $\tilde{\rho}^{SA}=\rho^{SA}$.
Denoting $YS_c$ by $\hat{S}_c$ and $\hat{S}_cS_r$ by $\hat{S}$, the data processing inequality yields
\alg{
I(S:B)_{\ca{N}(\rho)}
&
\leq
I(\hat{S}:B)_{\ca{N}(\tilde{\rho})},
\\
-H(S_r|BS_c)_{\ca{N}(\rho)}
&
\leq
-H(S_r|B\hat{S}_c)_{\ca{N}(\tilde{\rho})},
}
in addition to
\alg{
&
H(S_c)_\rho-H(S|B)_{\ca{N}(\rho)}
\nn\\
&=
I(S_c:B)_{\ca{N}(\rho)}-H(S_r|BS_c)_{\ca{N}(\rho)}
\\
&\leq
I(\hat{S}_c:B)_{\ca{N}(\tilde{\rho})}-H(S_r|B\hat{S}_c)_{\ca{N}(\tilde{\rho})}
\\
&
=
H(\hat{S}_c)_{\tilde{\rho}}-H(\hat{S}|B)_{\ca{N}(\tilde{\rho})}.
}
Combining these inequalities with \req{proogo21}-\req{proogo23},
we have $(C,Q,E)\in\Lambda(\ca{N},\tilde{\rho})$, which implies $\Lambda(\ca{N},\rho)\subseteq\Lambda(\ca{N},\tilde{\rho})$.
By taking the union over all $\rho$, we arrive at $\Lambda(\ca{N})\subseteq\Lambda_p(\ca{N})$ and complete the proof.
\QED

\bibliographystyle{IEEEtran}
\bibliography{/Users/eyuriwakakuwa/Dropbox/DropTop/latexfiles/bibbib.bib}

%

\end{document}